\documentclass[12pt,english]{article}
\usepackage[T1]{fontenc}
\usepackage[latin9]{inputenc}
\usepackage{refstyle}
\usepackage{rotfloat}
\usepackage{amsmath}
\usepackage{graphicx}
\usepackage[numbers,super,sort&compress]{natbib}

\makeatletter

\AtBeginDocument{\providecommand\eqref[1]{\ref{eq:#1}}}
\AtBeginDocument{\providecommand\figref[1]{\ref{fig:#1}}}
\AtBeginDocument{\providecommand\enuref[1]{\ref{enu:#1}}}
\AtBeginDocument{\providecommand\secref[1]{\ref{sec:#1}}}
\AtBeginDocument{\providecommand\tabref[1]{\ref{tab:#1}}}
\providecommand{\tabularnewline}{\\}
\RS@ifundefined{subsecref}
  {\newref{subsec}{name = \RSsectxt}}
  {}
\RS@ifundefined{thmref}
  {\def\RSthmtxt{theorem~}\newref{thm}{name = \RSthmtxt}}
  {}
\RS@ifundefined{lemref}
  {\def\RSlemtxt{lemma~}\newref{lem}{name = \RSlemtxt}}
  {}

\newcommand{\lyxaddress}[1]{
	\par {\raggedright #1
	\vspace{1.4em}
	\noindent\par}
}
\newcommand{\lyxrightaddress}[1]{
	\par {\raggedleft \begin{tabular}{l}\ignorespaces
	#1
	\end{tabular}
	\vspace{1.4em}
	\par}
}

\usepackage{braket}
\usepackage[version=3]{mhchem}
\usepackage{url}
\usepackage{notes2bib}

\@ifundefined{showcaptionsetup}{}{%
 \PassOptionsToPackage{caption=false}{subfig}}
\usepackage{subfig}
\makeatother

\usepackage{babel}
\begin{document}
\title{Fully numerical Hartree--Fock and density functional calculations.
I. Atoms}
\author{Susi Lehtola}
\maketitle

\lyxaddress{Department of Chemistry, University of Helsinki, P.O. Box 55 (A.
I. Virtasen aukio 1), FI-00014 University of Helsinki, Finland}

\lyxrightaddress{susi.lehtola@alumni.helsinki.fi}
\begin{abstract}
Although many programs have been published for fully numerical Hartree--Fock
(HF) or density functional (DF) calculations on atoms, we are not
aware of any that support hybrid DFs, which are popular within the
quantum chemistry community due to their better accuracy for many
applications, or that can be used to calculate electric properties.
Here, we present a variational atomic finite element solver in the
\textsc{HelFEM} program suite that overcomes these limitations. A
basis set of the type $\chi_{nlm}(r,\theta,\phi)=r^{-1}B_{n}(r)Y_{l}^{m}(\hat{\boldsymbol{r}})$
is used, where $B_{n}(r)$ are finite element shape functions and
$Y_{l}^{m}$ are spherical harmonics, which allows for an arbitrary
level of accuracy.

\textsc{HelFEM} supports nonrelativistic HF and DF including hybrid
functionals, which are not available in other commonly available program
packages. Hundreds of functionals at the local spin density approximation
(LDA), generalized gradient approximation (GGA), as well as the meta-GGA
levels of theory are included through an interface with the \textsc{Libxc}
library. Electric response properties are achievable via finite field
calculations.

We introduce an alternative grid that yields faster convergence to
the complete basis set than commonly used alternatives. We also show
that high-order Lagrange interpolating polynomials yield the best
convergence, and that excellent agreement with literature HF limit
values for electric properties, such as static dipole polarizabilities,
can be achieved with the present approach. Dipole moments and dipole
polarizabilities at finite field are reported with the PBE, PBEh,
TPSS, and TPSSh functionals. Finally, we show that a recently published
Gaussian basis set is able to reproduce absolute HF and DF energies
of neutral atoms, cations, as well as anions within a few dozen microhartrees. 
\end{abstract}
\global\long\def\ERI#1#2{(#1|#2)}%
\global\long\def\bra#1{\Bra{#1}}%
\global\long\def\ket#1{\Ket{#1}}%
\global\long\def\braket#1{\Braket{#1}}%

\newcommand*\citeref[1]{ref. \citenum{#1}}
\newcommand*\citerefs[1]{refs. \citenum{#1}} 

\section{Introduction\label{sec:Introduction}}

Thanks to decades of development in approximate functionals and computational
approaches, density functional theory (DFT) \citep{Hohenberg1964,Kohn1965}
has become the main workhorse of computational chemistry,\citep{Becke2014,Jones2015,Mardirossian2017a}
with several new density functionals still being published every year.
As atoms are the basic building block of molecules, the first test
of a new density functional often is -- and should be -- its performance
on atoms at the complete basis set (CBS) limit. However, although
multiple programs are available for fully numerical Hartree--Fock
(HF) and post-HF calculations on atoms, as we have recently reviewed
elsewhere,\citep{Lehtola2019c} the situation is not as good for DFT.

As far as we know, there are no publicly available programs for performing
fully numerical DFT calculations on atoms with hybrid and/or meta-GGA
functionals, especially in the presence of an electric field. This
is a problem, since the ability to generate accurate data on atoms
for new density functionals would greatly facilitate their assessment
and development. Accurate atomic calculations may also have other
uses: many density functionals have been fitted, fully or in part,
to \emph{ab initio }data on atoms. In addition, we have shown recently
that fully numerical atomic density functional calculations can be
used to fashion accurate and efficient initial guesses for self-consistent
field (SCF) calculations on molecules, even if the molecular calculations
are done with Gaussian basis sets.\citep{Lehtola2019}

In the present work, we will describe the implementation of an atomic
finite element solver for HF and DFT calculations, also with hybrid
and meta-GGA functionals. The program called \textsc{HelFEM},\citep{HelFEM}
where the first part stands both for the electronic Hamiltonian 
\begin{equation}
\hat{H}_{el}=-\frac{1}{2}\sum_{i}\nabla_{i}^{2}-\sum_{i}\frac{Z}{r_{i}}+\sum_{i>j}\frac{1}{r_{ij}}\label{eq:atH}
\end{equation}
as well as the city and university of Helsinki where the present author
is situated, is open source (GNU General Public License), is written
in object-oriented C++, and takes advantage of a number of recently
published open source algorithms and libraries for its capabilities.
Most importantly, \textsc{HelFEM} is interfaced with the \textsc{Libxc}
library\citep{Lehtola2018} that offers access to hundreds of exchange-correlation
functionals published in the literature\textsc{.} \textsc{HelFEM}
supports pure and hybrid\citep{Becke1993a} density functionals at
the local spin-density approximation\citep{Kohn1965} (LDA), generalized-gradient
approximation\citep{Langreth1980} (GGA) as well as meta-GGA\citep{Perdew1999}
levels of theory. Range-separation is not supported in \textsc{HelFEM}
at present due to reasons that will become obvious later in the manuscript.
The orbitals can be fully spin-restricted, spin-restricted open-shell,
or fully spin-unrestricted.

The data layout in \textsc{HelFEM} is deliberately similar to what
is used in typical quantum chemistry programs employing Gaussian basis
sets. The rationale for this is the following. First, if one wants
to use the program to study symmetry breaking effects in HF and DFT,
the program cannot employ symmetries, meaning that the basis set must
explicitly span all angular degrees of freedom. Second, although the
basis set is local, the exchange matrix is dense because the HF exchange
interaction is non-local. Furthermore, as evaluations of the total
energy require access to all of the elements of the density and exchange
matrices, this means that the full density and Fock matrices will
anyhow be necessary. Third, by the use of full, dense matrices alike
Gaussian-basis programs, many functionalities, such as the DIIS\citep{Pulay1980,Pulay1982}
and ADIIS\citep{Hu2010} SCF convergence accelerators can be adopted
directly from the \textsc{Erkale} program.\citep{erkale,Lehtola2012}
Furthermore, as many powerful open-source quantum chemistry programs
have recently become available, interfaces to \emph{e.g.} \textsc{Psi4}\citep{Parrish2017}
or \textsc{PySCF}\citep{Sun2018} could be implemented in the future
for post-HF treatments, including multiconfigurational methods, configuration
interaction, and coupled-cluster theories, thanks to the easy data
interface.

We present two applications of the novel code. The first application
is the calculation of atoms in finite electric fields. Finite electric
field calculations allow, for instance, the extraction of atomic static
dipole polarizabilities, which are a well-known challenge for theoretical
methods\citep{Schwerdtfeger2006} and the best values for which have
been recently reviewed by Schwerdtfeger and Nagle.\citep{Schwerdtfeger2018}
Atomic static dipole polarizabilities are related to global softness
and the Fukui function.\citep{Vela1990} As the molecule with the
lowest static dipole polarizability tends to be the chemically most
stable,\citep{Ghanty1996,Hohm2000,Gomez2003} the accuracy of static
dipole polarizabilities can be considered a proxy for thermochemical
accuracy. Various density functionals have been shown to outperform
HF for molecular static dipole polarizabilities with hybrid functionals
yielding the best results,\citep{Calaminici1998,Salek2005,Suponitsky2008,Hait2018,Fuentealba1997}
as the error in polarizabilities typically arises from the exchange
part.\citep{Fuentealba1997} Fully numerical all-electron HF results
for atoms\citep{Voegel1979,Stiehler1995,Koch2011,Koch2013} and density
functional results for molecules\citep{Dickson1996} have been reported
in the literature, whereas post-HF and relativistic DFT results have
been calculated using Gaussian basis sets.\citep{Woon1994,Soldan2001,Bast2008,Parmar2013,Parmar2014}
In our application, we study the \ce{Li+} and \ce{Sr^{2+}} ions
with HF and show that we are able to reproduce the fully numerical
HF limit values from \citeref{Kobus2015}. In addition, we report
dipole moments and polarizabilities with the LDA,\citep{Bloch1929,Dirac1930,Perdew1992a}
PBE,\citep{Perdew1996,Perdew1997} PBEh,\citep{Adamo1999,Ernzerhof1999}
TPSS,\citep{Tao2003,Perdew2004} and TPSSh\citep{Staroverov2003}
functionals.

Our second application is the benchmark of Gaussian basis set energies
for a variety of neutral, cationic, and anionic species with HF and
the BHHLYP\citep{Becke1993a} functional. Atomic anions are especially
challenging to model with DFT.\citep{Galbraith1996,Rosch1997,Jarecki1999,Jensen2010a,Anderson2017}
For instance, it has been shown that calculations on the well-bound
\ce{F-} anion may require extremely diffuse basis functions with
exponents as small as(!) $\alpha=6.9\times10^{-9}$ to achieve converged
results.\citep{Jarecki1999} The use of such small exponents requires
extensive modifications to the used Gaussian-basis quantum chemistry
program to ensure sufficient numerical accuracy.\citep{Jarecki1999,Anderson2017}
In contrast, the finite element method has none of these issues: because
the basis set has local support and is never ill-conditioned, calculations
are extremely stable numerically. We will show below that the absolute
energies reproduced by the large Gaussian basis set used in \citerefs{Jensen2010, Anderson2017}
are too large by several microhartrees for most systems. The second
part of the present series presents analogous applications to diatomic
molecules, where the deficiencies of Gaussian basis sets are considerably
more noticeable.\citep{Lehtola2019b}

The layout of the article is the following. Next, in the Theory section,
we provide a brief presentation of the finite element method as it
is unfamiliar to most quantum chemists as well as summarize the variational
approach, and then proceed with the calculation of various matrix
elements that are necessary for HF and DFT. The Theory section is
followed by a Computational details section, which describes the present
implementation and details various convergence parameters that were
used for the calculations. Then, the Results section begins with extensive
studies of the convergence properties of the finite-element expansion
for HF calculations on the noble elements, and presents applications
of the program to electric properties, and to the study of the accuracy
of Gaussian basis set calculations at the HF, LDA, GGA, and meta-GGA
levels of theory, including hybrid functionals. The article ends with
a brief Summary and Conclusions section. Atomic units are used, unless
specified otherwise. The Einstein summation convention is employed,
meaning implied summations over repeated indices.

\section{Theory\label{sec:Theory}}

\subsection{Finite elements\label{subsec:Finite-elements}}

As the finite element method is not well known in computational chemistry
-- to our best knowledge only one book exists on the application
of the method to quantum mechanics at an accessible level\citep{Ram-Mohan2002}
-- we will briefly describe the one-dimensional finite element method,
which is used here and in the second part of the series.\citep{Lehtola2019b} 

\begin{figure}
\begin{centering}
\subfloat[Two-node\label{fig:L2}]{\begin{centering}
\includegraphics[width=0.33\textwidth]{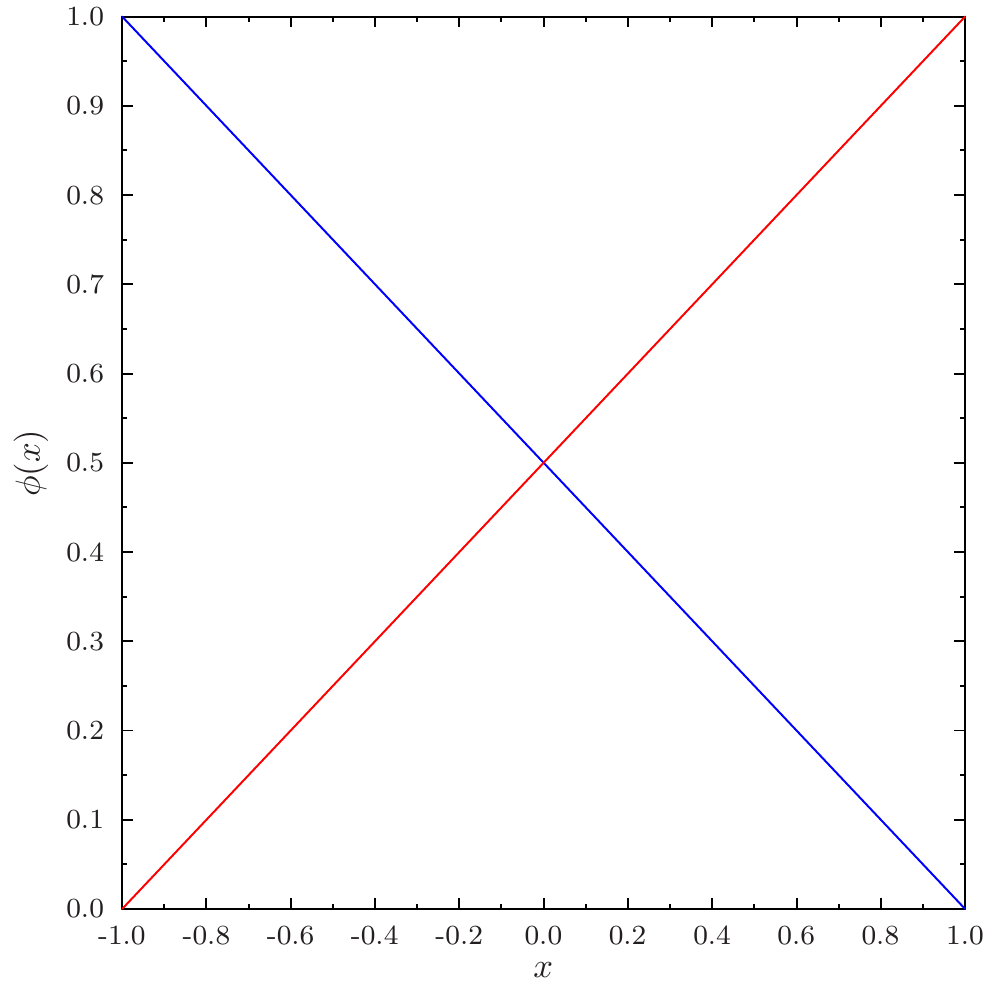}
\par\end{centering}
}\subfloat[Three-node\label{fig:L3}]{\begin{centering}
\includegraphics[width=0.33\textwidth]{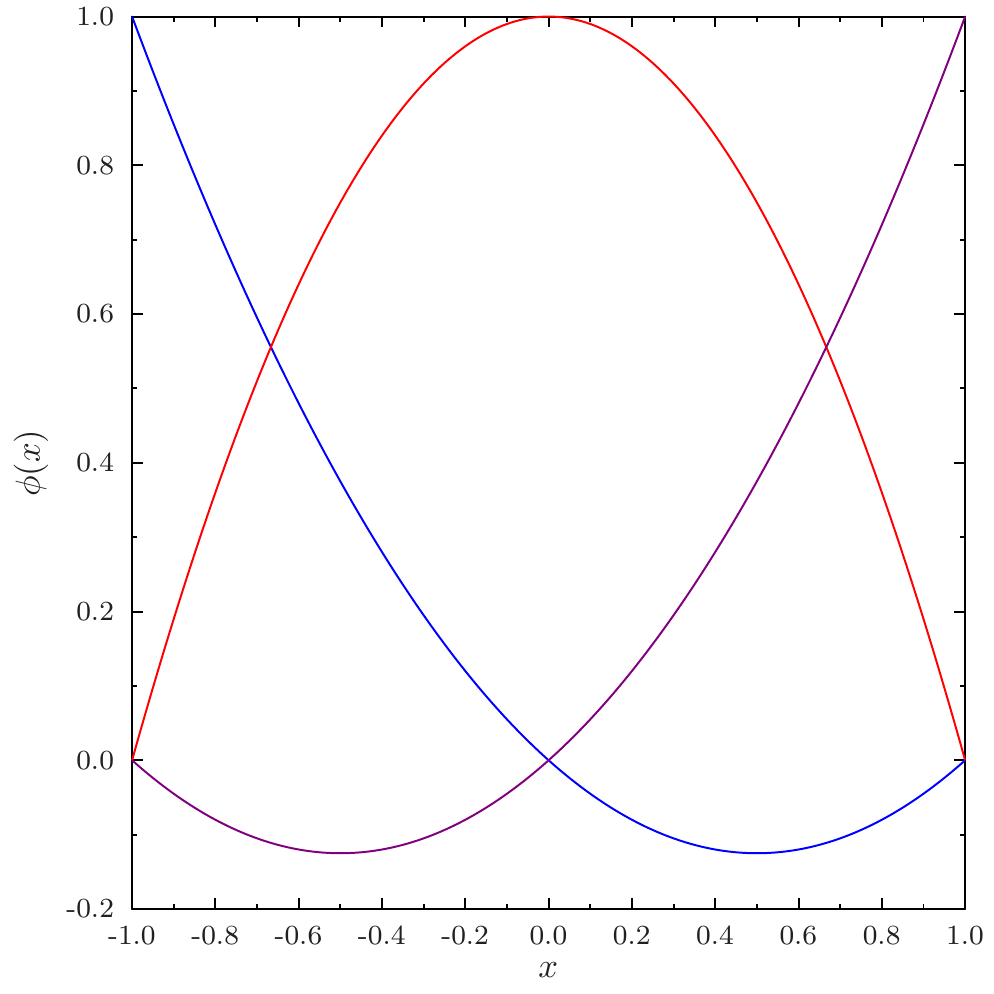}
\par\end{centering}
}
\par\end{centering}
\caption{LIP elements with two and three uniformly spaced nodes.\label{fig:Lagrange}}
\end{figure}

In the one-dimensional finite element method (FEM), the problem of
the global description of a function $f(r)$ is split into a number
of easier problems, that is, the description of $f(r)$ within line
segments $r\in[r_{\text{min}},r_{\text{max}}]$ called elements. Within
each element, the value of any function $f(r)$ can be approximated
using $n$ element-specific basis functions $\phi_{i}(r)$ also known
as shape functions as
\begin{equation}
f(r)\approx\sum_{i=1}^{n}f_{i}\phi_{i}(r).\label{eq:fem-desc}
\end{equation}
The shape functions are traditionally chosen by specifying $n$ control
points called nodes uniformly in the element including all its edges,
and demanding that each of the $n$ basis functions correspond to
the value of the function $f$ at one of these points 
\begin{equation}
f(r_{i})=\sum_{j}f_{j}\phi_{j}(r_{i})=f_{i};\label{eq:f-fem}
\end{equation}
the condition of \eqref{f-fem} can be equally written in the form
\begin{equation}
\phi_{i}(r_{j})=\delta_{ij}.\label{eq:LIP-cond}
\end{equation}
\Eqref{LIP-cond} yields the well-known Lagrange interpolating polynomials
(LIPs), which can also be written in closed form as
\begin{equation}
\phi_{i}(r)=\prod_{j=0,j\neq i}^{n-1}\frac{r-r_{j}}{r_{i}-r_{j}}.\label{eq:LIP}
\end{equation}
Two- and three-node LIPs are shown in \figref{Lagrange}.

In addition to LIPs, also Hermite interpolating polynomials (HIPs)
can be used. First-order HIPs are defined by
\begin{align}
\phi_{2i}(r_{j})=\delta_{ij},\;\; & \phi_{2i+1}(r_{j})=0,\label{eq:hip-f}\\
\phi_{2i}'(r_{j})=0,\;\; & \phi_{2i+1}'(r_{j})=\delta_{ij},\label{eq:hip-d}
\end{align}
that is, the even and odd-numbered basis functions describe the values
of $f(r)$ and $f'(r)$ at the nodes, respectively, guaranteeing continuity
both of the function and its derivative across element boundaries.
It has been claimed that due to this added flexibility, HIPs yield
better results for quantum mechanical problems than LIPs.\citep{Ram-Mohan1990,Ram-Mohan2002}
Analogous expressions to \eqref{hip-f,hip-d} can be developed for
higher order HIPs that guarantee continuity of the derivative up to
the $n^{\text{th}}$ order; LIPs being equivalent to $0^{\text{th}}$
order HIPs.

In order to derive expressions for HIPs, we shall follow the style
of traditional finite element textbooks such as \citeref{Ram-Mohan2002},
and write the basis functions in terms of primitive polynomials as
\begin{equation}
\phi_{i}(r)=c_{i,0}+c_{i,1}r+\dots+c_{i,n-1}r^{n-1}.\label{eq:primpol}
\end{equation}
The expansion for LIPs can be obtained by writing out \eqref{LIP-cond}
as a matrix equation
\begin{equation}
\left(\begin{array}{ccccc}
1 & r_{0} & r_{0}^{2} & \cdots & r_{0}^{n-1}\\
1 & r_{1} & r_{1}^{2} & \cdots & r_{1}^{n-1}\\
\vdots & \vdots & \vdots & \ddots & \vdots\\
1 & r_{n-1} & r_{n-1}^{2} & \cdots & r_{n-1}^{n-1}
\end{array}\right)\times\left(\begin{array}{ccccc}
c_{00} & c_{01} & c_{02} & \cdots & c_{0,n-1}\\
c_{10} & c_{11} & c_{12} & \cdots & c_{1,n-1}\\
\vdots & \vdots & \vdots & \ddots & \vdots\\
c_{n-1,0} & c_{n-1,1} & c_{n-1,2} & \cdots & c_{n-1,n-1}
\end{array}\right)=\boldsymbol{1}.\label{eq:mateq}
\end{equation}
Denoting the first matrix in \eqref{mateq} as $\boldsymbol{R}$ and
the second matrix containing the primitive coefficients as $\boldsymbol{C}$,
the primitive coefficients can be solved with $\boldsymbol{C}=\boldsymbol{R}^{-1}$.
HIPs can be solved in terms of primitive polynomials with a matrix
equation similar to \eqref{mateq}; HIPs of an arbitrary order are
supported in \textsc{HelFEM}.

\begin{figure*}
\centering{}\subfloat[Six-node LIP element with Lobatto nodes.\label{fig:Six-node-LIP-element}]{\begin{centering}
\includegraphics[width=0.33\textwidth]{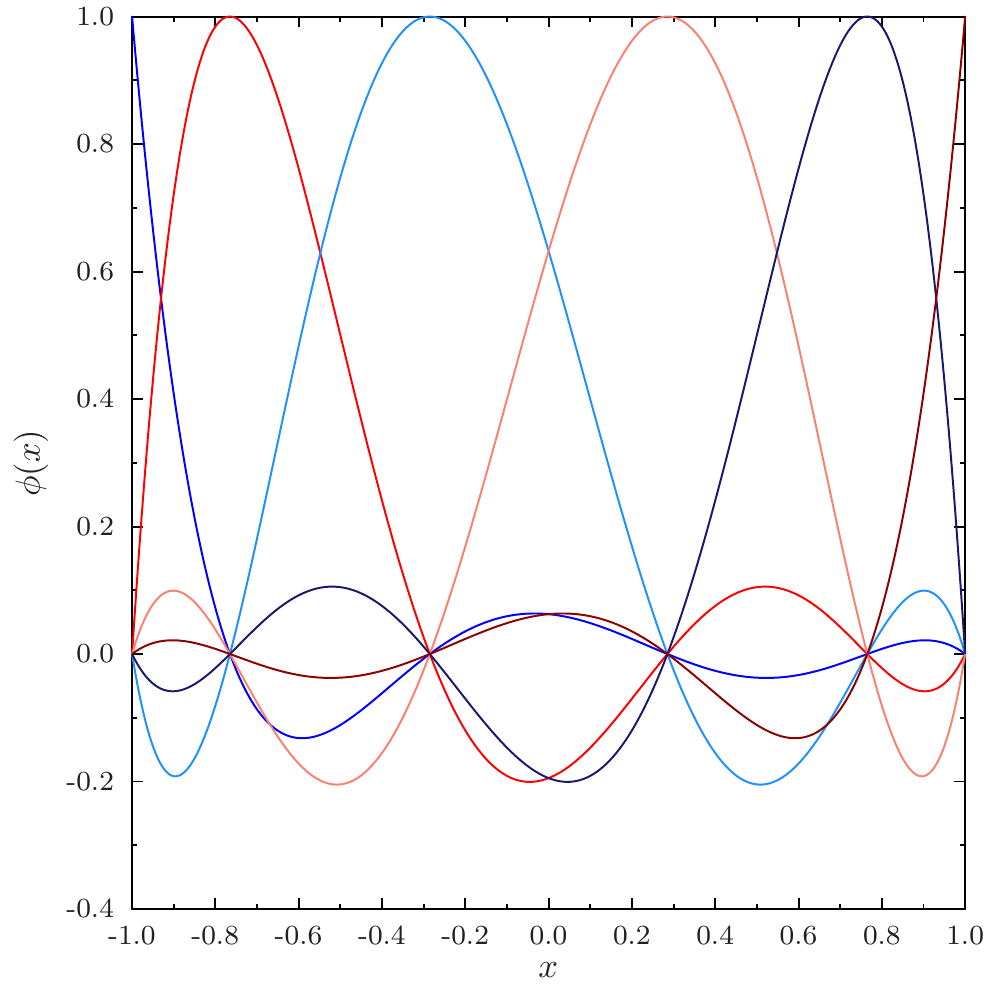}
\par\end{centering}
}\subfloat[Six-node Legendre element.\label{fig:Six-node-Legendre}]{\begin{centering}
\includegraphics[width=0.33\textwidth]{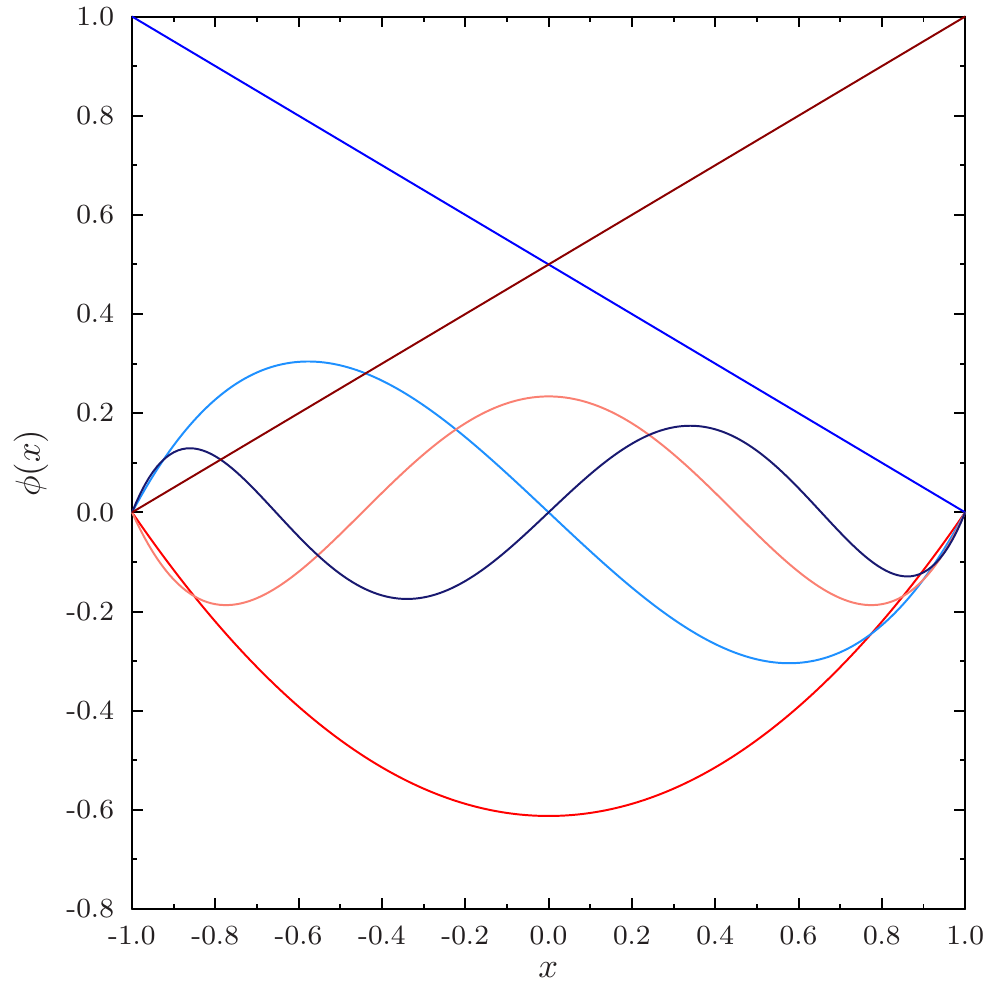}
\par\end{centering}
}\caption{Numerically stable elements.\label{fig:Spectral-elements.}}
\end{figure*}

Despite the use of primitive polynomial expansions in most finite
element textbooks, the resulting matrix equations of the type of \eqref{mateq}
become numerically unstable for high orders due to the Runge phenomenon,
limiting one to polynomials of a low order, such as five or six. However,
numerically stable alternatives can be fashioned\emph{ }by the use
of orthogonal polynomials. For instance, although \eqref{LIP} is
unstable with uniformly spaced nodes, it can be made stable to high
orders by switching to the use of non-uniformly spaced nodes. Choosing
the locations of the nodes based on a quadrature rule such as Gauss--Lobatto
as in the spectral element method\citep{Patera1984} yields an especially
powerful approach. An example of a six-node LIP element with Lobatto
nodes is shown in \figref{Six-node-LIP-element}. We have also implemented
another numerically stable primitive basis, similarly allowing the
use of high-order elements, by following Flores \emph{et al}.\citep{Flores1989,Flores1989a}
and using Legendre polynomials $P_{n}(x)$ in terms of the shape functions
\begin{align}
\phi_{j}(x)= & \frac{1}{\sqrt{4j+2}}\left(P_{j+1}(x)-P_{j-1}(x)\right),j\in[1,N-2]\label{eq:P-shape}
\end{align}
that vanish at the boundaries, the first and last basis functions
that guarantee continuity of the wave function across element boundaries
being given by 
\begin{align}
\phi_{0}(x)= & \frac{1}{2}\left(P_{0}(x)-P_{1}(x)\right),\label{eq:P-0}\\
\phi_{N-1}(x)= & \frac{1}{2}\left(P_{0}(x)+P_{1}(x)\right).\label{eq:P-N}
\end{align}
An example of the Legendre basis is shown in \figref{Six-node-Legendre}.
The lowest-order Legendre element given by \eqref{P-0,P-N} is equivalent
to the 2-node LIP element, whereas higher orders describe variations
at smaller and smaller scales.

As only one function contributes to the value at a given node, boundary
conditions can be easily implemented in the finite element approach.
Orbitals can be made to vanish at the origin by removing the first
basis function from the first radial element. Likewise, the vanishing
boundary condition at infinity is achieved by removing the last basis
function in the last radial element. Note, however, that for a HIP
basis, the boundary condition of origin only applies to the function
value, not any of its derivatives; at infinity both the function and
its derivatives are set to zero.

\subsubsection{Finite element matrices\label{subsec:Finite-element-matrices}}

Although the basis functions are only defined within a single element,
nodes at the element boundary are shared between the functions of
the two elements touching at the boundary. As a node defines a basis
function, this means that the basis functions sharing the node at
the boundary must be identified with each other. That is, while given
three three-node LIP elements, a naïve evaluation of \emph{e.g.} the
overlap matrix would read
\begin{equation}
\boldsymbol{S}^{\text{naïve}}=\left(\begin{array}{ccccccccc}
S_{11}^{\text{el 1}} & S_{12}^{\text{el 1}} & S_{13}^{\text{el 1}}\\
S_{21}^{\text{el 1}} & S_{22}^{\text{el 1}} & S_{23}^{\text{el 1}}\\
S_{31}^{\text{el 1}} & S_{32}^{\text{el 1}} & S_{33}^{\text{el 1}}\\
 &  &  & S_{11}^{\text{el 2}} & S_{12}^{\text{el 2}} & S_{13}^{\text{el 2}}\\
 &  &  & S_{21}^{\text{el 2}} & S_{22}^{\text{el 2}} & S_{23}^{\text{el 2}}\\
 &  &  & S_{31}^{\text{el 2}} & S_{32}^{\text{el 2}} & S_{33}^{\text{el 2}}\\
 &  &  &  &  &  & S_{11}^{\text{el 3}} & S_{12}^{\text{el 3}} & S_{13}^{\text{el 3}}\\
 &  &  &  &  &  & S_{21}^{\text{el 3}} & S_{22}^{\text{el 3}} & S_{23}^{\text{el 3}}\\
 &  &  &  &  &  & S_{31}^{\text{el 3}} & S_{32}^{\text{el 3}} & S_{33}^{\text{el 3}}
\end{array}\right)\label{eq:S-naive}
\end{equation}
where the superscript denotes the element in which the functions reside
and the subscripts contain the local function indices, identification
of the bordering functions results in an overlaying of the element
matrices 
\begin{equation}
\boldsymbol{S}^{\text{real}}=\left(\begin{array}{ccccccc}
S_{11}^{\text{el 1}} & S_{12}^{\text{el 1}} & S_{13}^{\text{el 1}}\\
S_{21}^{\text{el 1}} & S_{22}^{\text{el 1}} & S_{23}^{\text{el 1}}\\
S_{31}^{\text{el 1}} & S_{32}^{\text{el 1}} & S_{33}^{\text{el 1}}+S_{11}^{\text{el 2}} & S_{12}^{\text{el 2}} & S_{13}^{\text{el 2}}\\
 &  & S_{21}^{\text{el 2}} & S_{22}^{\text{el 2}} & S_{23}^{\text{el 2}}\\
 &  & S_{31}^{\text{el 2}} & S_{32}^{\text{el 2}} & S_{33}^{\text{el 2}}+S_{11}^{\text{el 3}} & S_{12}^{\text{el 3}} & S_{13}^{\text{el 3}}\\
 &  &  &  & S_{21}^{\text{el 3}} & S_{22}^{\text{el 3}} & S_{23}^{\text{el 3}}\\
 &  &  &  & S_{31}^{\text{el 3}} & S_{32}^{\text{el 3}} & S_{33}^{\text{el 3}}
\end{array}\right).\label{eq:S-real}
\end{equation}
The use of a HIP basis results in a larger amount of overlaying. While
\eqref{S-real} shows the true form of the matrix, the ``naïve''
representation of \eqref{S-naive} is useful due to its simplicity:
it is easy and efficient to contract matrices in the naïve representation,
which is the approach adopted in \textsc{HelFEM}.

\subsection{Basis set\label{subsec:Basis-set}}

Due to the high amount of symmetry present in the atomic case, polar
spherical coordinates are used, whereby the orbitals separate into
a radial part and and angular part. A trivial choice for the basis
set in an atomic calculation would thus be
\begin{align}
\chi_{nlm}^{\text{trivial}}= & B_{n}(r)Y_{l}^{m}(\theta,\phi),\label{eq:trivialbasis}
\end{align}
where $B_{n}(r)$ are finite element shape functions. Instead, following
tradition,\citep{FroeseFischer1992} we choose the basis set as 
\begin{equation}
\chi_{nlm}=r^{-1}B_{n}(r)Y_{l}^{m}(\theta,\phi),\label{eq:basis}
\end{equation}
since including the asymptotic $r^{-1}$ factor leads to much faster
convergence in the radial basis. This also has the benefit that in
\eqref{basis} all $B_{n}(0)$ have to vanish, whereas in \eqref{trivialbasis}
only the non-$s$-state radial functions vanish at the nucleus, as
will be detailed below.

Although the basis set (\eqref{basis}) could in principle employ
different radial grids for the various angular momentum channels,
the same radial finite element basis set is typically employed for
all values of $l$ and $m$. This approach is also chosen in the present
work, as the use of a common radial grid simplifies the implementation,
as will also be seen below.

In the following, basis functions $i$ carry both an angular part
$l_{i}$, $m_{i}$, and a radial part $B_{i}$. The full dimension
of the basis set is given by the number of radial functions times
the number of angular functions.

\subsection{Variational approach\label{subsec:Variational-approach}}

As is well known, the variational solution of the HF equations within
a basis set leads to the Roothaan\citep{Roothaan1951} or Pople--Nesbet
equations\citep{Pople1954}
\begin{equation}
\boldsymbol{F}_{\sigma}\boldsymbol{C}_{\sigma}=\boldsymbol{S}\boldsymbol{C}_{\sigma}\boldsymbol{\epsilon}_{\sigma};\label{eq:roothaan}
\end{equation}
analogous equations are also obtained in the case of Kohn--Sham DFT.\citep{Pople1992,Neumann1996}
Here, $\boldsymbol{F}_{\sigma}$ is the (Kohn--Sham) Fock operator
of spin $\sigma$, $\boldsymbol{C}_{\sigma}$ are the canonical molecular
orbital coefficients, and $\boldsymbol{\epsilon}_{\sigma}$ is a diagonal
matrix holding the corresponding orbital energies.

Although we have chosen the basis functions to be complex, we are
free to choose the coefficients to be real in the absence of a magnetic
field. Note that complex coefficients may be necessary in some approaches
even in the absence of an magnetic field, see \emph{e.g.} \citerefs{Ostlund1972, Edwards1988, Lehtola2014a, Small2015a, Lehtola2016, Lehtola2016a}.
In contrast, the coefficients can be chosen to be real even \emph{in
the presence} of a magnetic field in the case of atoms and diatomic
molecules in a parallel field; see \citeref{Lehtola2019d}.

The Roothaan / Pople--Nesbet equations are solved in the present
work by symmetric orthonormalization:\citep{Lowdin1950} writing the
unknown orbital coefficient in terms of a transformation matrix $\boldsymbol{X}$
as 
\begin{equation}
\boldsymbol{C}=\tilde{\boldsymbol{C}}\boldsymbol{X}\label{eq:CX}
\end{equation}
and left-multiplying \eqref{roothaan} with $\boldsymbol{X}^{\text{T}}$,
one obtains the equation
\begin{equation}
\boldsymbol{X}^{\text{T}}\boldsymbol{F}\boldsymbol{X}\tilde{\boldsymbol{C}}=\boldsymbol{X}^{\text{T}}\boldsymbol{S}\boldsymbol{X}\tilde{\boldsymbol{C}}\boldsymbol{\epsilon}.\label{eq:xroothaan}
\end{equation}
Setting $\boldsymbol{X}=\boldsymbol{S}^{-1/2}$, \eqref{xroothaan}
simplifies into a normal eigenvalue equation
\begin{equation}
\tilde{\boldsymbol{F}}\tilde{\boldsymbol{C}}=\tilde{\boldsymbol{C}}\boldsymbol{\epsilon},\label{eq:roothaan-trans}
\end{equation}
where the transformed Fock matrix is given by
\begin{equation}
\tilde{\boldsymbol{F}}=\left(\boldsymbol{S}^{-1/2}\right)^{\text{T}}\boldsymbol{F}\boldsymbol{S}^{-1/2}.\label{eq:F-trans}
\end{equation}
The molecular orbital coefficients in the original basis can be obtained
from the solution of \eqref{roothaan-trans} with \eqref{CX}.

Because the finite element basis set is never ill-conditioned, $\boldsymbol{S}^{-1/2}$
for \eqref{xroothaan} can be constructed in \textsc{HelFEM} with
either Cholesky factorization
\begin{align}
\boldsymbol{S}= & \boldsymbol{L}\boldsymbol{L}^{\text{T}},\label{eq:cholesky-L}\\
\boldsymbol{S}^{-1/2}= & \boldsymbol{L}^{-1},\label{eq:inv-cholesky}
\end{align}
or an eigendecomposition 
\begin{align}
\boldsymbol{S} & =\boldsymbol{Q}\boldsymbol{\Lambda}\boldsymbol{Q}^{\text{T}},\label{eq:eigendec}\\
\boldsymbol{S}^{-1/2} & =\boldsymbol{Q}\boldsymbol{\Lambda}^{-1/2}\boldsymbol{Q}^{\text{T}}.\label{eq:sinvh-eig}
\end{align}
The orthonormalization based on \eqref{cholesky-L,inv-cholesky} or
\eqref{eigendec,sinvh-eig} is performed in terms of normalized basis
functions, as this turns out to be necessary for the numerical stability
of the procedure. However, the basis functions themselves are not
normalized in \textsc{HelFEM}; the normalization is stored in the
rows of $\boldsymbol{S}^{-1/2}$. As symmetry with respect to the
$m$ quantum number is used by default in \textsc{HelFEM}, the eigendecomposition
can be blocked by $m$ channel in solving \eqref{roothaan-trans, eigendec};
this makes the diagonalizations fast even in large basis sets.

\subsection{One-electron matrix elements\label{subsec:One-electron-matrix-elements}}

To be able to solve the SCF equation (\eqref{roothaan}), we need
to compute several matrix elements. The Fock operator corresponding
to spin $\sigma$ is given by
\begin{equation}
\boldsymbol{F}_{\sigma}=\boldsymbol{T}+\boldsymbol{V}^{\text{nuc}}+\boldsymbol{J}(\boldsymbol{P})+\boldsymbol{K}(\boldsymbol{P}_{\sigma}),\label{eq:fmat}
\end{equation}
where $\boldsymbol{T}$ is the kinetic energy, $\boldsymbol{V}^{\text{nuc}}$
is the nuclear attraction, and $\boldsymbol{J}$ and $\boldsymbol{K}$
are the Coulomb repulsion and exchange(-correlation) matrices that
contain two-electron interactions. $\boldsymbol{P}_{\sigma}$ is the
density matrix for spin $\sigma$
\begin{equation}
\boldsymbol{P}_{\sigma}=\sum_{i\text{ occupied}}\boldsymbol{C}_{i\sigma}\boldsymbol{C}_{i\sigma}^{\dagger}\label{eq:denmat}
\end{equation}
and $\boldsymbol{P}$ is the total density matrix
\begin{equation}
\boldsymbol{P}=\boldsymbol{P}_{\alpha}+\boldsymbol{P}_{\beta}.\label{eq:Pmat}
\end{equation}

\subsubsection{Overlap\label{subsec:Overlap}}

The overlap matrix elements are simply
\begin{align}
S_{ij}= & \braket{i|j}=\delta_{l_{i},l_{j}}\delta_{m_{i},m_{j}}\int B_{i}(r)B_{j}(r){\rm d}r,\label{eq:overlap}
\end{align}
where $\delta_{ij}$ is the Kronecker delta symbol. Here and in the
following, integration over $r\in[0,\infty)$ is implied for brevity;
however, the finite support of the basis functions truncate the integrals
to a finite interval.

\subsubsection{Kinetic energy\label{subsec:Kinetic-energy}}

The evaluation of the kinetic energy matrix is slightly more complicated.
In spherical coordinates, the gradient of a function $f$ is given
by
\begin{align}
\nabla f= & \frac{\partial f}{\partial r}\hat{r}+\frac{1}{r}\frac{\partial f}{\partial\theta}\hat{\theta}+\frac{1}{r\sin\theta}\frac{\partial f}{\partial\phi}\hat{\phi},\label{eq:grad}
\end{align}
where $\hat{r}$, $\hat{\theta}$, and $\hat{\phi}$ are unit vectors
in the direction of the coordinates $r$, $\theta$, and $\phi$,
respectively. The Laplacian reads
\begin{align}
\nabla^{2}f= & \frac{1}{r^{2}}\frac{\partial}{\partial r}\left(r^{2}\frac{\partial f}{\partial r}\right)+\frac{1}{r^{2}\sin\theta}\frac{\partial}{\partial\theta}\left(\sin\theta\frac{\partial f}{\partial\theta}\right)+\frac{1}{r^{2}\sin^{2}\theta}\frac{\partial^{2}f}{\partial\phi^{2}},\label{eq:lapl}
\end{align}
which can be rewritten as

\begin{align}
\nabla^{2}f= & \frac{1}{r^{2}}\frac{\partial}{\partial r}\left(r^{2}\frac{\partial f}{\partial r}\right)-\frac{1}{r^{2}}\hat{L}^{2}f\label{eq:lapl2}
\end{align}
where $\hat{L}^{2}$ is the angular momentum operator. Thus, the kinetic
energy is
\begin{align}
T_{ij}= & \braket{i|-\frac{\nabla^{2}}{2}|j}\label{eq:T0-1}\\
= & -\frac{1}{2}\delta_{l_{i},l_{j}}\delta_{m_{i},m_{j}}\int\frac{B_{i}(r)}{r}\left[\frac{1}{r^{2}}\frac{\partial}{\partial r}\left(r^{2}\frac{\partial}{\partial r}\left[\frac{B_{j}(r)}{r}\right]\right)-\frac{l(l+1)}{r^{2}}\frac{B_{j}(r)}{r}\right]r^{2}{\rm d}r\label{eq:T0-2}\\
= & -\frac{1}{2}\delta_{l_{i},l_{j}}\delta_{m_{i},m_{j}}\int\frac{B_{i}(r)}{r}\left[\frac{\partial}{\partial r}\left(rB_{j}'(r)-B_{j}(r)\right)-\frac{l(l+1)B_{j}(r)}{r}\right]{\rm d}r\label{eq:T0-3}\\
= & -\frac{1}{2}\delta_{l_{i},l_{j}}\delta_{m_{i},m_{j}}\int\frac{B_{i}(r)}{r}\left[rB_{j}''(r)-\frac{l(l+1)B_{j}(r)}{r}\right]{\rm d}r\label{eq:T0}
\end{align}
Using partial integration to move the first derivative 
\begin{align}
\int B_{i}(r)\frac{\partial^{2}}{\partial r^{2}}B_{j}(r){\rm d}r= & \Bigg|B_{i}(r)\frac{\partial}{\partial r}\left[B_{j}(r)\right]-\int\frac{\partial}{\partial r}\left[B_{i}(r)\right]\frac{\partial}{\partial r}\left[B_{j}(r)\right]{\rm d}r,\label{eq:partial}
\end{align}
where the substitution term (first term in \eqref{partial}) vanishes
since the basis functions and their derivatives are zero at the end
points, one obtains the final expression

\begin{equation}
T_{ij}=\frac{1}{2}\delta_{l_{i},l_{j}}\delta_{m_{i},m_{j}}\left[\int B_{i}'(r)B_{j}'(r){\rm d}r+l(l+1)\int r^{-2}B_{i}(r)B_{j}(r){\rm d}r\right].\label{eq:T}
\end{equation}

The $l(l+1)/r^{2}$ term in \eqref{T} implies that non-$s$ states
must vanish at the origin, as otherwise the kinetic energy would go
to infinity. However, as the $r^{-1}$ factor has been included explicitly
in the basis set (\eqref{basis}), we must require that $B_{n}(r)$
has to go to zero at the origin also for $s$ states -- meaning the
radial basis set is identical for all values of $l$ and $m$ --
as otherwise the value of the orbital $r^{-1}B_{n}(r)$ would diverge
at the nucleus.

The $l(l+1)/r^{2}$ term is also responsible for the energy ordering
of atomic shells. As discussed in \citeref{Lehtola2019}, the term
prevents $p$, $d$, and $f$ orbitals from seeing the less-screened
regions of the nuclear potential close to the nucleus, thereby causing
the orbitals with $l>1$ to lie higher in energy than what would be
statically expected just from the $l(l+1)/r^{2}$ term itself.

\subsubsection{Nuclear attraction\label{subsec:Nuclear-attraction}}

The nuclear attraction matrix for a point nucleus is
\begin{align}
V_{ij}^{\text{nuc}}= & \braket{i|-\frac{Z}{r}|j}\label{eq:Vnuc-0}\\
= & -Z\delta_{l_{i},l_{j}}\delta_{m_{i},m_{j}}\int r^{-1}B_{i}(r)B_{j}(r){\rm d}r.\label{eq:Vnuc}
\end{align}

\subsection{Quadrature\label{subsec:Quadrature}}

Although the integrals can in principle computed analytically in a
primitive polynomial basis, the polynomials would need to be translated
to the location of the element, which we have found to be numerically
problematic. Furthermore, as the use of primitive polynomials is numerically
unstable, alike other FEM programs, we choose to calculate the integrals
using quadrature, as this allows the basis functions to be chosen
freely. Gauss--Chebyshev quadrature on the primitive interval $x\in[-1,1]$
is employed in \textsc{HelFEM}, as the integration nodes and weights
have closed-form expressions. The necessary coordinate transformation
from $r\in[r_{\min},r_{\max}]$ to $x\in[-1,1]$ is given by
\begin{align}
r= & r_{0}+\lambda x\label{eq:rquad}
\end{align}
where 
\begin{equation}
r_{0}=\frac{r_{\max}+r_{\min}}{2}\label{eq:r0}
\end{equation}
 is the midpoint of the interval and 
\begin{equation}
\lambda=\frac{r_{\max}-r_{\min}}{2}\label{eq:lambda}
\end{equation}
 is its length. Using the transformation in \eqref{rquad}, the necessary
quadrature rules are obtained as
\begin{align}
\int r^{n}B_{1}(r)B_{2}(r){\rm d}r\approx & \lambda\sum_{i}w_{i}r(x_{i})^{n}B_{1}(x_{i})B_{2}(x_{i})\label{eq:rnquad}\\
\int\frac{\partial B_{1}}{\partial r}\frac{\partial B_{2}}{\partial r}{\rm d}r\approx & \lambda^{-1}\sum_{i}w_{i}B_{1}'(x_{i})B_{2}'(x_{i})\label{eq:dquad}
\end{align}

\subsection{Two-electron integrals\label{subsec:Two-electron-integrals}}

The two-electron integrals
\begin{align}
(ij|kl)= & \int\frac{\chi_{i}(\boldsymbol{r})\chi_{j}^{*}(\boldsymbol{r})\chi_{k}(\boldsymbol{r}')\chi_{l}^{*}(\boldsymbol{r}')}{\left|\boldsymbol{r}-\boldsymbol{r}'\right|}{\rm d}^{3}r{\rm d}^{3}r'\label{eq:tei0}
\end{align}
can be evaluated with the help of the Laplace expansion
\begin{align}
\frac{1}{r_{12}}= & \frac{4\pi}{r_{>}}\sum_{L=0}^{\infty}\frac{1}{2L+1}\left(\frac{r_{<}}{r_{>}}\right)^{L}\sum_{M=-L}^{L}Y_{L}^{M}(\Omega_{1})\left(Y_{L}^{M}(\Omega_{2})\right)^{*},\label{eq:laplace}
\end{align}
where $r_{<}$ and $r_{>}$ denote the smaller and greater of $r_{1}$
and $r_{2}$, respectively, as
\begin{align}
(ij|kl)= & \int{\rm d}r_{1}{\rm d}r_{2}B_{i}(r_{1})B_{j}(r_{1})B_{k}(r_{2})B_{l}(r_{2})\nonumber \\
\times & \frac{4\pi}{r_{>}}\sum_{L=0}^{\infty}\frac{1}{2L+1}\left(\frac{r_{<}}{r_{>}}\right)^{L}\int d\Omega_{1}d\Omega_{2}\sum_{M=-L}^{L}Y_{L}^{M}(\Omega_{1})\left(Y_{L}^{M}(\Omega_{2})\right)^{*}\nonumber \\
\times & Y_{l_{i}}^{m_{i}}(\Omega_{1})\left(Y_{l_{j}}^{m_{j}}(\Omega_{1})\right)^{*}Y_{l_{k}}^{m_{k}}(\Omega_{2})\left(Y_{l_{l}}^{m_{l}}(\Omega_{2})\right)^{*}\label{eq:tei-exp}
\end{align}
Invoking the rule of complex conjugation of spherical harmonics
\begin{align}
\left(Y_{l}^{m}(\Omega)\right)^{*}= & \left(-1\right)^{m}Y_{l}^{-m}(\Omega)\label{eq:sph-conj}
\end{align}
and employing their closure relation
\begin{align}
Y_{l_{1}}^{m_{1}}(\Omega)Y_{l_{2}}^{m_{2}}(\Omega) & =\sum_{LM}G_{l_{1}l_{2},M}^{m_{1}m_{2},L}Y_{L}^{M}(\Omega)\label{eq:sph-close}
\end{align}
where an asymmetric definition for the Gaunt coefficient is used\citep{Lehtola2011}
\begin{align}
G_{l_{1}l_{2},M}^{m_{1}m_{2},L}= & \left(-1\right)^{M}\sqrt{\frac{\left(2L+1\right)\left(2l_{1}+1\right)\left(2l_{2}+1\right)}{4\pi}}\left(\begin{array}{ccc}
l_{1} & l_{2} & L\\
m_{1} & m_{2} & -M
\end{array}\right)\left(\begin{array}{ccc}
l_{1} & l_{2} & L\\
0 & 0 & 0
\end{array}\right)\label{eq:gaunt}
\end{align}
the two-electron integral is obtained in the form
\begin{align}
(ij|kl)= & \int{\rm d}r_{1}{\rm d}r_{2}B_{i}(r_{1})B_{j}(r_{1})B_{k}(r_{2})B_{l}(r_{2})\nonumber \\
\times & \frac{4\pi}{r_{>}}\sum_{L}\frac{1}{2L+1}\left(\frac{r_{<}}{r_{>}}\right)^{L}G_{Ll_{i},m_{j}}^{Mm_{i},l_{j}}G_{Ll_{l},m_{k}}^{Mm_{l},l_{k}}.\label{eq:tei}
\end{align}
From \eqref{tei}, it is seen that the integral is non-zero only if
\begin{equation}
L_{\min}\leq L\leq L_{\max}\label{eq:L-cond}
\end{equation}
and
\begin{equation}
m_{j}-m_{i}=M=m_{k}-m_{l},\label{eq:M-cond}
\end{equation}
where 
\begin{align}
L_{\min}= & \max\{\left|l_{i}-l_{j}\right|,\left|l_{k}-l_{l}\right|\},\label{eq:Lmin}\\
L_{\max}= & \min\{l_{i}+l_{j},l_{k}+l_{l}\}.\label{eq:Lmax}
\end{align}
Furthermore, as the spherical harmonic with quantum numbers $L$ and
$M$ must exist in order for the coupling to make sense, one obtains
the further condition
\begin{equation}
L_{\min}\geq|M|.\label{eq:LM-cond}
\end{equation}
The conditions in \eqrangeref{L-cond}{LM-cond} truncate the series
in \eqref{tei} to a finite number of terms. Thus, the repulsion integrals
reduce to the simple expression
\begin{equation}
(ij|kl)=\sum_{L_{\min}}^{L_{\max}}I_{ijkl}^{L}G_{Ll_{i},m_{j}}^{Mm_{i},l_{j}}G_{Ll_{l},m_{k}}^{Mm_{l},l_{k}}\label{eq:tei-ass}
\end{equation}
where $M$ is defined via \eqref{M-cond} and the primitive integrals
are defined as
\begin{equation}
I_{ijkl}^{L}=\frac{4\pi}{2L+1}\int{\rm d}r_{1}{\rm d}r_{2}B_{i}(r_{1})B_{j}(r_{1})B_{k}(r_{2})B_{l}(r_{2})\frac{r_{<}^{L}}{r_{>}^{L+1}}.\label{eq:I-prim}
\end{equation}

\subsubsection{Primitive integrals\label{subsec:Primitive-integrals}}

The primitive integrals can be split into two terms
\begin{align}
\int{\rm d}r_{1}{\rm d}r_{2}f(r_{1})g(r_{2})\frac{r_{<}^{L}}{r_{>}^{L+1}}= & \int_{0}^{\infty}{\rm d}r_{1}\int_{0}^{r_{1}}{\rm d}r_{2}f(r_{1})g(r_{2})\frac{r_{2}^{L}}{r_{1}^{L+1}}+\int_{0}^{\infty}{\rm d}r_{2}\int_{0}^{r_{2}}{\rm d}r_{1}f(r_{1})g(r_{2})\frac{r_{1}^{L}}{r_{2}^{L+1}}\label{eq:prim-split0}\\
= & \int_{0}^{\infty}{\rm d}r_{1}\int_{0}^{r_{1}}{\rm d}r_{2}f(r_{1})g(r_{2})\frac{r_{2}^{L}}{r_{1}^{L+1}}+\int_{0}^{\infty}{\rm d}r_{1}\int_{0}^{r_{1}}{\rm d}r_{2}f(r_{2})g(r_{1})\frac{r_{2}^{L}}{r_{1}^{L+1}}\label{eq:prim-split1}\\
= & \int_{0}^{\infty}{\rm d}r_{1}\int_{0}^{r_{1}}{\rm d}r_{2}\left[f(r_{1})g(r_{2})+f(r_{2})g(r_{1})\right]\frac{r_{2}^{L}}{r_{1}^{L+1}}\label{eq:prim-split}
\end{align}
as the integration over $r_{1}$ and $r_{2}$ can be divided into
integration over two triangles separated by the line $r_{1}=r_{2}$.
Substituting \eqref{prim-split} into \eqref{I-prim} yields
\begin{align}
I_{ijkl}^{L}= & \frac{4\pi}{2L+1}\int_{0}^{\infty}{\rm d}r_{1}r_{1}^{-L-1}B_{i}(r_{1})B_{j}(r_{1})\int_{0}^{r_{1}}{\rm d}r_{2}r_{2}^{L}B_{k}(r_{2})B_{l}(r_{2})\label{eq:tei-split0}\\
+ & \frac{4\pi}{2L+1}\int_{0}^{\infty}dr_{1}r_{1}^{-L-1}B_{k}(r_{1})B_{l}(r_{1})\int_{0}^{r_{1}}dr_{2}r_{2}^{L}B_{i}(r_{2})B_{j}(r_{2})\label{eq:tei-split}
\end{align}
As the basis functions have finite support, the functions $i$ and
$j$ have to reside in the same element, and the functions $k$ and
$l$ have to reside in the same element, as otherwise their product
vanishes.

If $ij$ and $kl$ are \emph{not} \emph{within} the same element,
then only a single term in \eqref{tei-split} survives 
\begin{align}
I_{ijkl}^{L}= & \frac{4\pi}{2L+1}\left[\int_{ij\text{ element}}{\rm d}r_{1}r_{1}^{-1-L}B_{i}(r_{1})B_{j}(r_{1})\right]\left[\int_{kl\text{ element}}{\rm d}r_{2}r_{2}^{L}B_{k}(r_{2})B_{l}(r_{2})\right]\label{eq:tei-prod}
\end{align}
and this interelement integral factorizes into two simple radial integrals
with indices $ij$ and $kl$. We have assumed in \eqref{tei-prod}
that $ij$ are farther from the origin than $kl$.

If $ij$ and $kl$ are \emph{within} the same element, one has to
evaluate the intraelement primitive integral from \eqref{tei-split}.
This proceeds in three steps:

\begin{align}
\phi_{kl}^{L}(r)= & \int_{0}^{r}{\rm d}r'r'^{L}B_{k}(r')B_{l}(r'),\label{eq:phi-int}\\{}
[ij|kl]^{L}= & \int_{0}^{\infty}{\rm d}rr^{-L-1}B_{i}(r)B_{j}(r)\phi_{kl}^{L}(r),\label{eq:aux-int}\\
I_{ijkl}^{L}= & \frac{4\pi}{2L+1}\left([ij|kl]^{L}+[kl|ij]^{L}\right).\label{eq:prim-int}
\end{align}
Note that the integral in \eqref{phi-int} does not range over the
whole element, \emph{i.e.} it only involves \emph{part} of the basis
functions. As the outer integral \eqref{aux-int} is performed using
quadrature with quadrature points $r_{i},i\in[1,N]$, the inner integral
\eqref{phi-int} is evaluated in slices by 
\begin{equation}
\phi_{kl}^{L}(r_{i};r_{i-1})=\int_{r_{i-1}}^{r_{i}}{\rm d}r'r'^{L}B_{k}(r')B_{l}(r')\label{eq:aux-int-slice}
\end{equation}
from which the full integral is recovered with

\begin{equation}
\phi_{kl}^{L}(r_{i})=\begin{cases}
\phi_{kl}^{L}(r_{1};0) & j=1\\
\phi_{kl}^{L}(r_{j};r_{j-1})+\phi_{kl}^{L}(r_{j-1}) & j>1
\end{cases}\label{eq:aux-int-1}
\end{equation}
Denoting the number of primitive basis functions per element as $N_{p}$
and the number of elements as $N_{\text{el}}$, the storage of the
two-electron integrals then requires $2(L_{\text{max}}+1)N_{p}^{2}N_{\text{el}}$
memory for the interelement integrals, and $(L_{\text{max}}+1)N_{p}^{4}N_{\text{el}}$
memory for the intraelement integrals, where the maximum possible
angular momentum is $L_{\max}=2l_{\max}$. Importantly, the scaling
of the storage cost is bilinear in the number of elements and in the
angular grid, implying that large expansions can be employed.

\subsubsection{Coulomb matrix\label{subsec:Coulomb-matrix}}

The Coulomb matrix is given by
\begin{equation}
J_{ij}=\sum_{kl}(ij|kl)P_{kl}.\label{eq:J}
\end{equation}
Insertion of the two-electron integrals (\eqref{tei-ass}) gives the
Coulomb matrix in the form 
\begin{align}
J_{ij}= & \sum_{L_{\min}}^{L_{\max}}G_{Ll_{i},m_{j}}^{Mm_{i},l_{j}}I_{ijkl}^{L}\left(P_{kl}G_{Ll_{l},m_{k}}^{Mm_{l},l_{k}}\right).\label{eq:J-fact}
\end{align}
Because the primitive integrals $I_{ijkl}^{L}$ only depend on the
radial part, the Coulomb matrix can be formed in three steps:
\begin{enumerate}
\item contract the density matrices into radial-only auxiliary matrices
${\displaystyle P_{kl}^{LM}=\sum_{kl}P_{kl}G_{Ll_{l},m_{k}}^{Mm_{l},l_{k}}}$
\item form primitive Coulomb integrals ${\displaystyle J_{ij}^{LM}=\sum_{kl}I_{ijkl}^{L}P_{kl}^{LM}}$\label{enu:form-elementary-Coulomb}
\item form the full Coulomb matrix ${\displaystyle J_{ij}=\sum_{ijLM}G_{Ll_{i},m_{j}}^{Mm_{i},l_{j}}J_{ij}^{LM}}$
\end{enumerate}
Step \enuref{form-elementary-Coulomb} above can be made computationally
efficient by employing the factorization of the primitive integrals,
reducing the scaling from $N_{p}^{4}$ to $N_{p}^{2}$, as well as
using matrix-vector products in the remaining $N_{p}^{4}$ step for
contracting the non-factorizable intraelement integrals with the density
matrix.

\subsubsection{Exchange matrix\label{subsec:Exchange-matrix}}

The exchange matrix is given by

\begin{equation}
K_{jk}^{\sigma}=\sum_{il}(ij|kl)P_{il}^{\sigma},\label{eq:K}
\end{equation}
which reduces to

\begin{equation}
K_{jk}^{\sigma}=\sum_{L_{\min}}^{L_{\max}}I_{ijkl}^{L}\left(P_{il}^{\sigma}G_{Ll_{i},m_{j}}^{Mm_{i},l_{j}}G_{Ll_{l},m_{k}}^{Mm_{l},l_{k}}\right).\label{eq:K-fact}
\end{equation}
As with the case of the Coulomb matrix above, it is beneficial to
construct auxiliary density matrices by performing the sums over the
angles in the first step, as this decreases the number of costly radial
contractions. However, in the case of the exchange, the angular parts
cannot be formed separately in the input and output indices, and so
separate auxiliary density matrices need to be built for every block
of the output $jk$.

The factorization of the interelement two-electron integrals can again
be exploited in the radial contractions, reducing the scaling from
$N_{p}^{4}$ to $N_{p}^{3}$. The intraelement integrals are made
more efficient by precomputing $i\leftrightarrow k$ permuted copies
of the intraelement two-electron integrals and storing them in memory,
which allows the use of efficient matrix-vector products for the contraction
($N_{p}^{4}$ cost).

\subsection{Electric field\label{subsec:Electric-field}}

Although electrons are formally unbound in the presence of a finite
field, in practice this is not a problem if the field is weak enough
microscopically -- macroscopically, such fields are still extremely
strong. Placing the atom in an electric dipole field in the $z$ direction
changes the Hamiltonian by 
\begin{equation}
\Delta H^{\text{dip}}=-\boldsymbol{\mu}\cdot\boldsymbol{E}=-\text{\ensuremath{\mu}}_{z}E_{z}=+zE_{z}\label{eq:Hdip}
\end{equation}
where the dipole matrix is given by
\begin{align}
\mu_{z;ij}= & 2\sqrt{\frac{\pi}{3}}G_{l_{j}1,l_{i}}^{m_{j}0,m_{i}}\int rB_{i}(r)B_{j}(r)dr\label{eq:dipole}
\end{align}
since
\begin{align}
\mu_{z}= & z=r\cos\theta,\label{eq:z}\\
\cos\theta & =2\sqrt{\frac{\pi}{3}}Y_{1}^{0}=2\sqrt{\frac{\pi}{3}}\left(Y_{1}^{0}\right)^{*}.\label{eq:costh}
\end{align}
For a quadrupole field we have 
\begin{equation}
\Delta H^{\text{quad}}=-\frac{1}{3}\Theta_{zz}E_{zz},\label{eq:Hquad}
\end{equation}
where the quadrupole operator is
\begin{equation}
\Theta_{zz}=\frac{1}{2}\left(3z^{2}-r^{2}\right)=(3\cos^{2}\theta-1)r^{2}\label{eq:quad}
\end{equation}
from which 
\begin{align}
\Theta_{zz;ij}= & \frac{2}{5}\sqrt{5\pi}G_{l_{j}2,l_{i}}^{m_{j}0,m_{i}}\int r^{2}B_{i}(r)B_{j}(r)dr.\label{eq:Equad}
\end{align}

\subsection{Radial expectation values\label{subsec:Radial-expectation-values}}

Radial expectation values of the wave function can be obtained simply
as 
\begin{align}
\braket{r^{n}}_{ij}= & \delta_{l_{i},l_{j}}\delta_{m_{i},m_{j}}\int B_{i}(r)r^{n}B_{j}(r){\rm d}r.\label{eq:radexp}
\end{align}

\subsection{Electron density at the nucleus\label{subsec:Electron-density-at}}

The inclusion of the $r^{-1}$ factor in the basis makes it slightly
non-trivial to calculate the electron density at the nucleus, as the
electron density in the slice $[r,r+dr]$ is given by
\begin{equation}
\overline{n}(r)=\sum_{\mu\nu}\int P_{\mu\nu}\chi_{\mu}^{*}(\boldsymbol{r})\chi_{\nu}(\boldsymbol{r})d\Omega=\sqrt{4\pi}G_{0l_{i},m_{j}}^{0m_{i},l_{j}}\sum_{\mu\nu}P_{\mu\nu}\frac{B_{\mu}(r)B_{\nu}(r)}{r^{2}}\label{eq:avedens}
\end{equation}
where at the nucleus both $B_{n}(r)\to0$ and $r\to0$. However, the
electron density at the nucleus is straightforwardly obtained using
two applications of l'Hôpital's rule as
\begin{align}
n_{0}= & \overline{n}(0)/4\pi=\frac{1}{\sqrt{4\pi}}G_{0l_{i},m_{j}}^{0m_{i},l_{j}}\sum_{\mu\nu}P_{\mu\nu}B_{\mu}'(0)B_{\nu}'(0)\label{eq:nucdens}
\end{align}
as $B_{\mu}(0)=0$ due to the boundary conditions.

\subsection{One-center expansions\label{subsec:One-center-expansions}}

Single-center expansions -- in which the electronic structure of
a polyatomic molecule is expanded in terms of functions on a single
center -- have been around in quantum chemistry for a long time.\citep{Huzinaga1956,Bishop1967,Desclaux1983}
While the single-center method is not employed in the present work,
for completeness we shall detail its use below, as it is also available
in \textsc{HelFEM} for calculations on diatomics XY or linear triatomics
XYX. An implementation of the single-center expansion for diatomic
molecules based on B-splines has been published recently with applications
to first- and second-period diatomics.\citep{Hu2014}

As the orbitals in linear molecules can be classified by their $m$
value, linear molecules are the most interesting use case for a one-center
expansion, since the $m$ component can be treated analytically as
for free atoms, while an expansion in $l$ is necessary as the spherical
symmetry of the system is broken by the off-center nuclear charges.
Using the Laplace expansion for the Coulomb interaction (\eqref{laplace})
the nuclear attraction matrix elements for a nucleus at $z=a$ can
be obtained as 
\begin{align}
V_{ij}= & \braket{i|-\frac{Z}{r_{a}}|j}=-Z\int B_{i}(r)B_{j}(r)\frac{4\pi}{r_{>}}\sum_{L=0}^{\infty}\frac{1}{2L+1}\left(\frac{r_{<}}{r_{>}}\right)^{L}\nonumber \\
\times & \sum_{M=-L}^{L}\left(Y_{l_{i}}^{m_{i}}(\Omega)\right)^{*}Y_{L}^{M}(\Omega)\left(Y_{L}^{M}(\Omega_{a})\right)^{*}Y_{l_{j}}^{m_{j}}(\Omega){\rm d}r{\rm d}\Omega\label{eq:offnuc0-1}\\
= & -Z\int B_{i}(r)B_{j}(r)\frac{4\pi}{r_{>}}\sum_{L=0}^{\infty}\frac{1}{2L+1}\left(\frac{r_{<}}{r_{>}}\right)^{L}\left(Y_{l_{i}}^{m_{i}}(\Omega)\right)^{*}Y_{L}^{0}(\Omega)\left(Y_{L}^{0}(\Omega_{a})\right)^{*}Y_{l_{j}}^{m_{j}}(\Omega){\rm d}r{\rm d}\Omega.\label{eq:offnuc0}
\end{align}
This simplifies to
\begin{align}
V_{ij}= & -Z\int B_{i}(r)B_{j}(r)\frac{4\pi}{r_{>}}\sum_{L=0}^{\infty}\frac{1}{2L+1}\left(\frac{r_{<}}{r_{>}}\right)^{L}G_{Ll_{j},m_{i}}^{0m_{j},m_{i}}\left(Y_{L}^{0}(\Omega_{a})\right)^{*}{\rm d}r\label{eq:offnuc-0}\\
= & -4\pi Z\sum_{L=0}^{\infty}\frac{1}{2L+1}G_{Ll_{j},m_{i}}^{0m_{j},l_{i}}\int B_{i}(r)B_{j}(r)\frac{1}{r_{>}}\left(\frac{r_{<}}{r_{>}}\right)^{L}\sqrt{\frac{2L+1}{4\pi}}P_{L}(\cos\theta_{a}){\rm d}r\label{eq:offnuc-1}\\
= & -Z\sum_{L=0}^{\infty}(\pm1)^{L}\sqrt{\frac{4\pi}{2L+1}}G_{Ll_{j},m_{i}}^{0m_{j},l_{i}}\int B_{i}(r)B_{j}(r)\frac{1}{r_{>}}\left(\frac{r_{<}}{r_{>}}\right)^{L}{\rm d}r\label{eq:offnuc}
\end{align}
where we have used $P_{L}^{0}(\pm1)=(\pm1)^{L}$.

As with the two-electron integrals above, the integral splits into
two cases, depending on the location of the element with respect to
the off-center nuclear charge. From this splitting, it is apparent
that element boundaries should be placed at the off-center nuclei,
as this makes the implementation simpler, and allows for a better
description of the nuclear cusp. A single radial grid is then no longer
sufficient; due to the additional nucleus, the radial grid should
first cover the region between the two nuclei $[0,a]$, and then the
region from the additional nucleus to the practical infinity $[a,R_{\infty}]$,
requiring that one converge the calculations with respect to both
parts of the grid.

Further challenges of this approach are seen in \eqref{offnuc}: the
various $l$ channels couple together via $L$, and the couplings
die off slowly. As the expansion in increasing $l$ describes smaller
and smaller features in the system -- especially around the off-center
nuclei -- the single-center expansion works best for light systems
with no tightly bound core orbitals. While the one-center approach
could be used for molecules with more than two atoms, the restriction
to linear molecules along with the difficulties describing heavy off-center
atoms in effect limits one to the treatment of hydrides, either of
the diatomic \ce{HX} form, or the triatomic \ce{HXH} form, where
\ce{X} is a heavy element. However, linear triatomic hydrides only
occur in the alkaline series (\ce{BeH2}, \ce{MgH2}, \dots), while
arbitrary diatomic molecules can be treated efficiently using the
prolate spheroidal coordinate system discussed in the second part
of the series.\citep{Lehtola2019b} In the prolate spheroidal coordinate
system the singularities at the nuclei vanish in the integration of
the nuclear potential matrices, guaranteeing fast convergence to the
CBS limit, at variance to the single-center expansion.

\subsection{Density functional theory\label{subsec:Density-functional-theory}}

The implementation of density functional theory in \textsc{HelFEM}
is done exactly the same way as in our Gaussian-basis program, \textsc{Erkale}.\citep{Lehtola2012,erkale}
Given an expression for the exchange-correlation energy at the LGA,
GGA or meta-GGA level
\begin{equation}
E_{xc}=\int f_{xc}(n_{\alpha},n_{\beta},\gamma_{\alpha\alpha},\gamma_{\alpha\beta},\gamma_{\beta\beta},\tau_{\alpha},\tau_{\beta}){\rm d}^{3}r,\label{eq:Exc}
\end{equation}
where $n_{\sigma}$ is the spin-$\sigma$ density and the reduced
gradient and kinetic energy density are given by
\begin{align}
\gamma_{\sigma\sigma'}= & \sqrt{\nabla n_{\sigma}\cdot\nabla n_{\sigma'}},\label{eq:gamma}\\
\tau= & \frac{1}{2}\sum_{i\text{ occ}}\left|\nabla\psi_{i}\right|^{2},\label{eq:tau}
\end{align}
respectively, the contribution to the Fock matrix is obtained as\citep{Pople1992,Neumann1996}
\begin{align}
K_{\mu\nu}^{xc;\sigma}= & \int\Bigg[\frac{\delta f_{\text{xc}}}{\delta n_{\sigma}\left(\mathbf{r}\right)}\phi_{\mu}\left(\mathbf{r}\right)\phi_{\nu}\left(\mathbf{r}\right)+\frac{1}{2}\frac{\partial f_{\text{xc}}}{\partial\tau_{\sigma}}\nabla\chi_{i}\cdot\nabla\chi_{j}\nonumber \\
+ & \Bigg(2\frac{\delta f_{\text{xc}}}{\delta\gamma_{\sigma\sigma}\left(\mathbf{r}\right)}\nabla n_{\sigma}\left(\mathbf{r}\right)+\frac{\delta f_{\text{xc}}}{\delta\gamma_{\sigma\sigma'}\left(\mathbf{r}\right)}\nabla n_{\sigma'}\left(\mathbf{r}\right)\Bigg)\cdot\nabla\left(\phi_{\mu}\left(\mathbf{r}\right)\phi_{\nu}\left(\mathbf{r}\right)\right)\Bigg]{\rm d}^{3}r\label{eq:F-DFT}
\end{align}
The quadrature in \eqref{F-DFT} is formulated efficiently employing
matrix-matrix products.

Due to the strict locality of the radial elements, it makes sense
to do the integrals element by element, as the resulting Fock matrix
is banded diagonal. \Eqref{F-DFT} contains three quadratures: one
radial, and two angular ($\theta$ and $\phi$). The same Gauss--Chebyshev
radial quadrature is used for the radial part as for all the preceding
matrix elements. However, the angular part is performed differently.
Gauss--Chebyshev quadrature is used for the $\theta$ part, while
a uniform grid is used for the $\phi$ part as it already yields exactness
properties.\citep{Murray1993} Note that in contrast to the general
molecular case, here the angular features of the electron density
are more restricted due to the finite $m$ expansion, and so the use
of a compound rule such as Lebedev quadrature\citep{Lebedev1975,Lebedev1976}
is less efficient. We have chosen $n_{\theta}=4l_{\max}+10$ and $n_{\phi}=4m_{\max}+5$
as the default values, which should guarantee sufficient accuracy
for the quadrature even for meta-GGA functionals.

A noteworthy difference in the DFT implementation from the Cartesian
case is that due to the curvilinear coordinate system, the dot products
are computed differently as
\begin{align}
\nabla f\cdot\nabla f= & \sum_{i}\left(\frac{\hat{\boldsymbol{e}}_{i}\cdot\hat{\boldsymbol{e}}_{i}}{h_{i}^{2}}\left(\frac{\partial f}{\partial q_{i}}\right)^{2}\right)=\sum_{i}\frac{1}{h_{i}^{2}}\left(\frac{\partial f}{\partial q_{i}}\right)^{2}\label{eq:dotprod}
\end{align}
where the scale factors for spherical polar coordinates are
\begin{align}
h_{r}= & 1,\label{eq:h-rad}\\
h_{\theta}= & r,\label{eq:h-theta}\\
h_{\phi}= & r\sin\theta.\label{eq:h-phi}
\end{align}

In range-separated exchange functionals, the two-electron Coulomb
operator $1/r_{12}$ is decomposed into a short-range and a long-range
part as\citep{Leininger1997}
\begin{equation}
\frac{1}{r_{12}}=\frac{\phi_{\text{sr}}(r_{12};\omega)}{r_{12}}+\frac{1-\phi_{\text{sr}}(r_{12};\omega)}{r_{12}},\label{eq:rangesep}
\end{equation}
where $\phi_{\text{sr}}(r_{12};\omega)$ is a screening function and
$\omega$ controls the speed of the screening. In almost all commonly
used range-separated functionals, such as CAM-B3LYP;\citep{Yanai2004b}
the range-separated Minnesota functionals M11,\citep{Peverati2011b}
N12-SX,\citep{Peverati2012a} and MN12-SX;\citep{Peverati2012a} as
well as the Head-Gordon group's $\omega$B97,\citep{Chai2008} $\omega$B97X,\citep{Chai2008},
$\omega$B97X-V,\citep{Mardirossian2014a} and $\omega$B97M-V\citep{Mardirossian2016}
functionals, the weight function is chosen as 
\begin{equation}
\phi_{\text{sr}}(r;\omega)=\text{erfc}(r;\omega),\label{eq:range-erfc}
\end{equation}
as this choice is extremely convenient for implementation in programs
employing Gaussian basis sets.\citep{Adamson1999,Ahlrichs2006} The
implementation of the range-separated functionals in the present approach
would require the calculation of a Laplace expansion alike \eqref{laplace}
for $\phi_{\text{sr}}(r_{12};\omega)/r_{12}$, which is outside the
scope of the present work.

\section{Computational details\label{sec:Computational-details}}

The equations presented above in \secref{Theory} have been implemented
\textsc{HelFEM} in C++, employing the \textsc{Armadillo} library for
linear algebra.\citep{Sanderson2016,Sanderson2018} Efficient basic
linear algebra subroutine (BLAS) libraries are used for the matrix
operations with \textsc{Armadillo}. \textsc{OpenMP} parallellization
is used throughout the program.

The one-electron and primitive two-electron integrals are computed
once at the beginning of the calculation, and stored in memory. Radial
integrals are evaluated with $5N_{p}$ points, which we have estimated
to be sufficient even for the highly non-linear integrals in DFT,
$N_{p}$ being the number of shape functions per element. The memory
requirements for the integrals are small, as instead of the full two-electron
integral tensor, only the auxiliary integrals are stored. Furthermore,
only the intraelement auxiliary integrals are stored as a rank-4 tensor,
whereas the interelement integrals are stored in factorial form, which
also allows for faster formation of the Coulomb and exchange matrices
as was described above in the Theory section.

The \textsc{Libxc} library\citep{Lehtola2018} is used to evaluate
all exchange-correlation functionals. The core guess, \emph{i.e.}
eigenvectors of $\boldsymbol{H}_{0}=\boldsymbol{T}+\boldsymbol{V}$
are used for initialization of the SCF calculations, and the Aufbau
principle is employed to determine orbital occupations during the
SCF cycle, unless the occupied orbital symmetries have been explicitly
specified. Convergence of the SCF procedure is accelerated with a
combination of the DIIS and ADIIS accelerators.\citep{Pulay1980,Pulay1982,Hu2010}
Unless otherwise stated, the calculations have been converged to an
orbital gradient \emph{i.e.} DIIS error of $10^{-7}$.

Calculations can be performed in \textsc{HelFEM} with fully spin-restricted
orbitals, restricted open-shell orbitals via the constrained unrestricted
HF update,\citep{Tsuchimochi2010a,Tsuchimochi2011} or fully spin-unrestricted
orbitals. The orbitals are updated by full diagonalization. Depending
on the targeted orbital symmetry, the diagonalization can be performed
in angular subblocks: by default, the diagonalization splits by $m$
block, a symmetry which is maintained even under an electric field
unless the orbitals break symmetry.

\section{Results\label{sec:Results}}

\subsection{Choice of element type and radial grid\label{subsec:Choice-of-element-type}}

To apply the new \textsc{HelFEM} program to calculations, we must
first establish the best way to use it. As in the atomic case the
angular basis is determined by the occupied orbital symmetries that
are typically known in advance, the only remaining question is the
radial basis. As FEM calculations can be converged to the basis set
limit either by increasing the \emph{number} of elements, or by increasing
their \emph{order}, the question is which approach yields the fastest
convergence for a given number of basis functions. We shall first
tackle the question of the radial grid, which has long been recognized
as crucial to the efficiency of real-space approaches.\citep{Gazquez1977}
In contrast to finite-difference approaches that typically use a logarithmic
radial coordinate,\citep{Froese1963} the present implementation employs
an untransformed $r$ coordinate; thus, in analogy to previously published
B-spline implementations,\citep{Saito2003,FroeseFischer2011} the
optimal element spacing is probably not an uniform one. 

Although adaptive approaches could be used to determine the most efficient
element grid -- see \emph{e.g.} \citeref{Romanowski2009} and \citerefs{Flores1989, Flores1989a}
for \emph{h}-adaptive and \emph{p}-adaptive approaches, respectively
-- it is evident that such an approach, while certainly possible,
is not necessary given the high amount of symmetry present in the
atomic problem. As the only problem is to determine a suitably accurate
radial grid, and as atomic calculations are not computationally costly
even with large grids, it suffices to just pick a grid large enough
to yield a fully converged result. We will thus focus on universal
optimizations of the element grid by global parametrizations of the
placement of the elements in order to yield efficient grids for all
atoms. The question is thus: what is the optimal way to arrange the
elements?

The radial elements span the range $[0,r_{\infty}]$ where $r_{\infty}=40a_{0}$
typically yields converged results, whereas larger values of $r_{\infty}$
may be required for loosely bound anions.\citep{Saito2003,Saito2009}
We have studied the problem by using $N_{\text{el}}$ elements with
uniform node spacing within the element, and varied the size distribution
of the elements. The elements are defined by the placement of the
borders between the elements, defined by the array $r_{i}$, with
the $i$:th element ranging from $r_{i}$ to $r_{i+1}$, with the
numbering starting from 0. We have chosen to study four different
types of element spacings:
\begin{enumerate}
\item a linear grid\label{enu:linear-grid}
\begin{equation}
r_{i}=\frac{i}{N}R_{\infty}\label{eq:lingrid}
\end{equation}
\emph{i.e.} $N$ uniform elements,
\item a quadratic grid\label{enu:quadratic-grid}
\begin{equation}
r_{i}=\frac{i^{2}}{N^{2}}R_{\infty}\label{eq:quadgrid}
\end{equation}
which places leads to a denser grid near the nucleus and which has
been previously suggested to be optimal for atoms,\citep{Schweizer1999}
\item a generalized polynomial grid\label{enu:a-generalized-polynomial}
\begin{equation}
r_{i}=\frac{i^{x}}{N^{x}}R_{\infty}\label{eq:polygrid}
\end{equation}
\emph{i.e.} a generalization of the linear and quadratic grids to
arbitrary order, resulting in a denser grid near the nucleus for higher
$x$ values, where $x$ is a constant defining the grid,
\item an exponential grid\label{enu:a-generalized-logarithmic}
\begin{equation}
r_{i}=\left(1+R_{\infty}\right)^{i^{x}/N^{x}}-1\label{eq:exploggrid}
\end{equation}
which leads to even denser grids near the nucleus than the generalized
polynomial grid above.
\end{enumerate}
Note that $x$ in \eqref{polygrid,exploggrid} has nothing to do with
the primitive coordinate system used in the quadrature in \eqrangeref{rquad}{dquad}.

Because the generalized polynomial grid yields the linear and quadratic
grids with $x=1$ and $x=2$, respectively, it suffices to study the
performance of the generalized polynomial and exponential grids in
the following. While a larger value of $x$ results in more points
in the energy-sensitive regions near the nucleus, it also results
in less points \emph{i.e.} a poorer description in the valence region,
implying that $x$ cannot be chosen arbitrarily large.

The radial element grid turns out to be sensitive to the type of the
used elements (LIP, HIP, 2$^{\text{nd}}$ order HIP, \emph{etc.}),
necessitating separate grid analyses for each element type. Despite
claims to the contrary,\citep{Ram-Mohan1990,Ram-Mohan2002} we have
found LIPs to outperform HIPs by a wide margin. Choosing $r_{\infty}=40a_{0}$,
\figref{eltest} shows scans for the optimal element grid for argon
for calculations with six-node uniform LIP elements, three-node uniform
HIP elements, and two-node $2^{\text{nd}}$ order uniform HIP elements,
all corresponding to the use of a fifth-order primitive expansion.
Because the larger number of functions overlayed across elements in
HIP calculations leads to a fewer number of basis functions than in
LIP calculations, the number of elements for the HIP calculations
have been adjusted so that the number of HIP and LIP functions match
as closely as possible.

\Figref{eltest} shows that the best result in the exponential grid
is orders of magnitude better than the best result in the polynomial
grids, which include the commonly used linear and quadratic element
grids. This result holds regardless of the element type: for LIPs,
for HIPs of the first order, and for HIPs of the second order. The
results for other noble atoms are similar to \figref{eltest} (not
shown).

Interestingly enough, even though the HIP elements yield significantly
better results than the LIP elements when a linear or an exponential
element grid with $x=1$ is employed, with the $2^{\text{nd}}$ degree
HIP outperforming the ($1^{\text{st}}$ degree) HIP, this ranking
changes radically when the element grid is optimized. The HIPs have
a sharp minimum around $x=1$ with the exponential grid, whereas for
LIPs the performance can be significantly improved by tuning the value
of $x$ with the exponential grid. The polynomial grid yields worse
results for all three kinds of elements. Note that even though the
LIP basis does not explicitly enforce continuity of the derivative
\begin{equation}
\phi_{i}'(r)=\sum_{j}\frac{1}{r_{i}-r_{j}}\prod_{k=0,k\neq i,k\neq j}^{n-1}\frac{r-r_{k}}{r_{i}-r_{k}},\label{eq:LIP-der}
\end{equation}
at the element boundaries in contrast to HIPs, this does not mean
that the derivatives will be non-continuous across the boundary for
LIPs. Namely, given the freedom of \eqref{LIP-der}, the variational
principle will strive to make the derivative continuous across element
boundaries even for LIPs, as a non-continuous derivative implies a
higher kinetic energy.
\begin{center}
\begin{figure}
\begin{centering}
\subfloat[Six-node LIP: 5 elements (blue), 10 elements (red), and 20 elements
(cyan), yielding 24, 49, and 99 radial functions.]{\begin{centering}
\includegraphics[width=0.33\textwidth]{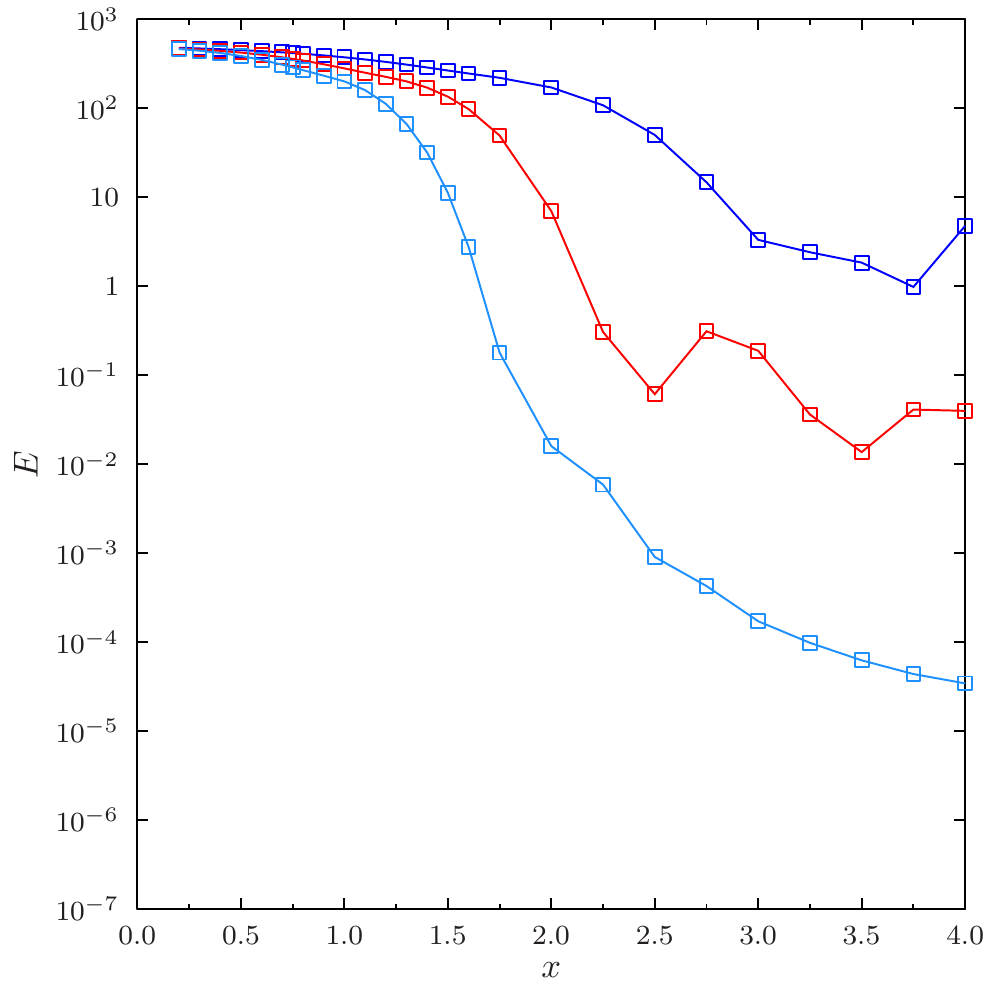}\includegraphics[width=0.33\textwidth]{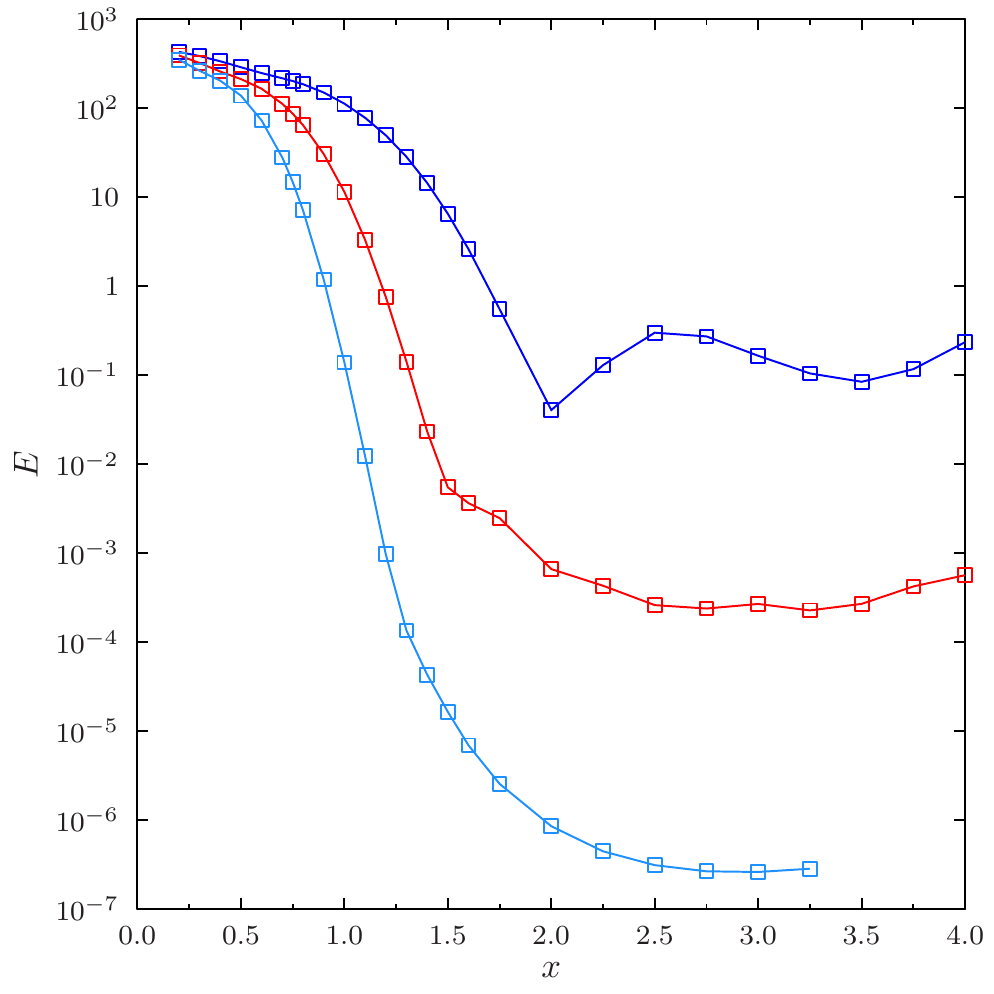}
\par\end{centering}
\centering{}}
\par\end{centering}
\begin{centering}
\subfloat[Three-node HIP: 6 elements (blue), 12 elements (red), and 24 elements
(cyan), yielding 23, 47, and 95 radial functions.]{\begin{centering}
\includegraphics[width=0.33\textwidth]{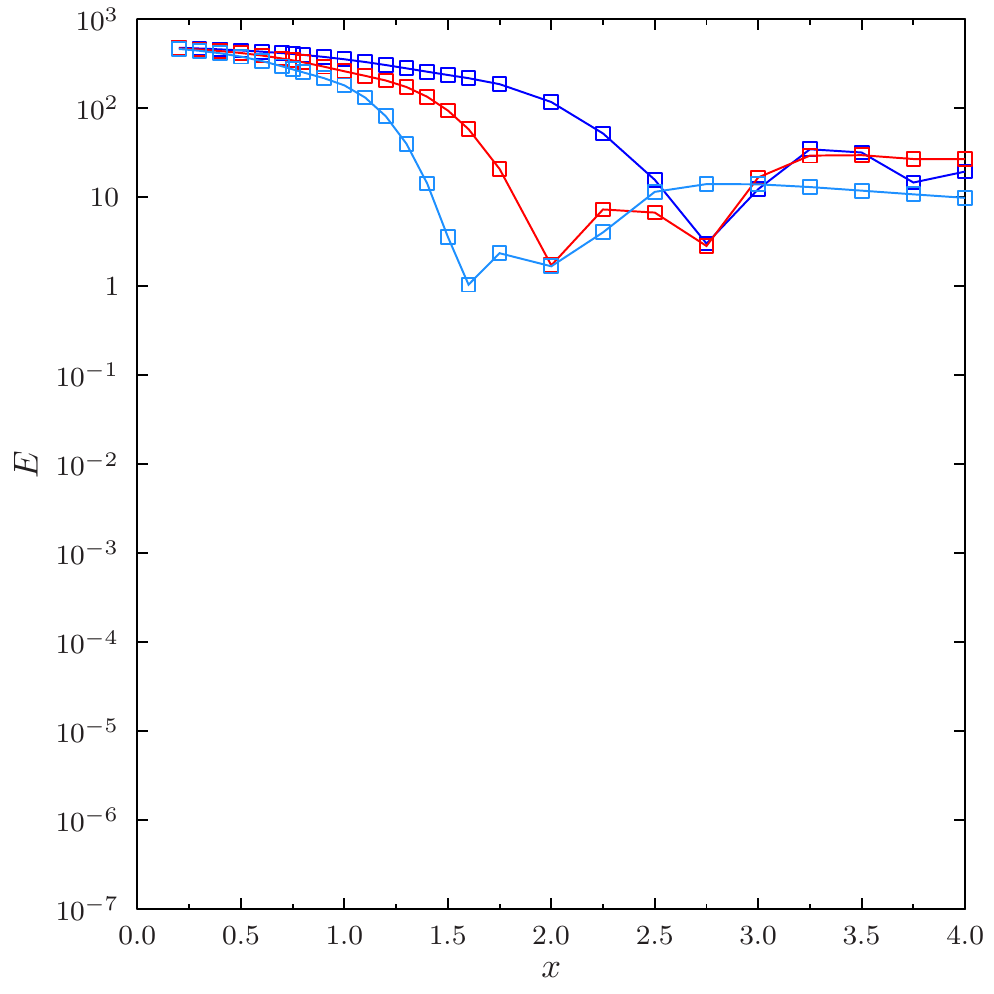}\includegraphics[width=0.33\textwidth]{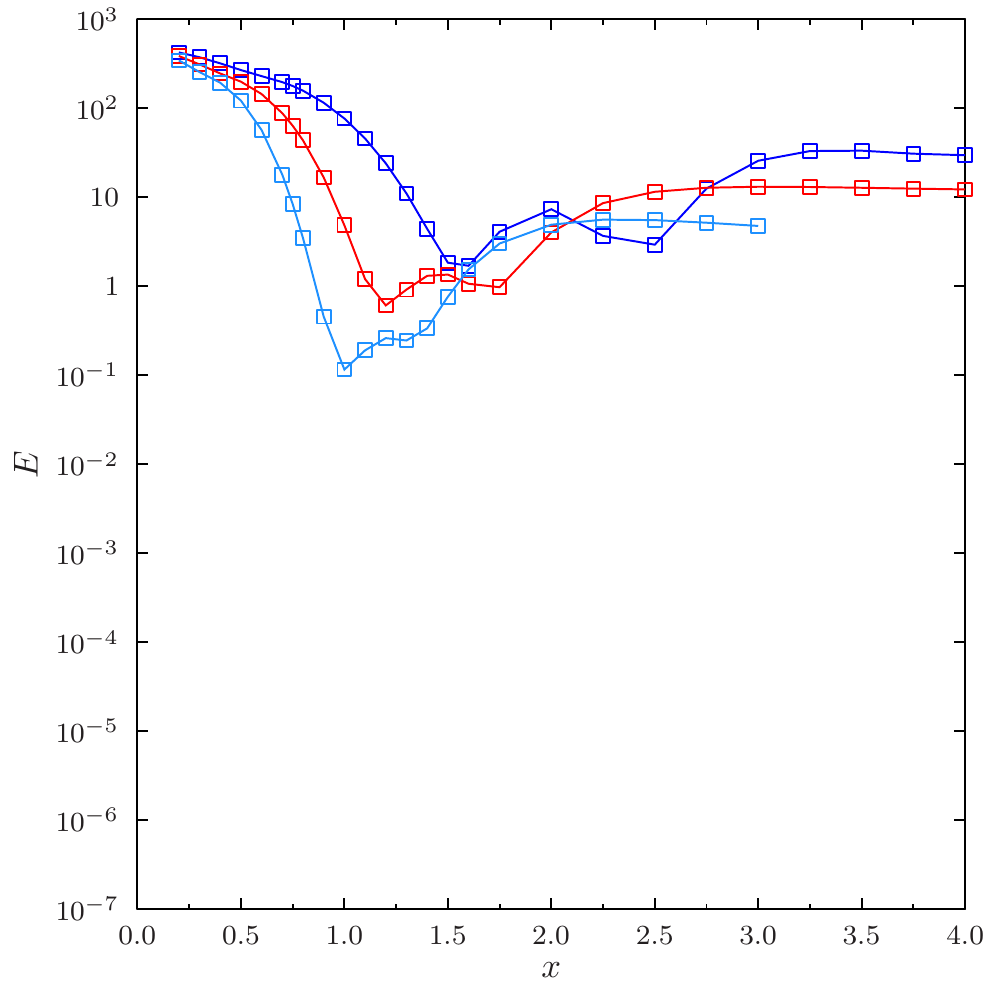}
\par\end{centering}
\centering{}}
\par\end{centering}
\begin{centering}
\subfloat[Two-node $2^{\text{nd}}$ order HIP: 8 elements (blue), 16 elements
(red), and 32 elements (cyan), yielding 23, 47, and 95 radial functions.]{\begin{centering}
\includegraphics[width=0.33\textwidth]{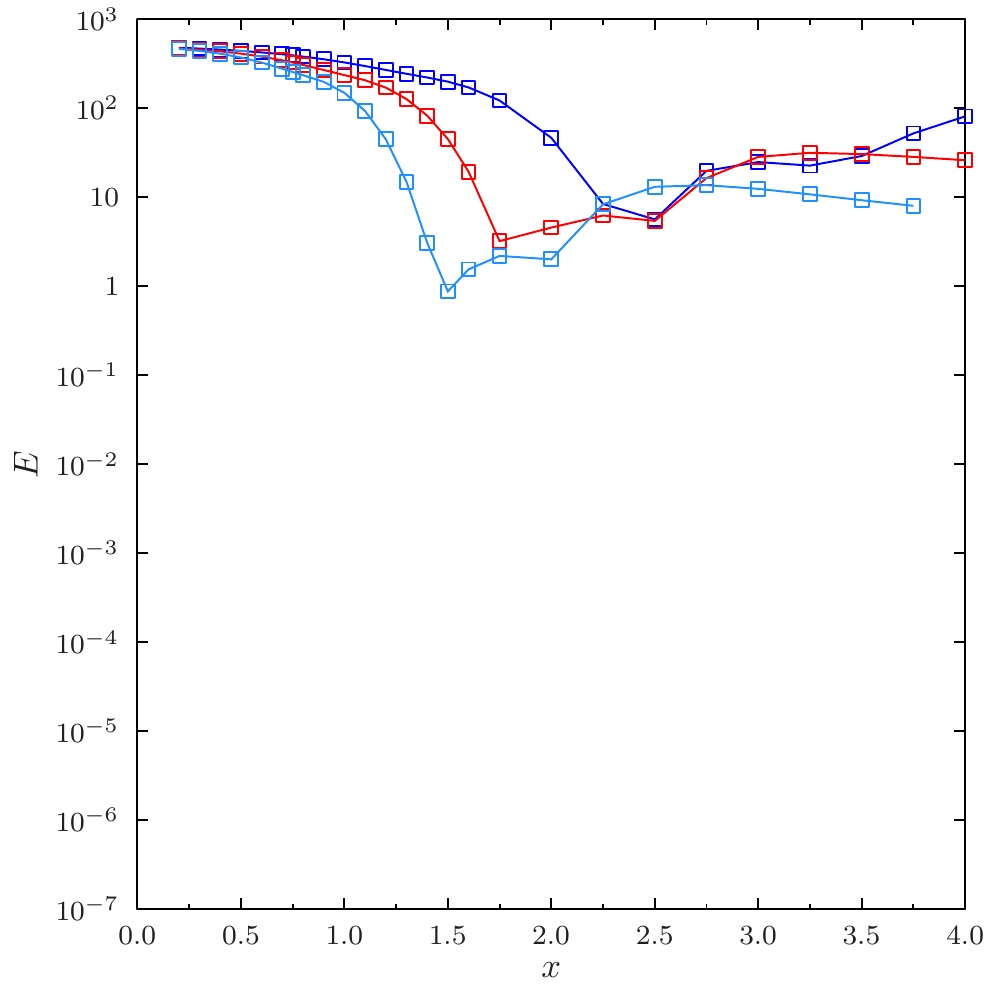}\includegraphics[width=0.33\textwidth]{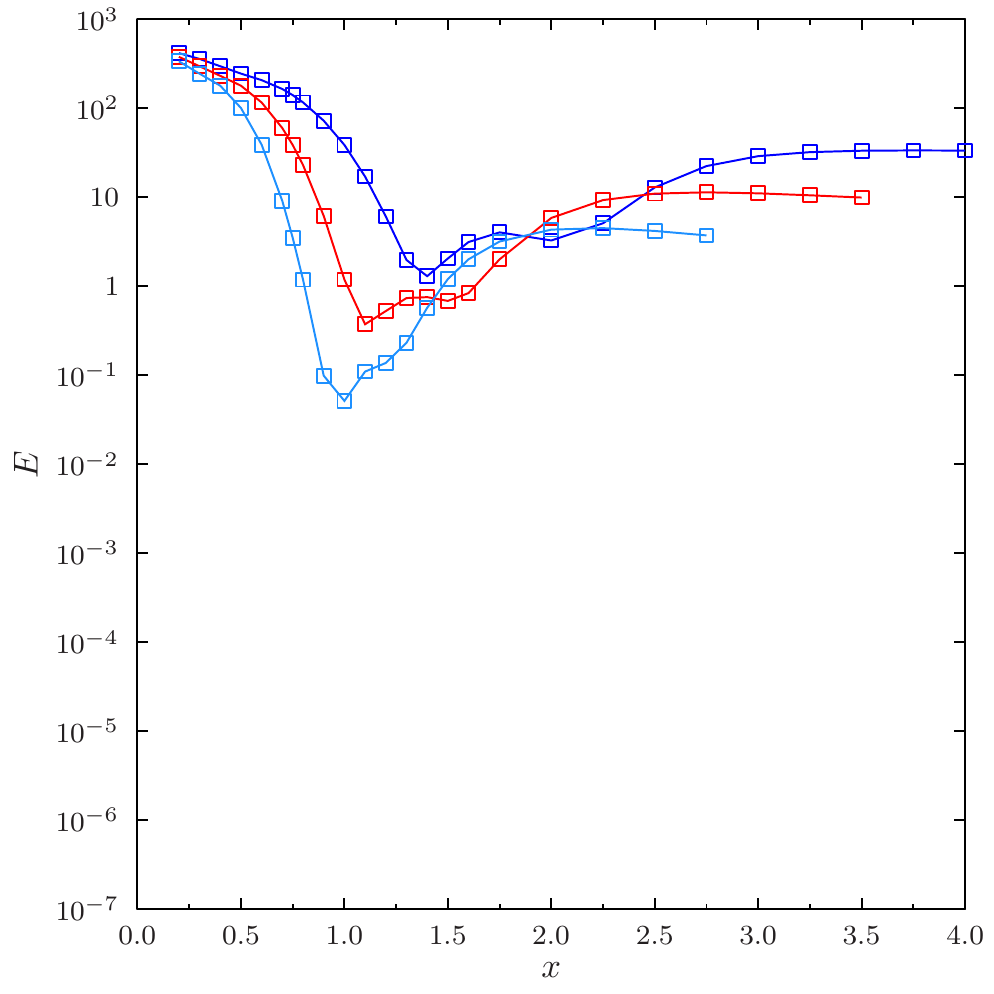}
\par\end{centering}
\centering{}}
\par\end{centering}
\caption{Error in HF energy for argon with a polynomial (left) or exponential
(right) grid, employing a different types of elements. The reference
energy is\citep{Saito2009} -526.817512803.\label{fig:eltest}}
\end{figure}
\par\end{center}

\begin{figure}
\begin{centering}
\includegraphics[width=0.33\textwidth]{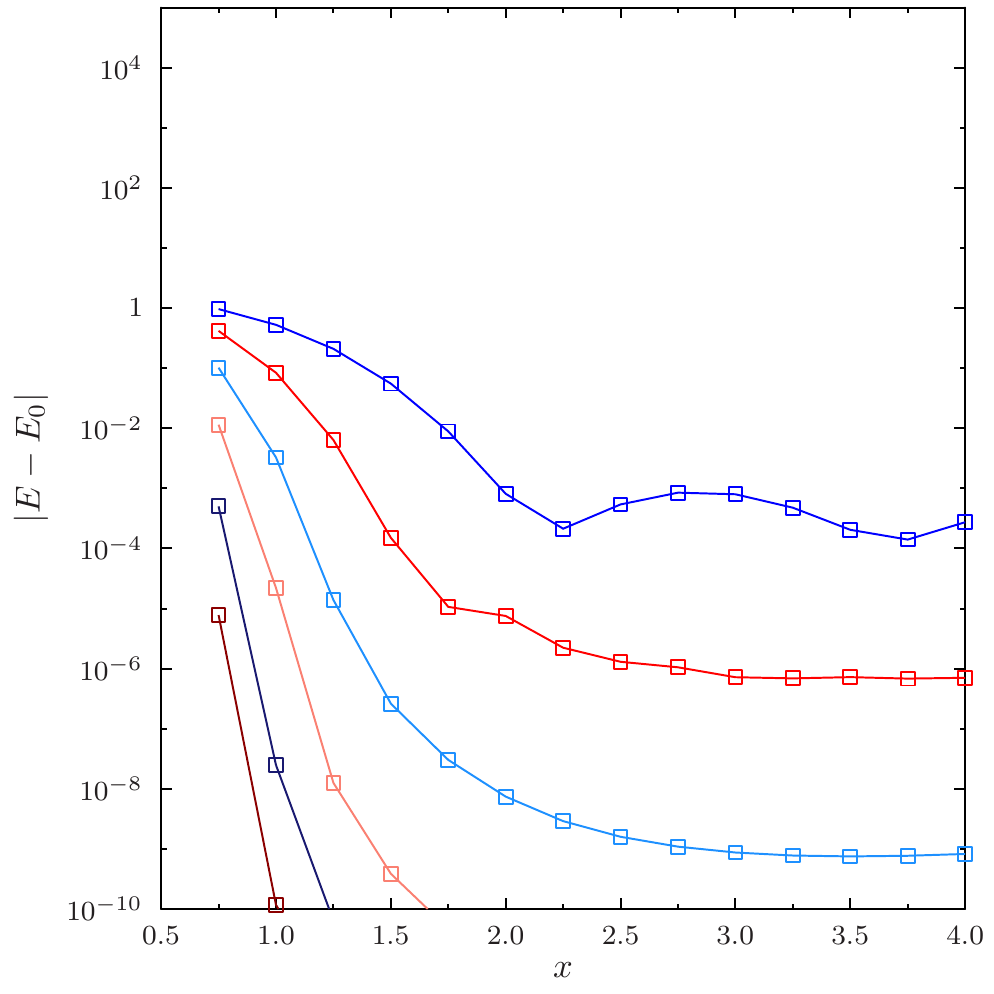}\includegraphics[width=0.33\textwidth]{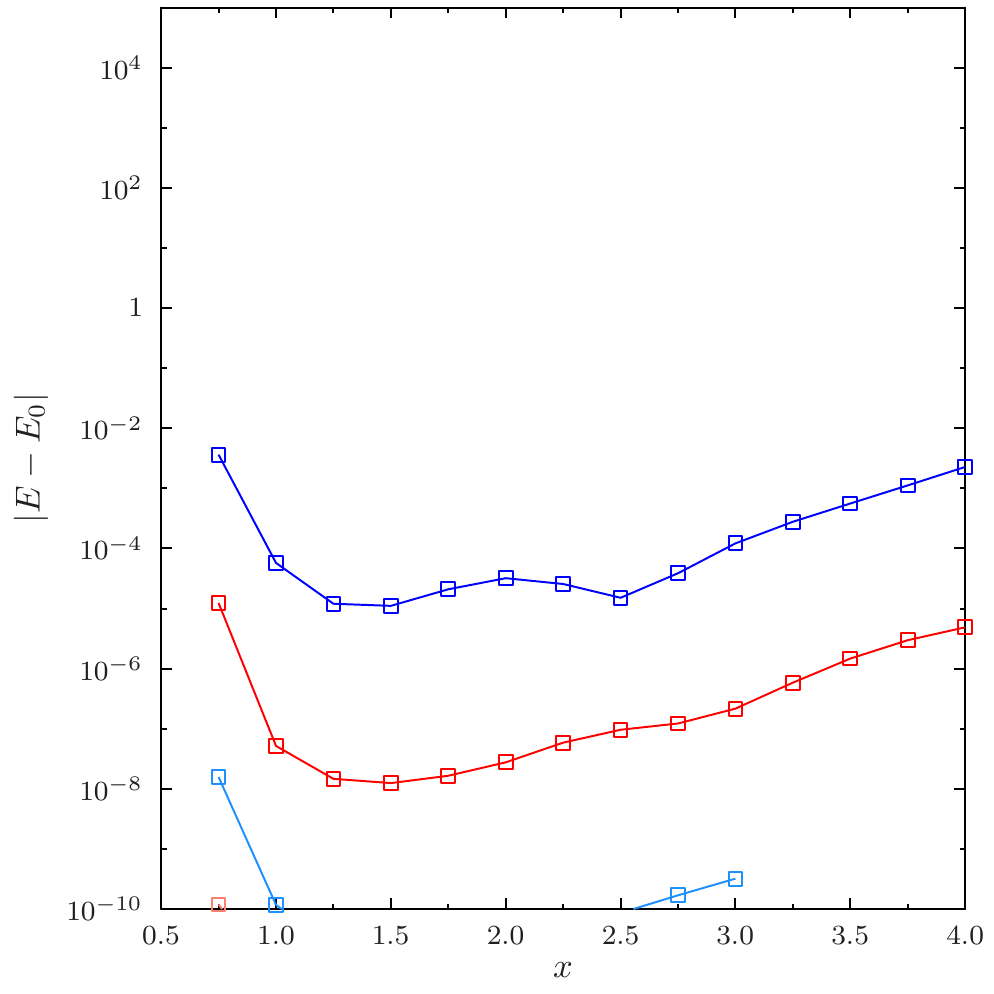}
\par\end{centering}
\caption{Error in HF energy for helium with a polynomial (left) or exponential
(right) grid, employing 5 (blue), 10 (red), 20 (light blue), 40 (salmon),
80 (dark blue), or 160 elements (dark red) employing six-node uniform
LIPs. The reference energy is\citep{Saito2009} -2.8616799956.\label{fig:He}}
\end{figure}

\begin{figure}
\begin{centering}
\includegraphics[width=0.33\textwidth]{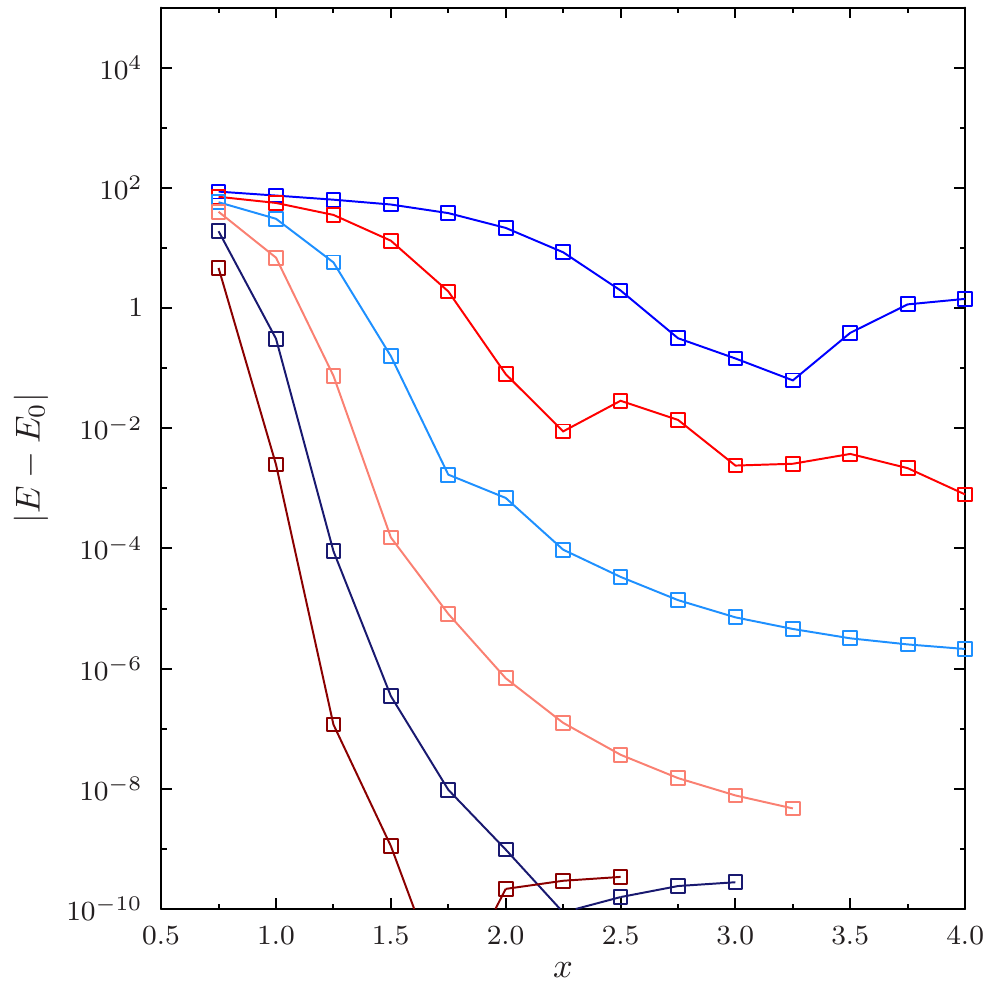}\includegraphics[width=0.33\textwidth]{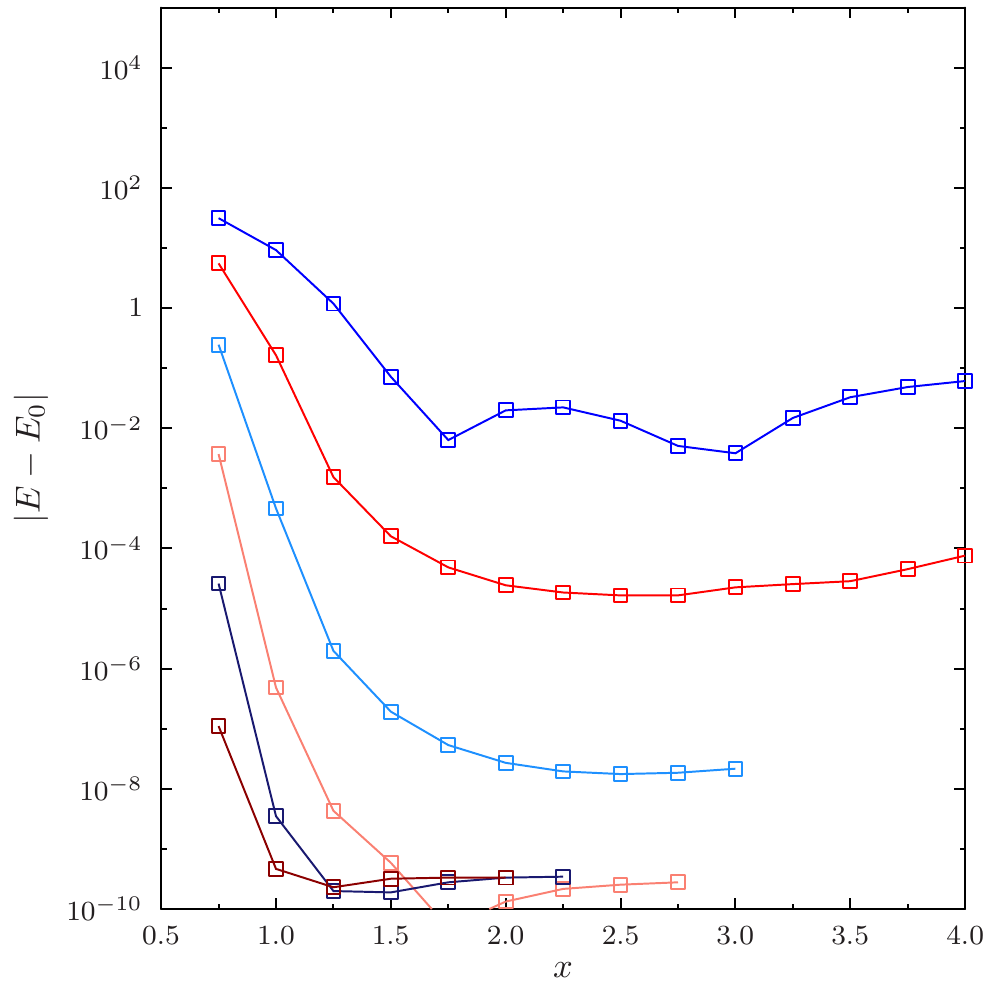}
\par\end{centering}
\caption{Error in HF energy for neon with a polynomial (left) or exponential
(right) grid, employing 5 (blue), 10 (red), 20 (light blue), 40 (salmon),
80 (dark blue), or 160 elements (dark red) employing six-node uniform
LIPs. The reference energy is\citep{Saito2009} -128.547098109.\label{fig:Ne}}
\end{figure}

\begin{figure}
\begin{centering}
\includegraphics[width=0.33\textwidth]{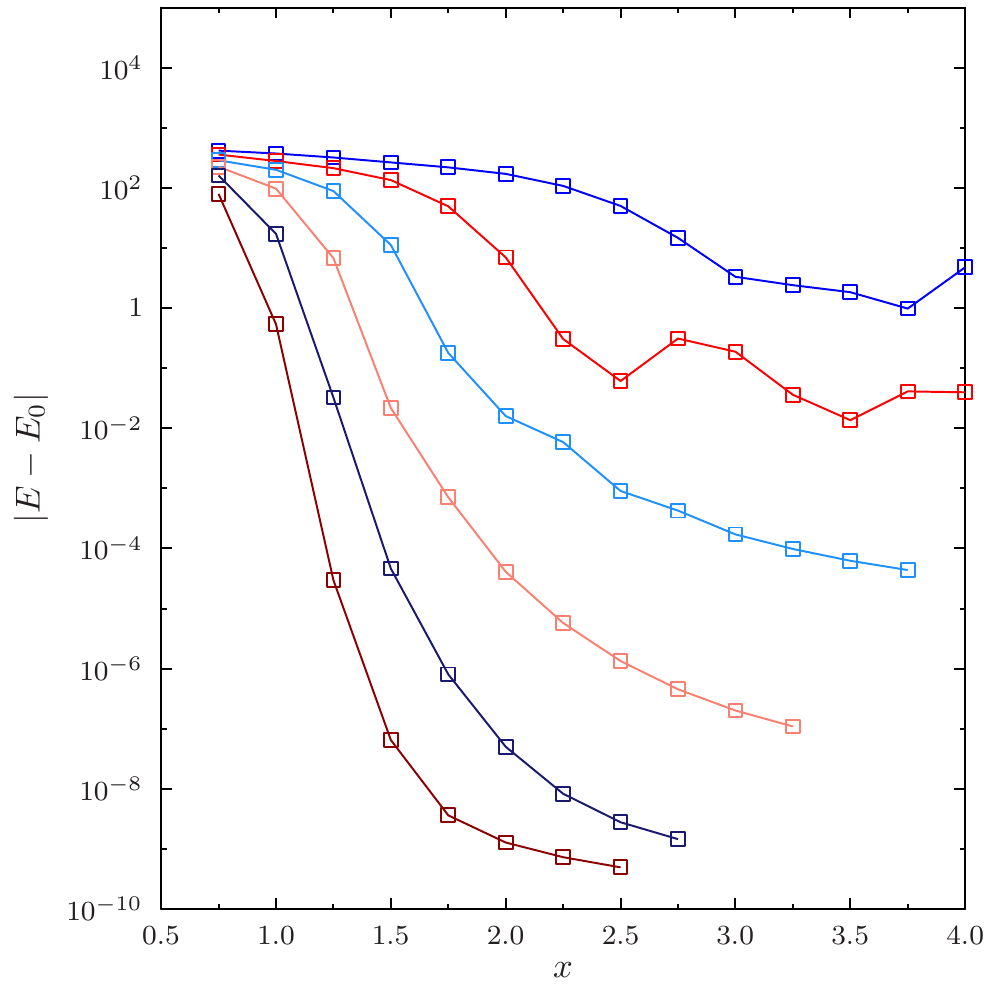}\includegraphics[width=0.33\textwidth]{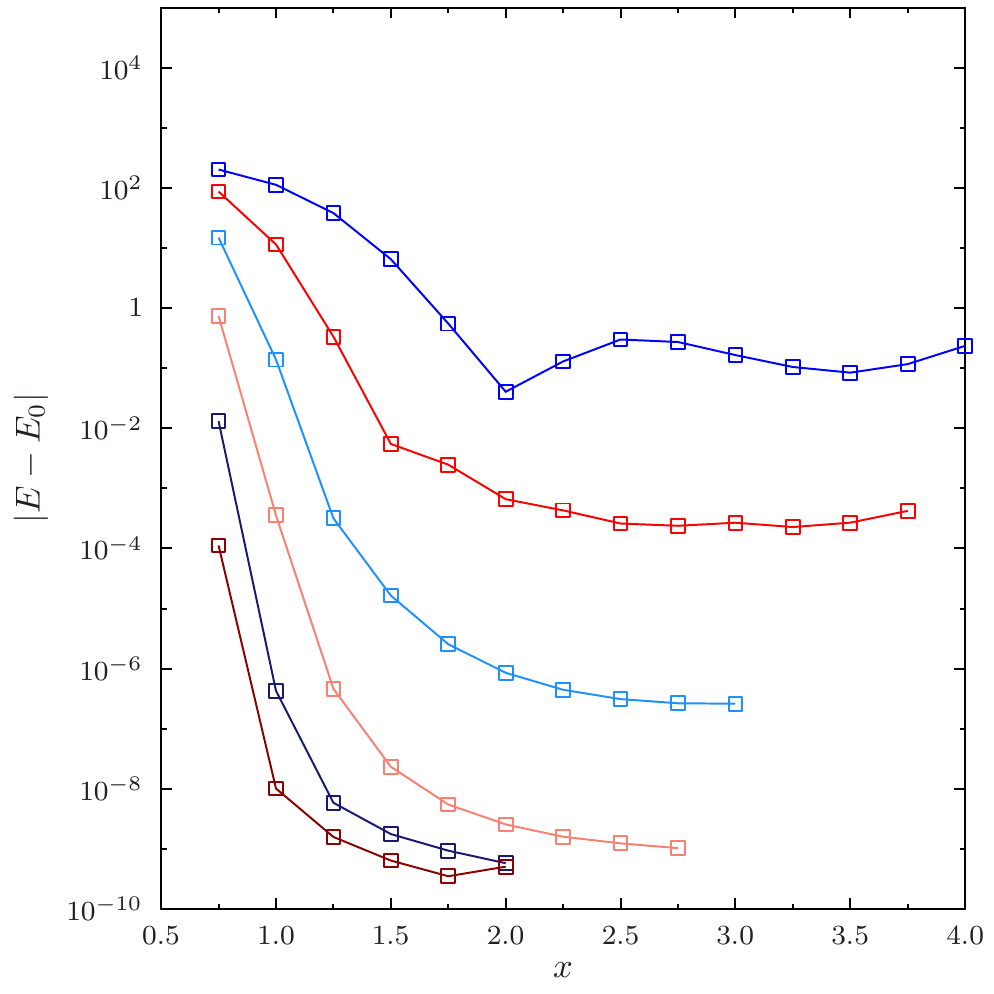}
\par\end{centering}
\caption{Error in HF energy for argon with a polynomial (left) or exponential
(right) grid, employing 5 (blue), 10 (red), 20 (light blue), 40 (salmon),
80 (dark blue), or 160 elements (dark red) employing six-node uniform
LIPs. The reference energy is\citep{Saito2009} -526.817512803.\label{fig:Ar}}
\end{figure}

\begin{figure}
\begin{centering}
\includegraphics[width=0.33\textwidth]{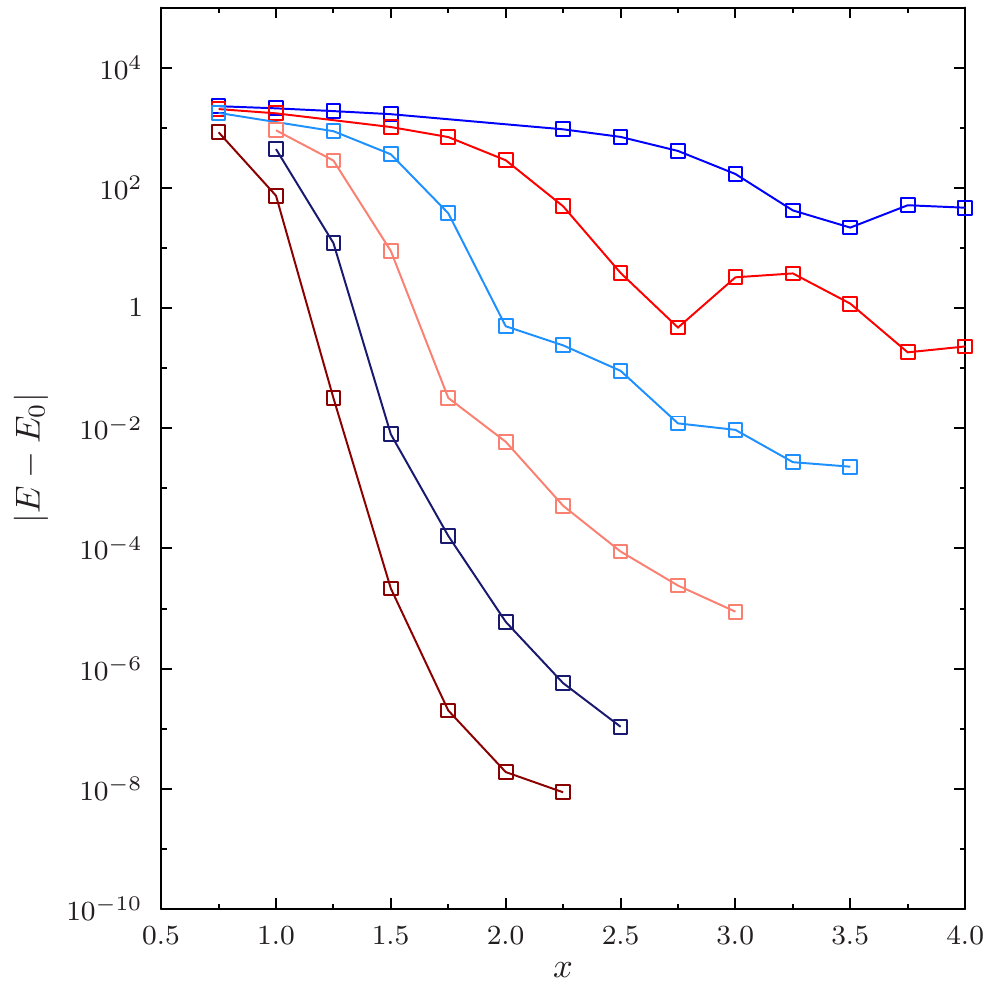}\includegraphics[width=0.33\textwidth]{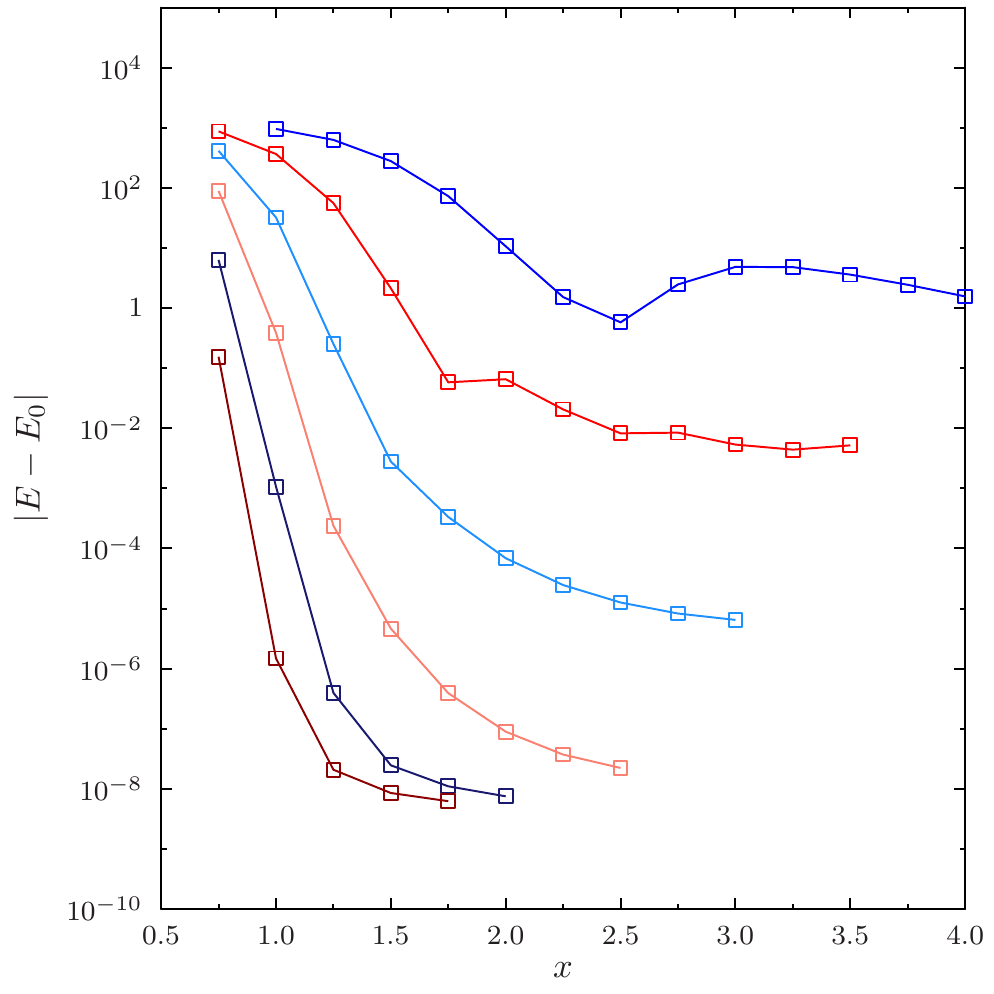}
\par\end{centering}
\caption{Error in HF energy for krypton with a polynomial (left) or exponential
(right) grid, employing 5 (blue), 10 (red), 20 (light blue), 40 (salmon),
80 (dark blue), or 160 elements (dark red) employing six-node uniform
LIPs. The reference energy is\citep{Saito2009} -2752.05497735.\label{fig:Kr}}
\end{figure}

\begin{figure}
\begin{centering}
\includegraphics[width=0.33\textwidth]{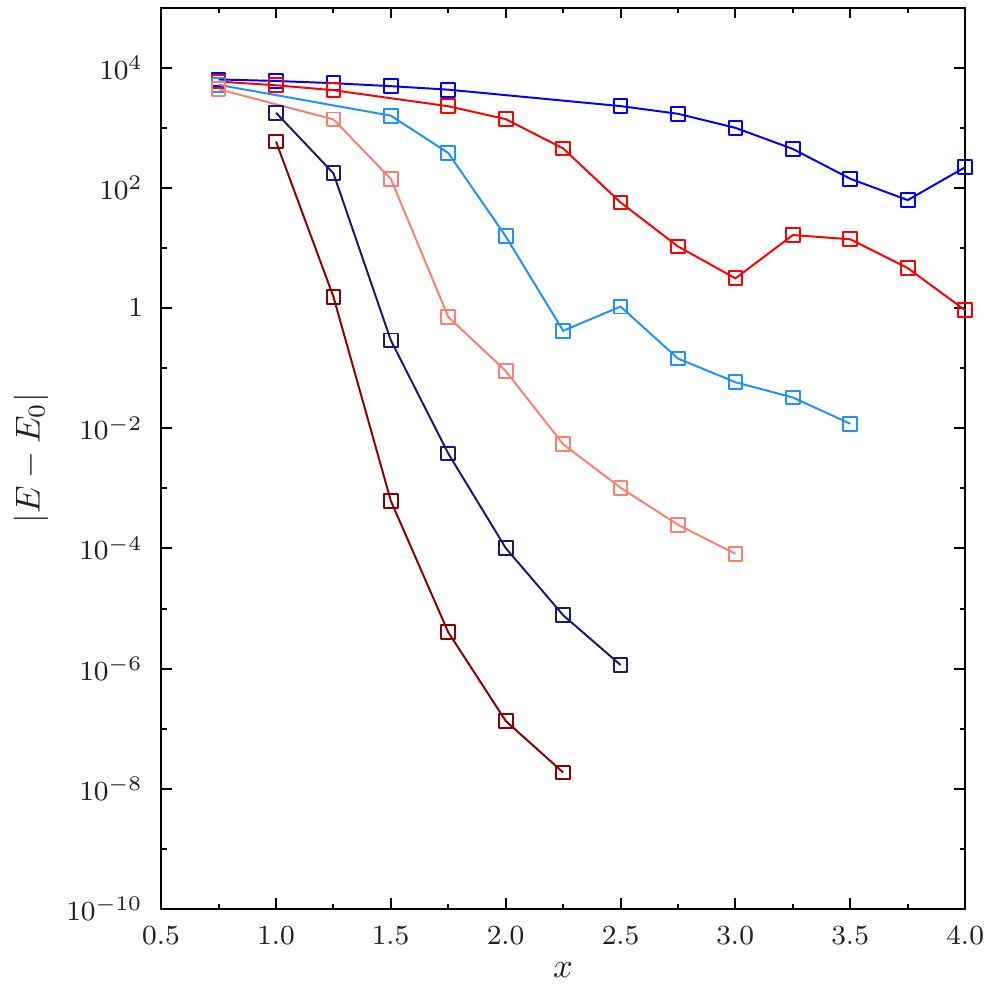}\includegraphics[width=0.33\textwidth]{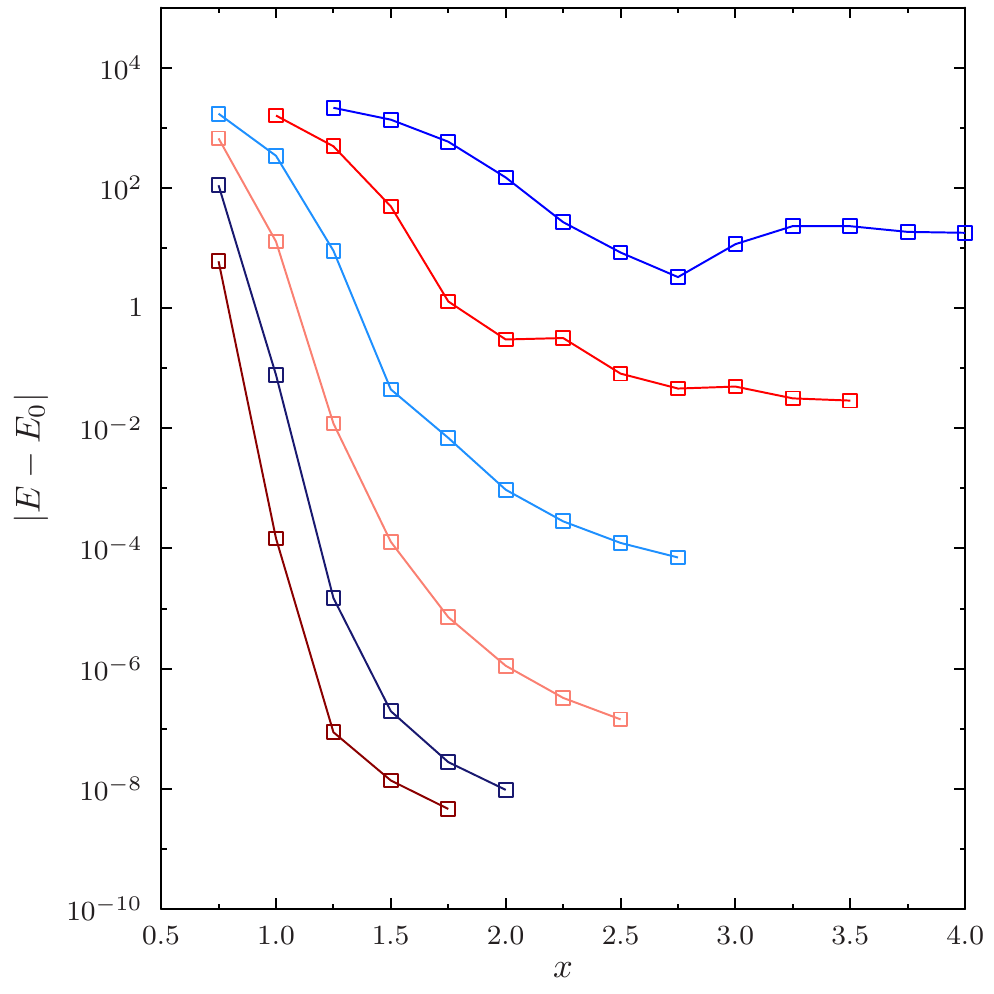}
\par\end{centering}
\caption{Error in HF energy for xenon with a polynomial (left) or exponential
(right) grid, employing 5 (blue), 10 (red), 20 (light blue), 40 (salmon),
80 (dark blue), or 160 elements (dark red) employing six-node uniform
LIPs. The reference energy is\citep{Saito2009} -7232.13836387.\label{fig:Xe}}
\end{figure}

\begin{figure}
\begin{centering}
\includegraphics[width=0.33\textwidth]{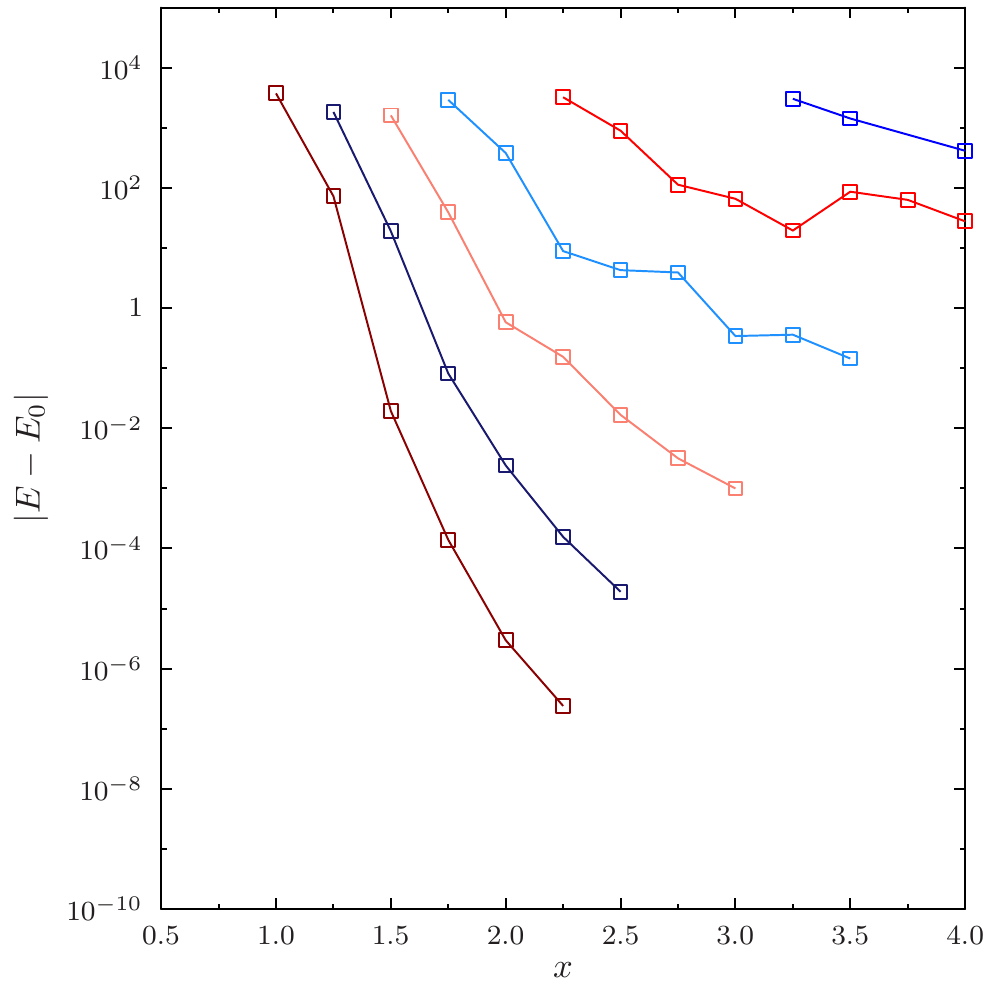}\includegraphics[width=0.33\textwidth]{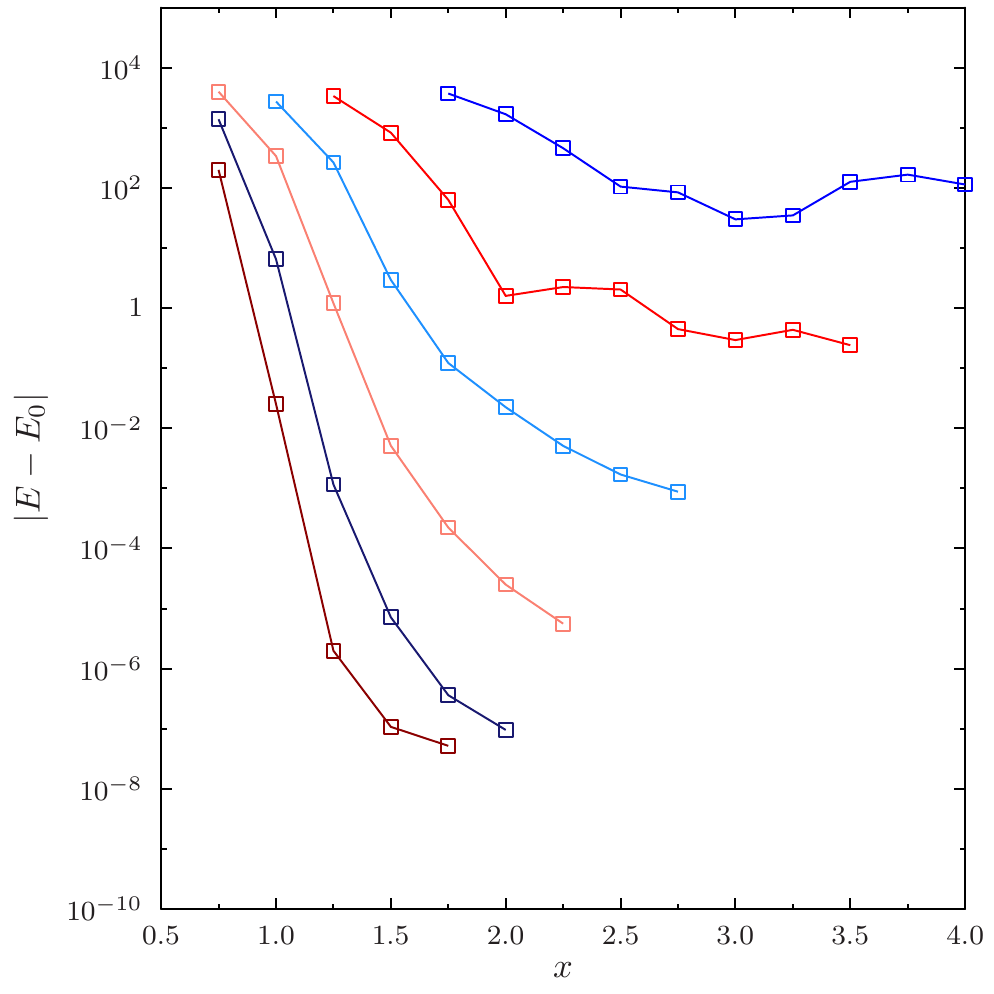}
\par\end{centering}
\caption{Error in HF energy for radon with a polynomial (left) or exponential
(right) grid, employing 5 (blue), 10 (red), 20 (light blue), 40 (salmon),
80 (dark blue), or 160 elements (dark red) employing six-node uniform
LIPs. The reference energy is\citep{Saito2009} -21866.7722409.\label{fig:Rn}}
\end{figure}

\begin{figure}
\begin{centering}
\includegraphics[width=0.33\textwidth]{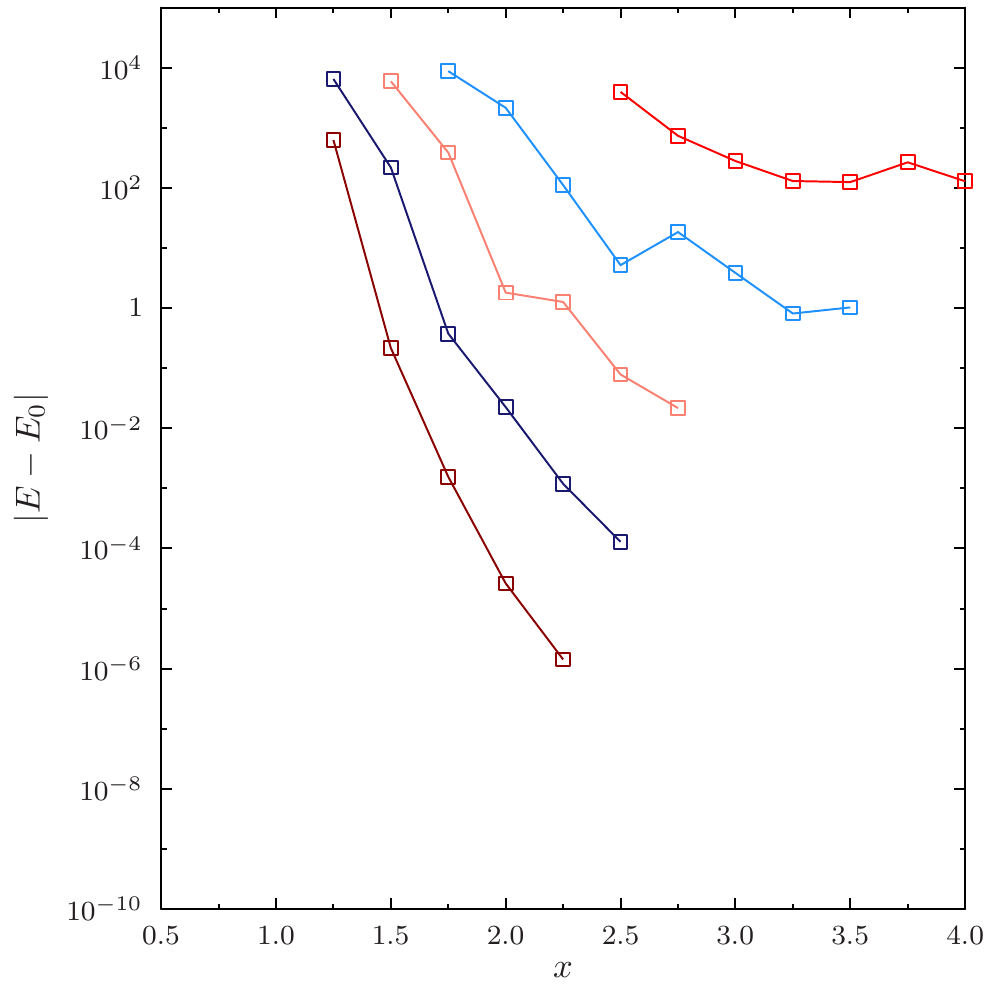}\includegraphics[width=0.33\textwidth]{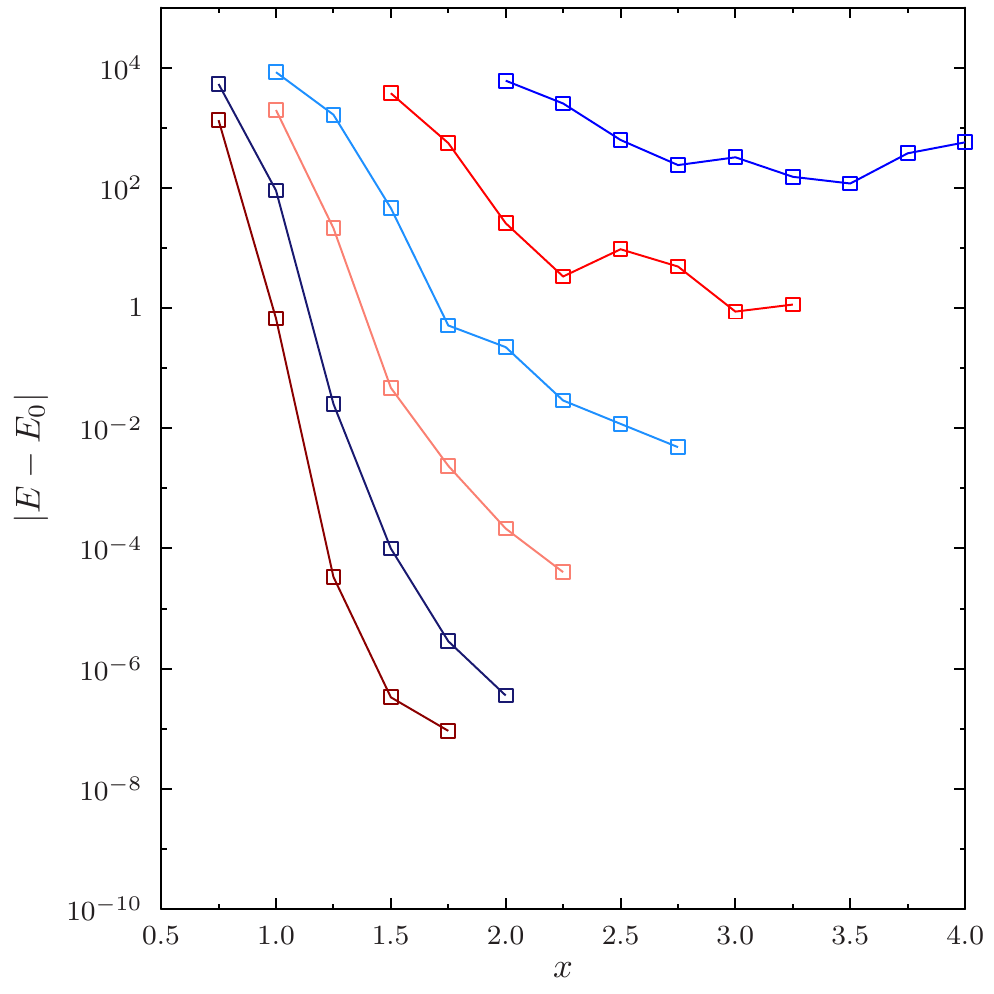}
\par\end{centering}
\caption{Error in HF energy for oganesson with a polynomial (left) or exponential
(right) grid, employing 5 (blue), 10 (red), 20 (light blue), 40 (salmon),
80 (dark blue), or 160 elements (dark red) employing six-node uniform
LIPs. The reference energy is\citep{Saito2009} -46324.3558151.\label{fig:Og}}
\end{figure}

Having determined that LIPs are a better basis than HIPs, we continue
by determining the optimal element grid. Employing $N=5$, $N=10$,
$N=20$, $N=40$, $N=80$, and $N=160$ elements with 6-node uniform
LIPs, we obtain the errors in the HF energies of the noble elements
compared to literature values (\citeref{Saito2009}) shown in \figrangeref{He}{Og},
with values of $x$ ranging from 0.75 to 4.0 with a spacing of 0.25.
Points not shown on the plots failed to converge to the used threshold,
indicating the grid offers a poor description of the wave function.

As can be seen from the results, sublinear grids $x<1$ yield poor
results even for helium, while increasing the value of $x$ dramatically
improves the basis set. Even though the quadratic grid (polynomial
with $x=2$) is indeed better than the linear grid (polynomial with
$x=1$) as suggested in \citeref{Schweizer1999}, we find that much
better results are obtained with the exponential grid (\eqref{exploggrid}),
which is also less sensitive to the chosen value of $x$ than the
polynomial grid (\eqref{polygrid}). Based on these results, we have
chosen the default grid for atomic calculations to be the exponential
one with $x=2$, which appears to offer the best compromise between
convergence and stability.

\subsection{Choice of element order\label{subsec:Choice-of-element-order}}

The supremacy of LIPs is very convenient for calculations, as LIPs
can be made numerically stable even at high orders, as was discussed
above in the Theory section. We now proceed by studying the efficiency
of LIPs with various numbers of nodes. For low numbers of nodes, the
primitive expansion (\eqref{mateq}) with uniform node spacing, the
analytical LIP expressions (\eqref{LIP}) with Lobatto node spacing,
and Legendre polynomials (\eqrangeref{P-shape}{P-N}) all yield similar
results (not shown). For higher numbers of nodes, the primitive expansion
is no longer numerically stable, but the Lobatto scheme and Legendre
polynomials still yield similar results (not shown). Thus, we have
chosen the Lobatto elements as the default, as they can be easily
obtained, and employ them to study the speed of convergence to the
basis set limit.

\begin{figure}
\begin{centering}
\includegraphics[width=0.33\textwidth]{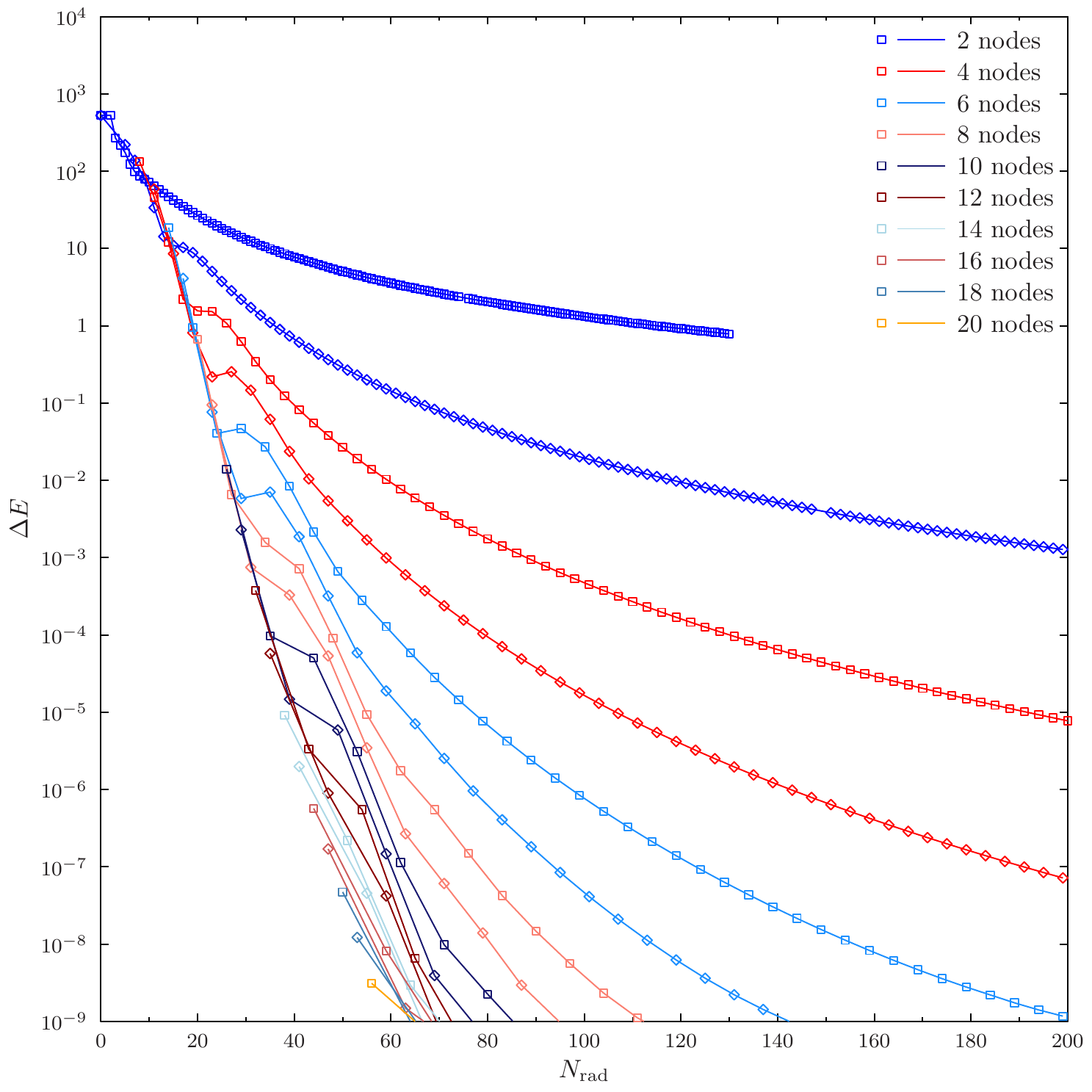}\includegraphics[width=0.33\textwidth]{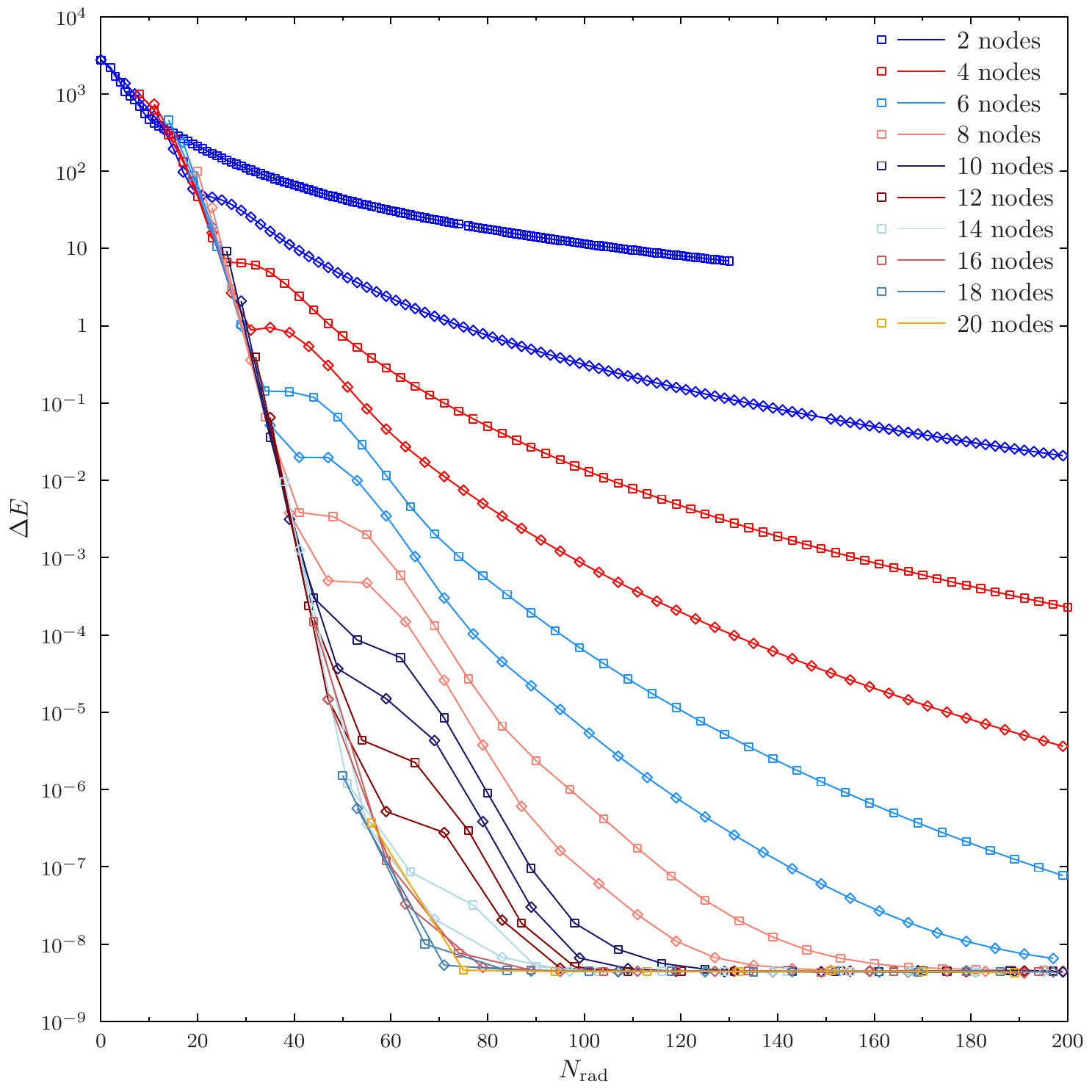}\includegraphics[width=0.33\textwidth]{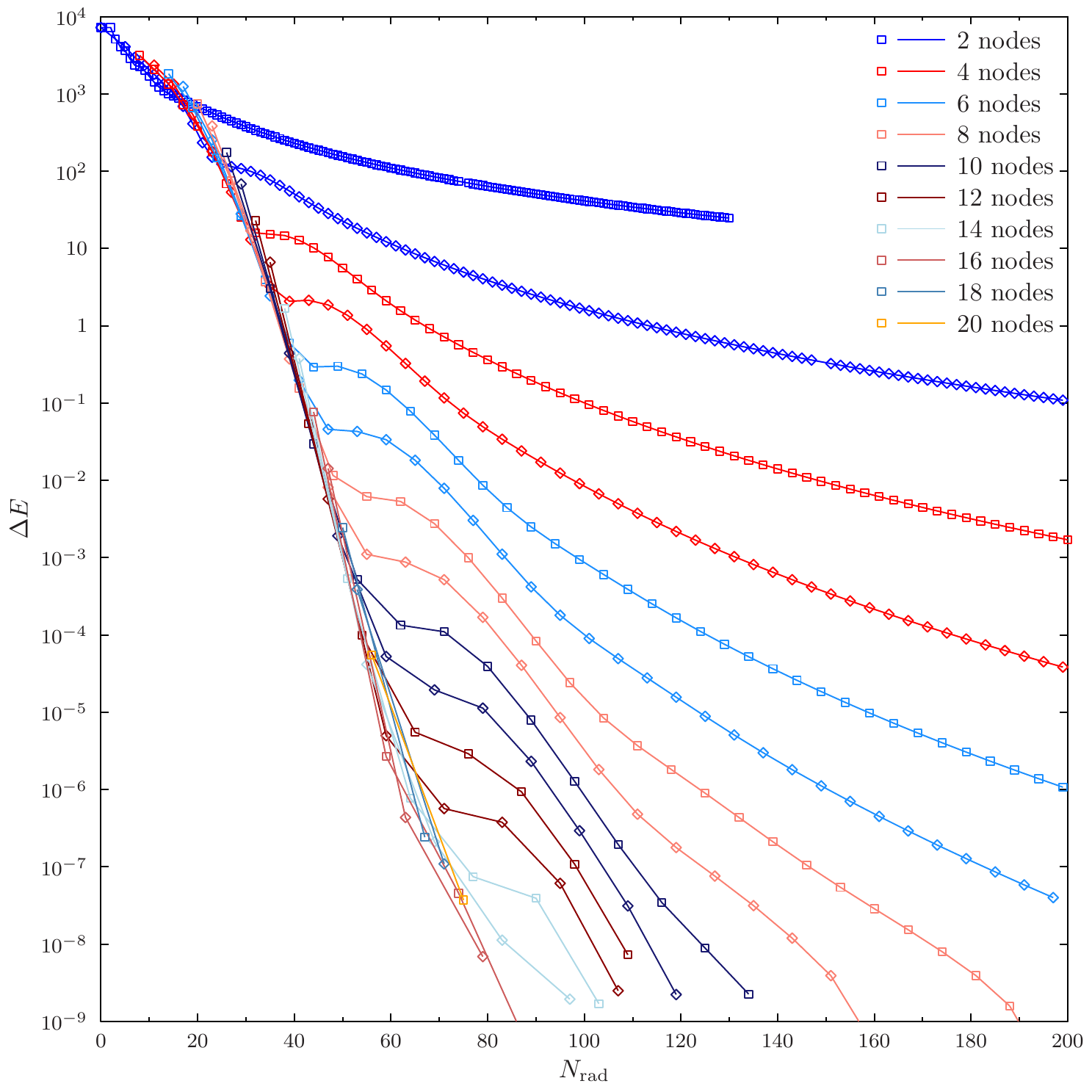}
\par\end{centering}
\caption{Error in HF energy for argon, krypton, and xenon as function of size
of radial basis set with LIP elements employing 2 to 20 Lobatto nodes.
The legend shows the colors for even numbers of nodes (square markers),
whereas the consecutive odd-number node result is shown in the same
color with diamond markers. The topmost two curves correspond to 2
and 3 node elements, correspondingly. \label{fig:elementtype}}
\end{figure}

The calculations we will shortly present employ the exponential grid
with $x=2$, which was tuned above for 6-node uniform LIP elements.
One might imagine that this choice of grid would be biased towards
the 6-node elements, or that the use of the non-linear grid would
favor using more elements with fewer nodes instead of fewer elements
with more nodes. However, these speculations are emphatically rejected
by the results shown in \figref{elementtype} for the errors in the
total energy of the argon and krypton atoms: the use of high-order
elements drastically improves convergence, yielding orders of magnitude
more accuracy for the same number of basis functions. For example,
while the energy for Ar is converged to the accuracy $O(10^{-9})$
of the reference result\citep{Saito2009} with $\sim$80 radial basis
functions using 12-node LIPs, the similar-size calculation with 6-node
LIPs only has an accuracy of $O(10^{-5})$. Although clearly the basis
set limit can be reached with any of the primitive basis sets --
provided enough elements -- the higher order polynomials provide
an astounding speedup in convergence. Similar results have been obtained
recently for three-dimensional meshes in molecular calculations.\citep{Motamarri2013}

The drawback of the high-order elements is not only that higher-order
quadrature rules are needed, but also that the storage costs of the
primitive two-electron integrals is increased. However, based on the
amazing accuracy benefits of higher-order polynomials, we have chosen
15-node LIPs as the default in \textsc{HelFEM}, which are also used
for the rest of the manuscript.

\subsection{Electric response\label{subsec:Electric-response}}

We demonstrate the code with electric response calculations on \ce{Li+}
and \ce{Sr^{2+}} for which HF benchmark values are available in \citeref{Kobus2015}.
As the perturbation caused by the electric field is most strongly
felt by the valence orbitals, accurate calculation of the electric
response requires a fine representation of the valence region, whereas
the core orbitals are mostly unaffected. Since the radial grid we
have chosen emphasizes the core region over the valence region, a
large number of elements may be necessary to converge the electric
properties. It is possible that more accurate electric properties
could be reproduced by re-evaluating the emphases of the radial grid
by sacrificing accuracy in the inert core region for more flexibility
in the valence region. However, as we wish to reproduce both the absolute
energies and electric properties exactly for comparison with \citeref{Kobus2015},
a large radial grid with 10 elements \emph{i.e.} 139 radial functions
will be used. For these calculations we set $r_{\infty}=40a_{0}$,
and an orbital gradient convergence threshold of $10^{-8}$. In this
radial grid, the atomic energies are converged to beyond nanohartree
accuracy, as can be verified by doubling the number of elements (not
shown).

Because the dipole field has a $l$ component (see \eqref{Hdip,z,costh}),
it generates higher $l$ components in an atomic wave function that
would otherwise lack them. In order to calculate, for instance, static
dipole polarizabilities with the present approach, it is first necessary
to determine how well the expansion converges. Because the field was
chosen to be weak, using numerical values determined in \citeref{Kobus2015},
the response of the wave function should be linear, and that of the
energy quadratic. 

It is instructive to begin the analysis from \ce{Li+}, as its electronic
configuration at zero field is simply $1s^{2}$. The values for the
energy, dipole moment and quadrupole moment of \ce{Li+} for increasing
sizes of the basis set are given in \tabref{Li-conv}. No angular
freedom exists in the atomic basis set consisting only of $Y_{0}^{0}$,
and so the energy is constant and the dipole and quadrupole moments
vanish for $l_{\max}=0$. Adding the first polarization shell decreases
the energy at finite fields noticeably, but the energy appears already
to have reached converge. In contrast, the dipole and quadrupole moments
change noticeably upon the addition of a second polarization shell,
as well. While the effect is small for the dipole moment, for the
quadrupole moment the first and second polarization shells appear
to be of equal importance. The addition of a third polarization shell
appears insignificant.

Next, we move on to \ce{Sr^{2+}}, which has the electronic configuration
$1s^{2}2s^{2}2p^{6}3s^{2}3p^{6}4s^{2}3d^{10}4p^{6}$, and the values
for the energy, dipole moment and quadrupole moment for increasing
sizes of the basis set are given in \tabref{Sr-conv}. Based on the
experience with \ce{Li+} above, as the atomic basis set already contains
full flexibility for polarizing the $ns$ levels, partial flexibility
for polarizing the $np$ levels, but no flexibility for the $3d$
level, it can be assumed that the results should be close to converged
with the atomic basis set, as the $3d$ orbitals are considerably
more bound ($\epsilon=-6.1856E_{h}$) than the $4s$ ($\epsilon=-2.3755E_{h}$)
or $4p$ ($\epsilon=-1.5786$) levels. Indeed, it can be seen that
in addition to the energy, also the dipole and quadrupole moments
converge to numerical precision (with the used field strengths) with
a single additional polarization shell in this case. Having established
that the results for \ce{Li+} and \ce{Sr^{2+}} are converged with
$l_{\max}=2$ and $l_{\max}=3$, respectively, we can proceed by comparison
of the field-dependent energy, dipole moment and quadrupole moment
against literature data from \citeref{Kobus2015}. These results are
shown in \tabref{Electric-properties}.

In the case of \ce{Li+}, the energies are in perfect agreement for
the 11 first decimals. For the dipole moments, discrepancies can be
seen in the sixth decimal, meaning that the first six digits are converged,
while the quadrupole moment appears to carry a five-digit accuracy,
with discrepancies seen in the fourth decimal.

There is a constant 7 nanohartree difference \ce{Sr^{2+}} at all
values of the finite field, with the value of \citeref{Kobus2015}
undercutting the value obtained in the present work. As stated above,
we have checked that our basis is accurate at least to nanohartree
level. However, contrary to the present work, the approach used in
\citeref{Kobus2015} is non-variational. The relative error estimated
for the total energy of the \ce{Sr^{2+}} calculation in \citeref{Kobus2015}
was $2\times10^{-13}$, which translates to $0.6\ \text{n}E_{h}$,
ten times less than the observed difference. Still, we are fairly
confident that this is the reason for the discrepance between the
results.

Differences between the dipole moments of \ce{Sr^{2+}} can be seen
in the fifth decimal, whereas differences in the quadrupole moment
appear already at the third decimal for the largest field. Overall,
the agreement is clearly satisfactory, while we again note that for
accurate applications of the present methodology to electric properties,
the choice of the radial grid could be investigated in more detail.

\begin{table}
\begin{centering}
\subfloat[Energy]{\begin{centering}
\begin{tabular}{r@{\extracolsep{0pt}.}lr@{\extracolsep{0pt}.}lr@{\extracolsep{0pt}.}lr@{\extracolsep{0pt}.}lr@{\extracolsep{0pt}.}lr@{\extracolsep{0pt}.}l}
\multicolumn{2}{c}{$E_{z}$} & \multicolumn{2}{c}{$l_{\max}=0$} & \multicolumn{2}{c}{$l_{\max}=1$} & \multicolumn{2}{c}{$l_{\max}=2$} & \multicolumn{2}{c}{$l_{\max}=3$} & \multicolumn{2}{c}{$l_{\max}=4$}\tabularnewline
-0&006 & -7&236415201 & -3&41 (-6) & -6&63 (-12) & 2&52 (-13) & 1&27 (-13)\tabularnewline
-0&004 & -7&236415201 & -1&52 (-6) & -2&12 (-13) & -6&42 (-13) & -7&90 (-14)\tabularnewline
-0&002 & -7&236415201 & -3&79 (-7) & -3&20 (-14) & 2&07 (-13) & -9&50 (-14)\tabularnewline
0&000 & -7&236415201 & 3&55 (-13) & 2&42 (-13) & -3&45 (-13) & 4&34 (-13)\tabularnewline
0&002 & -7&236415201 & -3&79 (-7) & -3&20 (-14) & 2&07 (-13) & -9&50 (-14)\tabularnewline
0&004 & -7&236415201 & -1&52 (-6) & -2&12 (-13) & -6&42 (-13) & -7&90 (-14)\tabularnewline
0&006 & -7&236415201 & -3&41 (-6) & -6&63 (-12) & 2&52 (-13) & 1&27 (-13)\tabularnewline
\end{tabular}
\par\end{centering}
}
\par\end{centering}
\begin{centering}
\subfloat[Dipole moment]{\begin{centering}
\begin{tabular}{r@{\extracolsep{0pt}.}lr@{\extracolsep{0pt}.}lr@{\extracolsep{0pt}.}lr@{\extracolsep{0pt}.}lr@{\extracolsep{0pt}.}lr@{\extracolsep{0pt}.}l}
\multicolumn{2}{c}{$E_{z}$} & \multicolumn{2}{c}{$l_{\max}=0$} & \multicolumn{2}{c}{$l_{\max}=1$} & \multicolumn{2}{c}{$l_{\max}=2$} & \multicolumn{2}{c}{$l_{\max}=3$} & \multicolumn{2}{c}{$l_{\max}=4$}\tabularnewline
-0&006 & 0&00 (0) & 1&14 (-3) & 4&34 (-9) & 3&62 (-14) & -4&15 (-13)\tabularnewline
-0&004 & 0&00 (0) & 7&58 (-4) & 1&28 (-9) & -1&57 (-14) & 1&52 (-14)\tabularnewline
-0&002 & 0&00 (0) & 3&79 (-4) & 1&61 (-10) & 4&95 (-15) & -1&82 (-14)\tabularnewline
0&000 & 0&00 (0) & 0&00 (0) & 0&00 (0) & 0&00 (0) & 0&00 (0)\tabularnewline
0&002 & 0&00 (0) & -3&79 (-4) & -1&61 (-10) & -4&95 (-15) & 1&82 (-14)\tabularnewline
0&004 & 0&00 (0) & -7&58 (-4) & -1&28 (-9) & 1&57 (-14) & -1&52 (-14)\tabularnewline
0&006 & 0&00 (0) & -1&14 (-3) & -4&34 (-9) & -3&62 (-14) & 4&15 (-13)\tabularnewline
\end{tabular}
\par\end{centering}
}
\par\end{centering}
\begin{centering}
\subfloat[Quadrupole moment]{\begin{centering}
\begin{tabular}{r@{\extracolsep{0pt}.}lr@{\extracolsep{0pt}.}lr@{\extracolsep{0pt}.}lr@{\extracolsep{0pt}.}lr@{\extracolsep{0pt}.}lr@{\extracolsep{0pt}.}l}
\multicolumn{2}{c}{$E_{z}$} & \multicolumn{2}{c}{$l_{\max}=0$} & \multicolumn{2}{c}{$l_{\max}=1$} & \multicolumn{2}{c}{$l_{\max}=2$} & \multicolumn{2}{c}{$l_{\max}=3$} & \multicolumn{2}{c}{$l_{\max}=4$}\tabularnewline
-0&006 & 0&00 (0) & 9&36 (-7) & 1&17 (-6) & 9&40 (-12) & -3&97 (-14)\tabularnewline
-0&004 & 0&00 (0) & 4&16 (-7) & 5&18 (-7) & 1&52 (-12) & -4&31 (-16)\tabularnewline
-0&002 & 0&00 (0) & 1&04 (-7) & 1&30 (-7) & -3&29 (-14) & -5&74 (-14)\tabularnewline
0&000 & 0&00 (0) & 0&00 (0) & 0&00 (0) & 0&00 (0) & 0&00 (0)\tabularnewline
0&002 & 0&00 (0) & 1&04 (-7) & 1&30 (-7) & -3&29 (-14) & -5&74 (-14)\tabularnewline
0&004 & 0&00 (0) & 4&16 (-7) & 5&18 (-7) & 1&52 (-12) & -4&31 (-16)\tabularnewline
0&006 & 0&00 (0) & 9&36 (-7) & 1&17 (-6) & 9&40 (-12) & -3&97 (-14)\tabularnewline
\end{tabular}
\par\end{centering}
}
\par\end{centering}
\caption{Convergence of the electric properties of \ce{Li+} in a finite field
HF calculation. The first column lists the field strength, the second
column gives the value of the property in the unmodified basis set
for the atom with minimal $l$ value, while the successive columns
describe the change of the property from the previous $l$ value.
The values in the parentheses indicate magnitude, $A(n)=A\times10^{n}$.
\label{tab:Li-conv}}
\end{table}

\begin{table}
\begin{centering}
\subfloat[Energy]{\begin{centering}
\begin{tabular}{r@{\extracolsep{0pt}.}lr@{\extracolsep{0pt}.}lr@{\extracolsep{0pt}.}lr@{\extracolsep{0pt}.}lr@{\extracolsep{0pt}.}lr@{\extracolsep{0pt}.}l}
\multicolumn{2}{c}{$E_{z}$} & \multicolumn{2}{c}{$l_{\max}=2$} & \multicolumn{2}{c}{$l_{\max}=3$} & \multicolumn{2}{c}{$l_{\max}=4$} & \multicolumn{2}{c}{$l_{\max}=5$} & \multicolumn{2}{c}{$l_{\max}=6$}\tabularnewline
-0&0018 & -3130&995692183 & -1&08 (-8) & -3&64 (-12) & -7&14 (-11) & 6&37 (-11)\tabularnewline
-0&0012 & -3130&995686893 & -4&85 (-9) & 4&55 (-11) & -9&64 (-11) & 8&55 (-11)\tabularnewline
-0&0006 & -3130&995683719 & -1&10 (-9) & -3&77 (-11) & -7&73 (-12) & 3&91 (-11)\tabularnewline
0&0000 & -3130&995682661 & 2&46 (-11) & -4&32 (-11) & 6&18 (-11) & -1&96 (-11)\tabularnewline
0&0006 & -3130&995683719 & -1&10 (-9) & -3&77 (-11) & -7&73 (-12) & 3&91 (-11)\tabularnewline
0&0012 & -3130&995686893 & -4&85 (-9) & 4&55 (-11) & -9&64 (-11) & 8&55 (-11)\tabularnewline
0&0018 & -3130&995692183 & -1&08 (-8) & -3&64 (-12) & -7&14 (-11) & 6&37 (-11)\tabularnewline
\end{tabular}
\par\end{centering}
}
\par\end{centering}
\begin{centering}
\subfloat[Dipole moment]{\begin{centering}
\begin{tabular}{r@{\extracolsep{0pt}.}lr@{\extracolsep{0pt}.}lr@{\extracolsep{0pt}.}lr@{\extracolsep{0pt}.}lr@{\extracolsep{0pt}.}lr@{\extracolsep{0pt}.}l}
\multicolumn{2}{c}{$E_{z}$} & \multicolumn{2}{c}{$l_{\max}=2$} & \multicolumn{2}{c}{$l_{\max}=3$} & \multicolumn{2}{c}{$l_{\max}=4$} & \multicolumn{2}{c}{$l_{\max}=5$} & \multicolumn{2}{c}{$l_{\max}=6$}\tabularnewline
-0&0018 & 1&06 (-2) & 1&21 (-5) & -8&32 (-11) & 1&50 (-11) & 3&71 (-12)\tabularnewline
-0&0012 & 7&05 (-3) & 8&05 (-6) & 1&54 (-10) & 9&81 (-12) & 1&23 (-11)\tabularnewline
-0&0006 & 3&53 (-3) & 4&02 (-6) & -2&73 (-11) & 1&13 (-11) & -3&43 (-11)\tabularnewline
0&0000 & -3&11 (-9) & 5&31 (-9) & -2&75 (-9) & -2&16 (-8) & 5&37 (-9)\tabularnewline
0&0006 & -3&53 (-3) & -4&02 (-6) & 2&73 (-11) & -1&13 (-11) & 3&43 (-11)\tabularnewline
0&0012 & -7&05 (-3) & -8&05 (-6) & -1&54 (-10) & -9&81 (-12) & -1&23 (-11)\tabularnewline
0&0018 & -1&06 (-2) & -1&21 (-5) & 8&32 (-11) & -1&50 (-11) & -3&71 (-12)\tabularnewline
\end{tabular}
\par\end{centering}
}
\par\end{centering}
\begin{centering}
\subfloat[Quadrupole moment]{\begin{centering}
\begin{tabular}{r@{\extracolsep{0pt}.}lr@{\extracolsep{0pt}.}lr@{\extracolsep{0pt}.}lr@{\extracolsep{0pt}.}lr@{\extracolsep{0pt}.}lr@{\extracolsep{0pt}.}l}
\multicolumn{2}{c}{$E_{z}$} & \multicolumn{2}{c}{$l_{\max}=2$} & \multicolumn{2}{c}{$l_{\max}=3$} & \multicolumn{2}{c}{$l_{\max}=4$} & \multicolumn{2}{c}{$l_{\max}=5$} & \multicolumn{2}{c}{$l_{\max}=6$}\tabularnewline
-0&0018 & 2&00 (-5) & 1&49 (-5) & 5&45 (-9) & 1&12 (-10) & -5&96 (-11)\tabularnewline
-0&0012 & 8&90 (-6) & 6&61 (-6) & 2&48 (-9) & -2&51 (-11) & 3&91 (-11)\tabularnewline
-0&0006 & 2&23 (-6) & 1&65 (-6) & 6&28 (-10) & 1&38 (-11) & -5&41 (-11)\tabularnewline
0&0000 & 4&52 (-10) & 2&97 (-9) & -4&98 (-9) & -1&37 (-9) & 7&39 (-9)\tabularnewline
0&0006 & 2&23 (-6) & 1&65 (-6) & 6&28 (-10) & 1&38 (-11) & -5&41 (-11)\tabularnewline
0&0012 & 8&90 (-6) & 6&61 (-6) & 2&48 (-9) & -2&51 (-11) & 3&91 (-11)\tabularnewline
0&0018 & 2&00 (-5) & 1&49 (-5) & 5&45 (-9) & 1&12 (-10) & -5&96 (-11)\tabularnewline
\end{tabular}
\par\end{centering}
}
\par\end{centering}
\caption{Convergence of the electric properties of \ce{Sr^{2+}} in a finite
field HF calculation. The notation is the same as in \tabref{Li-conv}.
\label{tab:Sr-conv}}
\end{table}

\begin{sidewaystable*}
\begin{centering}
\subfloat[\ce{Li+}\label{tab:Electric-properties.Li}]{\begin{centering}
\begin{tabular}{r@{\extracolsep{0pt}.}lr@{\extracolsep{0pt}.}lr@{\extracolsep{0pt}.}lr@{\extracolsep{0pt}.}lr@{\extracolsep{0pt}.}lr@{\extracolsep{0pt}.}lr@{\extracolsep{0pt}.}lr@{\extracolsep{0pt}.}l}
\multicolumn{2}{c}{} & \multicolumn{6}{c}{{\footnotesize{}present work}} & \multicolumn{2}{c}{} & \multicolumn{6}{c}{{\footnotesize{}Literature value}}\tabularnewline
\multicolumn{2}{c}{{\footnotesize{}$E_{z}$}} & \multicolumn{2}{c}{{\footnotesize{}Energy}} & \multicolumn{2}{c}{{\footnotesize{}Dipole}} & \multicolumn{2}{c}{{\footnotesize{}Quadrupole}} & \multicolumn{2}{c}{} & \multicolumn{2}{c}{{\footnotesize{}Energy}} & \multicolumn{2}{c}{{\footnotesize{}Dipole}} & \multicolumn{2}{c}{{\footnotesize{}Quadrupole}}\tabularnewline
{\footnotesize{}-0}&{\footnotesize{}004} & {\footnotesize{}-7}&{\footnotesize{}23641671725} & {\footnotesize{}-7}&{\footnotesize{}578993 (--4)} & {\footnotesize{}-9}&{\footnotesize{}341741 (-7)} & \multicolumn{2}{c}{} & {\footnotesize{}-7}&{\footnotesize{}23641671725} & {\footnotesize{}-7}&{\footnotesize{}579002 (-4)} & {\footnotesize{}-9}&{\footnotesize{}341812 (-7)}\tabularnewline
{\footnotesize{}-0}&{\footnotesize{}002} & {\footnotesize{}-7}&{\footnotesize{}23641558040} & {\footnotesize{}-3}&{\footnotesize{}789487 (-4)} & {\footnotesize{}-2}&{\footnotesize{}335419 (-7)} & \multicolumn{2}{c}{} & {\footnotesize{}-7}&{\footnotesize{}23641558040} & {\footnotesize{}-3}&{\footnotesize{}789492 (-4)} & {\footnotesize{}-2}&{\footnotesize{}335434 (-7)}\tabularnewline
{\footnotesize{} 0}&{\footnotesize{}000} & {\footnotesize{}-7}&{\footnotesize{}23641520145} & {\footnotesize{}0}&{\footnotesize{}0} & {\footnotesize{}0}&{\footnotesize{}0} & \multicolumn{2}{c}{} & {\footnotesize{}-7}&{\footnotesize{}23641520145} & {\footnotesize{}-1}&{\footnotesize{}393330 (-14)} & {\footnotesize{}-5}&{\footnotesize{}228457 (-15)}\tabularnewline
{\footnotesize{} 0}&{\footnotesize{}002} & {\footnotesize{}-7}&{\footnotesize{}23641558040} & {\footnotesize{}3}&{\footnotesize{}789487 (-4)} & {\footnotesize{}-2}&{\footnotesize{}335419 (-7)} & \multicolumn{2}{c}{} & {\footnotesize{}-7}&{\footnotesize{}23641558040} & {\footnotesize{}3}&{\footnotesize{}789492 (-4)} & {\footnotesize{}-2}&{\footnotesize{}335434 (-7)}\tabularnewline
{\footnotesize{} 0}&{\footnotesize{}004} & {\footnotesize{}-7}&{\footnotesize{}23641671725} & {\footnotesize{}7}&{\footnotesize{}578993 (-4)} & {\footnotesize{}-9}&{\footnotesize{}341741 (-7)} & \multicolumn{2}{c}{} & {\footnotesize{}-7}&{\footnotesize{}23641671725} & {\footnotesize{}7}&{\footnotesize{}579002 (-4)} & {\footnotesize{}-9}&{\footnotesize{}341812 (-7)}\tabularnewline
\end{tabular}
\par\end{centering}
}
\par\end{centering}
\begin{centering}
\subfloat[\ce{Sr^{2+}}\label{tab:Electric-properties.Sr}]{\begin{centering}
\begin{tabular}{r@{\extracolsep{0pt}.}lr@{\extracolsep{0pt}.}lr@{\extracolsep{0pt}.}lr@{\extracolsep{0pt}.}lr@{\extracolsep{0pt}.}lr@{\extracolsep{0pt}.}lr@{\extracolsep{0pt}.}lr@{\extracolsep{0pt}.}l}
\multicolumn{2}{c}{} & \multicolumn{6}{c}{{\footnotesize{}present work}} & \multicolumn{2}{c}{} & \multicolumn{6}{c}{{\footnotesize{}Literature value}}\tabularnewline
\multicolumn{2}{c}{{\footnotesize{}$E_{z}$}} & \multicolumn{2}{c}{{\footnotesize{}Energy}} & \multicolumn{2}{c}{{\footnotesize{}Dipole}} & \multicolumn{2}{c}{{\footnotesize{}Quadrupole}} & \multicolumn{2}{c}{} & \multicolumn{2}{c}{{\footnotesize{}Energy}} & \multicolumn{2}{c}{{\footnotesize{}Dipole}} & \multicolumn{2}{c}{{\footnotesize{}Quadrupole}}\tabularnewline
{\footnotesize{}-0}&{\footnotesize{}0012} & {\footnotesize{}-3130}&{\footnotesize{}995686898} & {\footnotesize{}-7}&{\footnotesize{}061234 (-3)} & {\footnotesize{}-1}&{\footnotesize{}550859 (-5)} & \multicolumn{2}{c}{} & {\footnotesize{}-3130}&{\footnotesize{}995686905} & {\footnotesize{}-7}&{\footnotesize{}061227 (-3)} & {\footnotesize{}-1}&{\footnotesize{}551909 (-5)}\tabularnewline
{\footnotesize{}-0}&{\footnotesize{}0006} & {\footnotesize{}-3130}&{\footnotesize{}995683720} & {\footnotesize{}-3}&{\footnotesize{}530623 (-3)} & {\footnotesize{}-3}&{\footnotesize{}866475 (-6)} & \multicolumn{2}{c}{} & {\footnotesize{}-3130}&{\footnotesize{}995683727} & {\footnotesize{}-3}&{\footnotesize{}530607 (-3)} & {\footnotesize{}-3}&{\footnotesize{}879769 (-6)}\tabularnewline
{\footnotesize{} 0}&{\footnotesize{}0000} & {\footnotesize{}-3130}&{\footnotesize{}995682661} & {\footnotesize{} 3}&{\footnotesize{}875110 (-9)} & {\footnotesize{} 5}&{\footnotesize{}749758 (-9)} & \multicolumn{2}{c}{} & {\footnotesize{}-3130}&{\footnotesize{}995682668} & {\footnotesize{}-9}&{\footnotesize{}387699 (-11)} & {\footnotesize{}-2}&{\footnotesize{}648433 (-11)}\tabularnewline
{\footnotesize{} 0}&{\footnotesize{}0006} & {\footnotesize{}-3130}&{\footnotesize{}995683720} & {\footnotesize{} 3}&{\footnotesize{}530623 (-3)} & {\footnotesize{}-3}&{\footnotesize{}866475 (-6)} & \multicolumn{2}{c}{} & {\footnotesize{}-3130}&{\footnotesize{}995683727} & {\footnotesize{}3}&{\footnotesize{}530607 (-3)} & {\footnotesize{}-3}&{\footnotesize{}879762 (-6)}\tabularnewline
{\footnotesize{} 0}&{\footnotesize{}0012} & {\footnotesize{}-3130}&{\footnotesize{}995686898} & {\footnotesize{} 7}&{\footnotesize{}061234 (-3)} & {\footnotesize{}-1}&{\footnotesize{}550859 (-5)} & \multicolumn{2}{c}{} & {\footnotesize{}-3130}&{\footnotesize{}995686905} & {\footnotesize{}7}&{\footnotesize{}061227 (-3)} & {\footnotesize{}-1}&{\footnotesize{}551909 (-5)}\tabularnewline
\end{tabular}
\par\end{centering}
}
\par\end{centering}
\caption{Electric properties of \ce{Li+} and \ce{Sr^{2+}} in a finite field,
compared to literature values from \citeref{Kobus2015} with a truncated
number of decimals. The values in the parentheses indicate magnitude,
$A(n)=A\times10^{n}$.\label{tab:Electric-properties}}
\end{sidewaystable*}

Next, static dipole polarizabilities
\begin{equation}
\alpha_{zz}=\left(\frac{{\rm d}\mu_{z}}{{\rm d}E_{z}}\right)_{E_{z}=0}\label{eq:staticpol}
\end{equation}
can be extracted from the data at finite fields given in \tabref{Electric-properties}
by employing finite difference approximations such as the two-point
rule
\begin{equation}
f'(x)\approx\frac{f(x+h)-f(x-h)}{2h}+O(h^{2})\label{eq:twopoint}
\end{equation}
or the four-point rule
\begin{equation}
f'(x)\approx\frac{-f(x+2h)+8f(x+h)-8f(x-h)+f(x-2h)}{12h}+O(h^{4});\label{eq:fourpoint}
\end{equation}
these results are shown in \tabref{DFT-pol}. The values for the polarizability
in \tabref{DFT-pol} have been obtained with \eqref{fourpoint}, whereas
the error estimate is simply the difference between the four-point
and two-point values given by \eqref{fourpoint,twopoint}, respectively.
The HF polarizability agrees well with literature values from \citeref{Kobus2015},
which are $1.89474455\times10^{-1}$ for \ce{Li+} (discrepance at
sixth decimal) and $5.8843416$ for \ce{Sr^{2+}} (full agreement).
\Tabref{DFT-pol} also presents analogous calculations at various
levels of DFT with the LDA,\citep{Bloch1929,Dirac1930,Perdew1992a}
PBE\citep{Perdew1996,Perdew1997} and TPSS\citep{Tao2003,Perdew2004}
functionals, and their hybrids; PBEh\citep{Adamo1999,Ernzerhof1999}
and TPSSh,\citep{Staroverov2003} respectively. All DFT functionals
predict a higher polarizability, \emph{i.e.} a more flexible electron
density than the one reproduced by HF.

\begin{sidewaystable*}
\begin{centering}
\subfloat[\ce{Li+}, $h=0.002$]{\begin{centering}
\begin{tabular}{lr@{\extracolsep{0pt}.}lr@{\extracolsep{0pt}.}lr@{\extracolsep{0pt}.}lr@{\extracolsep{0pt}.}lr@{\extracolsep{0pt}.}lr@{\extracolsep{0pt}.}lr@{\extracolsep{0pt}.}l}
method & \multicolumn{2}{c}{$\mu_{z}$, $E_{z}=-2h$} & \multicolumn{2}{c}{$\mu_{z}$, $E_{z}=-h$} & \multicolumn{2}{c}{$\mu_{z}$, $E_{z}=0$} & \multicolumn{2}{c}{$\mu_{z}$, $E_{z}=h$} & \multicolumn{2}{c}{$\mu_{z}$, $E_{z}=2h$} & \multicolumn{2}{c}{$\alpha_{zz}$} & \multicolumn{2}{c}{$\Delta\alpha_{zz}$}\tabularnewline
HF & -7&578993 (-4) & -3&789487 (-4) & 0&000000 (0) & 3&789487 (-4) & 7&578993 (-4) & 1&894742 (-1) & -1&52 (-7)\tabularnewline
LDA & -8&602004 (-4) & -4&300989 (-4) & 1&058460 (-16) & 4&300989 (-4) & 8&602004 (-4) & 2&150492 (-1) & -2&21 (-7)\tabularnewline
PBE & -8&317930 (-4) & -4&158951 (-4) & 1&433630 (-15) & 4&158951 (-4) & 8&317930 (-4) & 2&079473 (-1) & -2&25 (-7)\tabularnewline
TPSS & -8&110240 (-4) & -4&055108 (-4) & 1&139966 (-14) & 4&055108 (-4) & 8&110240 (-4) & 2&027552 (-1) & -2&05 (-7)\tabularnewline
PBEh & -8&096724 (-4) & -4&048350 (-4) & -6&712163 (-15) & 4&048350 (-4) & 8&096724 (-4) & 2&024173 (-1) & -2&00 (-7)\tabularnewline
TPSSh & -8&045716 (-4) & -4&022846 (-4) & 9&588448 (-15) & 4&022846 (-4) & 8&045716 (-4) & 2&011421 (-1) & -1&98 (-7)\tabularnewline
\end{tabular}
\par\end{centering}
}
\par\end{centering}
\begin{centering}
\subfloat[\ce{Sr^{2+}}, $h=0.0006$]{\begin{centering}
\begin{tabular}{lr@{\extracolsep{0pt}.}lr@{\extracolsep{0pt}.}lr@{\extracolsep{0pt}.}lr@{\extracolsep{0pt}.}lr@{\extracolsep{0pt}.}lr@{\extracolsep{0pt}.}lr@{\extracolsep{0pt}.}l}
method & \multicolumn{2}{c}{$\mu_{z}$, $E_{z}=-2h$} & \multicolumn{2}{c}{$\mu_{z}$, $E_{z}=-h$} & \multicolumn{2}{c}{$\mu_{z}$, $E_{z}=0$} & \multicolumn{2}{c}{$\mu_{z}$, $E_{z}=h$} & \multicolumn{2}{c}{$\mu_{z}$, $E_{z}=2h$} & \multicolumn{2}{c}{$\alpha_{zz}$} & \multicolumn{2}{c}{$\Delta\alpha_{zz}$}\tabularnewline
HF & -7&061227 (-3) & -3&530607 (-3) & -2&195314 (-9) & 3&530607 (-3) & 7&061227 (-3) & 5&884342 (0) & -3&55 (-6)\tabularnewline
LDA & -7&197242 (-3) & -3&598471 (-3) & 9&338964 (-9) & 3&598472 (-3) & 7&197242 (-3) & 5&997369 (0) & -8&30 (-5)\tabularnewline
PBE & -7&218082 (-3) & -3&608891 (-3) & -4&167875 (-10) & 3&608890 (-3) & 7&218082 (-3) & 6&014734 (0) & -8&37 (-5)\tabularnewline
TPSS & -7&148780 (-3) & -3&574260 (-3) & -1&918774 (-8) & 3&574260 (-3) & 7&148779 (-3) & 5&957027 (0) & -7&22 (-5)\tabularnewline
PBEh & -7&136971 (-3) & -3&568168 (-3) & -1&432482 (-10) & 3&568168 (-3) & 7&136970 (-3) & 5&946771 (0) & -1&76 (-4)\tabularnewline
TPSSh & -7&123533 (-3) & -3&561625 (-3) & -7&538647 (-10) & 3&561622 (-3) & 7&123535 (-3) & 5&935959 (0) & -7&97 (-5)\tabularnewline
\end{tabular}
\par\end{centering}
}
\par\end{centering}
\caption{Dipole moments $\mu_{z}$ at finite fields $E_{z}$, and the extracted
dipole polarizabilities $\alpha_{zz}$ and its estimated error $\Delta\alpha_{zz}$
for \ce{Li+} and \ce{Sr^{2+}} at HF, LDA, PBE, TPSS, PBEh, and TPSSh
levels of theory.\label{tab:DFT-pol}}
\end{sidewaystable*}

\subsection{Accuracy of Gaussian basis sets\label{subsec:Accuracy-of-Gaussian}}

In order to study the accuracy of the results obtained with an extended
Gaussian basis set in \citeref{Anderson2017}, we decided to repeat
the calculations in the finite element approach. We chose to study
the species \ce{H-}, \ce{He}, \ce{Li+}, \ce{Li-}, \ce{Be}, \ce{B+},
\ce{C-}, \ce{N}, \ce{O+}, \ce{F-}, \ce{Ne}, \ce{Na+}, \ce{Na-},
\ce{Mg}, \ce{Al+}, \ce{Si-}, \ce{P}, \ce{S+}, \ce{Cl-}, and
\ce{Ar}, as each of them has only fully filled subshells. 

Although systems with partially filled shells can also be computed
with the present approach, the corresponding minimal-energy solutions
are well-known to break symmetry unless spherical averaging is employed.
Thus, instead of the expected exactness of the SCF solution for second-period
atoms with $l_{\max}=1$, the energy is lowered by the addition of
functions with higher $l$; see for instance the discussion by Löwdin
in \citeref{Lykos1963}. In these cases, finding the lowest solution
within the FEM approach may be nontrivial, as convergence may occur
to any number of solutions. Surprisingly, symmetry breaking can sometimes
also be seen for cases with fully filled shells, such as in the case
of the \ce{Ne} atom and the \ce{F-} anion.\citep{Prat1972} We chose
the above systems for the present work, as the study by Anderson and
coworkers in \citeref{Anderson2017} explicitly considered broken
symmetry solutions by the use of wave function stability analysis,\citep{Seeger1977,Bauernschmitt1996}
which is not currently implemented in \textsc{HelFEM}.

Starting with HF, in agreement with \citeref{Saito2003} we find that
although the neutral atoms are converged with $r_{\infty}=40a_{0}$,
the extended anions \ce{Li-} and \ce{Na-} require a larger value
to be employed. Whereas \citeref{Saito2003} employed $r_{\infty}=60a_{0}$,
we chose $r_{\infty}=80a_{0}$ as this changes the nanohartree digit
of the energy of \ce{Na-}. Unlike \citeref{Anderson2017}, the symmetries
of the occupied orbitals were enforced, as this was found to speed
up the convergence of the SCF procedure. The energies were found to
converge to nanohartree accuracy with 10 radial elements.

The results for the HF calculations are shown in \tabref{Hartree=002013Fock-energies}.
For further reference, we have repeated the aug-pc-$\infty$ Gaussian-basis
calculations of \citeref{Anderson2017} with \textsc{Erkale}; these
results are also given in \tabref{Hartree=002013Fock-energies}. Tight
integral screening thresholds were used in \textsc{Erkale}. For comparison,
\tabref{Hartree=002013Fock-energies} also reports the energies given
in the supplementary information of \citeref{Anderson2017}. The aug-pc-$\infty$
basis of \citeref{Anderson2017} was originally developed by Jensen
in \citeref{Jensen2010a}.

The \textsc{HelFEM} and \textsc{Erkale} data are in excellent agreement:
the FEM calculations yield energies that are up to a few dozen microhartree
lower than the ones reproduced by the large Gaussian basis set. The
agreement between the calculations performed in the present work and
those of \citeref{Anderson2017} is also in general excellent, with
three notable exceptions: the extended, weakly bound anions \ce{Li-}
and \ce{Na-}, as well as the \ce{Be} atom. In these cases, the energies
reported in \citeref{Anderson2017} are considerably lower than the
energies we have computed using FEM or with \textsc{Erkale} using
the same basis set as \citeref{Anderson2017}. Comparison to literature
values\citealp{Koga1995,Koga1996,Saito2003} for these systems affirm
the accuracy of the values produced in the present work. Our FEM values
for the \ce{Li-}, \ce{Be}, and \ce{Na-} are in perfect agreement
with the literature values\citep{Koga1995,Koga1996,Saito2003} -7.428232061,
-14.57302317, and -161.8551260, respectively, with the Gaussian basis
values being consistently upper bounds to the converged values. We
can thus conclude that the values reported in \citeref{Anderson2017}
for \ce{Li-}, \ce{Be}, and \ce{Na-} represent symmetry broken solutions
of closed-shell species that have been discussed in \citeref{Prat1972}.

Repeating the calculations with the BHHLYP functional,\citealp{Becke1993a}
the results in \tabref{BHHLYP-energies} are obtained. A (250,770)
integration grid was employed in the \textsc{Erkale} calculations,
again with tight integral screening thresholds. The BHHLYP functional
binds \ce{Li-} and \ce{Na-} less strongly than HF, and the values
for these systems profit from the chosen large value for $r_{\infty}$.
Although the reported finite element energies for \ce{Li-} and \ce{Na-}
are still in error by tens of nanohartrees compared to a larger value
of $r_{\infty}$, the conclusions of our study are not affected. Namely,
the \textsc{HelFEM} and \textsc{Erkale} calculations are in excellent
agreement, the differences between the two approaches being again
in the microhartrees but somewhat smaller than in the case of the
HF calculations in \tabref{Hartree=002013Fock-energies}.

As in the HF calculations, while the agreement with the results of
\citeref{Anderson2017} is generally excellent, also here the values
for \ce{Li-} and \ce{Na-} stand out, undercutting the converged
complete basis set energy in the millihartree range. At variance to
\tabref{Hartree=002013Fock-energies}, the energy for \ce{Be} is
now in perfect agreement. Instead, the energy for \ce{H-} of \citeref{Anderson2017}
is too low by 1.6 m$E_{h}$, which is again likely caused by symmetry
breaking.

\begin{table}
\begin{centering}
\begin{tabular}{lrrrr}
 & finite element & Gaussian, \textsc{Erkale} & Gaussian, \citeref{Anderson2017} & difference ($\mu E_{h}$)\tabularnewline
\ce{H-} & -0.487929734 & -0.487929397 & -0.48793 &   -0.34\tabularnewline
\ce{He} & -2.861679996 & -2.861675168 & -2.86168 &   -4.83\tabularnewline
\ce{Li+} & -7.236415201 & -7.236414275 & -7.23641 &   -0.93\tabularnewline
\ce{Li-} & -7.428232061 & -7.428231023 & -7.43152 &   -1.04\tabularnewline
\ce{Be} & -14.573023168 & -14.573021658 & -14.57335 &   -1.51\tabularnewline
\ce{B+} & -24.237575184 & -24.237566607 & -24.23757 &   -8.58\tabularnewline
\ce{C-} & -37.710309470 & -37.710305344 & -37.71031 &   -4.13\tabularnewline
\ce{N} & -54.404548303 & -54.404543006 & -54.40454 &   -5.30\tabularnewline
\ce{O+} & -74.377133274 & -74.377123988 & -74.37712 &   -9.29\tabularnewline
\ce{F-} & -99.459453913 & -99.459442803 & -99.45944 &  -11.11\tabularnewline
\ce{Ne} & -128.547098109 & -128.547079874 & -128.54708 &  -18.24\tabularnewline
\ce{Na+} & -161.676962614 & -161.676950741 & -161.67695 &  -11.87\tabularnewline
\ce{Na-} & -161.855125996 & -161.855114256 & -161.85702 &  -11.74\tabularnewline
\ce{Mg} & -199.614636424 & -199.614623656 & -199.61462 &  -12.77\tabularnewline
\ce{Al+} & -241.674670465 & -241.674657663 & -241.67466 &  -12.80\tabularnewline
\ce{Si-} & -288.890058853 & -288.890044560 & -288.89004 &  -14.29\tabularnewline
\ce{P} & -340.719275268 & -340.719259261 & -340.71926 &  -16.01\tabularnewline
\ce{S+} & -397.173947455 & -397.173928130 & -397.17393 &  -19.33\tabularnewline
\ce{Cl-} & -459.576925268 & -459.576907117 & -459.57691 &  -18.15\tabularnewline
\ce{Ar} & -526.817512803 & -526.817490166 & -526.81749 &  -22.64\tabularnewline
\end{tabular}
\par\end{centering}
\caption{HF energies from a finite element calculation (present work, second
column) compared to a Gaussian basis calculation with \textsc{Erkale}
using the basis set from \citeref{Anderson2017} (present work, third
column). The fourth column shows the Gaussian basis set energies from
\citeref{Anderson2017}. The fifth column lists the energy difference
between finite element and Gaussian basis set calculations of the
present work in microhartree.\label{tab:Hartree=002013Fock-energies}}
\end{table}

\begin{table}
\centering{}%
\begin{tabular}{lrrrr}
 & finite element & Gaussian, \textsc{Erkale} & Gaussian, \citeref{Anderson2017} & difference ($\mu E_{h}$)\tabularnewline
\ce{H-} & -0.523455900 & N/C & -0.52501 & N/C\tabularnewline
\ce{He} & -2.905757890 & -2.905754694 & -2.90575 &   -3.20\tabularnewline
\ce{Li+} & -7.281288205 & -7.281287221 & -7.28129 &   -0.98\tabularnewline
\ce{Li-} & -7.500010377$^{a}$ & N/C & -7.50081 & N/C\tabularnewline
\ce{Be} & -14.664037985 & -14.664035679 & -14.66404 &   -2.31\tabularnewline
\ce{B+} & -24.339819186 & -24.339811410 & -24.33981 &   -7.78\tabularnewline
\ce{C-} & -37.887327417 & -37.887323754 & -37.88732 &   -3.66\tabularnewline
\ce{N} & -54.593153473 & -54.593148512 & -54.59315 &   -4.96\tabularnewline
\ce{O+} & -74.575815265 & -74.575805610 & -74.57581 &   -9.66\tabularnewline
\ce{F-} & -99.856395949 & -99.856387120 & -99.85639 &   -8.83\tabularnewline
\ce{Ne} & -128.948416397 & -128.948403201 & -128.94840 &  -13.20\tabularnewline
\ce{Na+} & -162.083275414 & -162.083265318 & -162.08327 &  -10.10\tabularnewline
\ce{Na-} & -162.293938337$^{a}$ & N/C & -162.29465 & N/C\tabularnewline
\ce{Mg} & -200.080754856 & -200.080744137 & -200.08074 &  -10.72\tabularnewline
\ce{Al+} & -242.159244333 & -242.159234002 & -242.15923 &  -10.33\tabularnewline
\ce{Si-} & -289.428472604 & -289.428460403 & -289.42846 &  -12.20\tabularnewline
\ce{P} & -341.278950750 & -341.278937694 & -341.27894 &  -13.06\tabularnewline
\ce{S+} & -397.752619769 & -397.752604276 & -397.75260 &  -15.49\tabularnewline
\ce{Cl-} & -460.295948313 & -460.295933897 & -460.29593 &  -14.42\tabularnewline
\ce{Ar} & -527.556251384 & -527.556235529 & -527.55624 &  -15.85\tabularnewline
\end{tabular}\caption{BHHLYP energies from a finite element calculation (present work) compared
to a previously reported Gaussian basis calculation (\citeref{Anderson2017}).
Cases where the \textsc{Erkale} calculations failed to converge are
marked with N/C. \protect \\
\protect \\
$^{a}$The energy still changed by $-1.6\times10^{-7}E_{h}$ going
from $R_{\infty}=60a_{0}$ to $R_{\infty}=80a_{0}$ (used value),
and would lower $\sim5\times10^{-8}E_{h}$ more by going from $R_{\infty}=80a_{0}$
to $R_{\infty}=100a_{0}$. \label{tab:BHHLYP-energies}}
\end{table}

\section{Summary and Conclusions\label{sec:Summary-and-Conclusions}}

We have described the implementation of a finite element program called
\textsc{HelFEM\citep{HelFEM}} for electronic structure calculations
on atoms in the framework of Hartree--Fock (HF) or Kohn--Sham density
functional theory. \textsc{HelFEM} is interfaced with the \textsc{Libxc}
library of exchange-correlation functionals,\citep{Lehtola2018} and
supports calculations at the local spin-density approximation (LDA),
generalized gradient approximation (GGA) and meta-GGA levels of theory,
including hybrid functionals. Calculations can be performed with fully
spin-restricted, spin-restricted open-shell, and spin-unrestricted
orbitals.

We have suggested an exponential radial grid for atomic calculations
that we have extensively tested in applications of the program on
noble elements. The exponential grid with $x=2$ was found to yield
faster convergence to the basis set limit than commonly used linear
or quadratic element grids.

Tests of the various kinds of elements supported by the program showed
that Lagrange interpolating polynomials (LIPs) or Legendre polynomials
outperform Hermite interpolating polynomials by a wide margin, and
the use of high-order Lagrange/Legendre polynomials yields the most
accurate results. 15-node LIPs with Lobatto nodes were chosen as the
default radial basis in \textsc{HelFEM}.

The capabilities of the program were demonstrated by calculations
of \ce{Li+} and \ce{Sr^{2+}} in an electric field, with the results
at the HF limit being in good agreement with literature values.\citealp{Kobus2015}
Furthermore, static dipole polarizabilities for \ce{Li+} and \ce{Sr^{2+}}
were reported with the LDA, PBE, PBEh, TPSS, and TPSSh functionals.

Finally, the program was used to study the accuracy of recently reported
atomic HF and DFT calculations employing Gaussian basis sets.\citealp{Anderson2017}
Cross-comparisons with results from the \textsc{Erkale} program\citealp{erkale,Lehtola2012}
showed that the errors in the Gaussian basis set are only up to a
few dozen microhartrees. Closed-shell symmetry-breaking effects were
identified in the calculations of \citeref{Anderson2017}, with energy
lowerings of several millihartrees.

\section*{Funding information}

This work has been supported by the Academy of Finland through project
number 311149. 

\section*{Acknowledgments}

I thank Dage Sundholm and Gregory Beylkin for discussions, and Dage
Sundholm and Pekka Pyykkö for comments on the manuscript. Computational
resources provided by CSC -- It Center for Science Ltd (Espoo, Finland)
and the Finnish Grid and Cloud Infrastructure (persistent identifier
urn:nbn:fi:research-infras-2016072533) are gratefully acknowledged.

\bibliographystyle{apsrev}
\bibliography{citations}

\end{document}